\documentclass[%
 reprint,
nofootinbib,
 amsmath,amssymb,
 aps,
]{revtex4-1}
\usepackage{graphicx}% Include figure files
\usepackage{dcolumn}% Align table columns on decimal point
\usepackage{bm}% bold math
\usepackage{hyperref}% add hypertext capabilities
\usepackage{booktabs}
\usepackage{cleveref}

\newcommand{\ie}{i.\,e.\ }
\newcommand{\eg}{e.\,g.\ }

\newcommand{\dd}{\textnormal{d}}

\usepackage[usenames, dvipsnames]{color}

\begin{document}

%\preprint{APS/123-QED}

\title{Improved Mixed Dark Matter Halo Model for Ultralight Axions}% Force line breaks with \\
%\thanks{A footnote to the article title}%

%\author{Ann Author}
% \altaffiliation[Also at ]{Physics Department, XYZ University.}%Lines break automatically or can be forced with \\
%\author{Second Author}%
% \email{Second.Author@institution.edu}
%\affiliation{%
% Authors' institution and/or address\\
% This line break forced with \textbackslash\textbackslash
%}%

%\collaboration{MUSO Collaboration}%\noaffiliation

\author{Sophie M. L. Vogt$^{1,2}$}
\email{s.vogt@physik.lmu.de}
\author{David J. E. Marsh$^3$}
\email{david.j.marsh@kcl.ac.uk}
\author{Alex Lagu\"e$^{4,5,6,7}$}
\email{alague@sas.upenn.edu}

\affiliation{
$^1$Institut f\"ur Astrophysik und Geophysik,  Universit\"at  G\"ottingen, Germany\\
$^2$Universit\"ats-Sternwarte, Fakult\"at für Physik, Ludwig-Maximilians-Universit\"at M\"unchen, Scheinerstr. 1, 81679 München, Germany \\
$^3$Theoretical Particle Physics and Cosmology, King’s College London, Strand, London WC2R 2LS, United Kingdom \\
$^4$Department of Physics and Astronomy, University of Pennsylvania, Philadelphia, PA 19104, USA\\
$^5$David A. Dunlap Department of Astronomy \& Astrophysics, University of Toronto, Canada\\
$^6$Canadian Institute for Theoretical Astrophysics, University of Toronto, Canada \\
$^7$Dunlap Institute for Astronomy and Astrophysics, University of Toronto, Canada}

\date{\today}% It is always \today, today,
             %  but any date may be explicitly specified

\begin{abstract}
We present a complete halo model for mixed dark matter composed of cold dark matter (CDM) and ultralight axion-like particles (ULAs). Our model treats ULAs as a biased tracer of CDM, in analogy to treatments of massive neutrinos and neutral hydrogen. The model accounts for clustering of ULAs around CDM host halos, and fully models the cross correlations of both components. The model inputs include the ULA Jeans scale, and soliton density profile. Our model can be used to predict the matter power spectrum, $P(k)$, on non-linear scales for sub-populations of ULAs across the mass range $10^{-33}\text{ eV}\leq m\leq 10^{-21}\text{ eV}$, and can be calibrated against future mixed DM simulations to improve its accuracy. The mixed DM halo model also allows us to assess the importance of various approximations. The code is available at \url{https://github.com/SophieMLV/axionHMcode}.

%\begin{description}
%\item[Usage]
%Secondary publications and information retrieval purposes.
%\item[Structure]
%You may use the \textsc{description} environment to structure your abstract;
%use the optional argument of the \verb+\item+ command to give the category of each item. 
%\end{description}
\end{abstract}

%Use showkeys class option if keyword
                              %display desired
\maketitle

%\tableofcontents

%%%%%%%%%%%%%%%%%%%%%%%%%%%%%%%%%%%%%%%%%%%%%%%%%%%%%%%%%%%%%%
%%%%%%%%%%%%%%%%%--------INTRODUCTION-------%%%%%%%%%%%%%%%%%%
%%%%%%%%%%%%%%%%%%%%%%%%%%%%%%%%%%%%%%%%%%%%%%%%%%%%%%%%%%%%%%
\section{\label{sec:intro}Introduction}

The standard $\Lambda$ cold dark matter ($\Lambda$CDM) cosmological model is highly successful at describing the Universe~\cite{Planck:2018nkj,Planck:2018vyg}, yet the microphysical nature of DM remains a mystery. DM candidates' masses span orders of magnitude from the heaviest primordial black holes to the lightest sub-eV particles (see Ref.~\cite{AlvesBatista:2021gzc}). Ultralight axion-like particles (ULAs) in the mass range $10^{-33}\text{ eV}\leq m\leq 10^{-21}\text{ eV}$ are the lightest DM candidates, yet they are highly constrained, with current data allowing only a few percent contribution to the energy density~\cite{Hlozek:2014lca,Hlozek:2017zzf,Kobayashi:2017jcf,Lague:2021frh}. On the other hand, such particles are abundant in the string theory landscape~\cite{Witten:1984dg,Svrcek:2006yi,Conlon:2006tq,Arvanitaki:2009fg}, as has been shown recently in increasingly explicit compactifications~\cite{Mehta:2021pwf,Demirtas:2021gsq,Cicoli:2021gss}. Thus, one expects a sub-population of the cosmic DM to be composed of ULAs. Next generation cosmological surveys will increase the precision of ULA searches by orders of magnitude, and could detect a sub-population of ULAs as small as $\mathcal{O}(0.1\%)$~\cite{Hlozek:2016lzm,bauer_biased_tracer_H1,Farren:2021jcd,CMB-S4:2016ple,Dvorkin:2022bsc,Flitter:2022pzf}. 

Exploiting to the full the next generation of cosmological data requires considering cosmological statistics beyond the linear regime, and parameter estimation from such data requires the non-linear physics to be computable rapidly. Non-linear physics can be modelled extremely accurately using $N$-body and hydrodynamical simulations (e.g. Refs.~\cite{sims_review}), but such methods are not appropriate for parameter estimation. Two methods that allow for fast estimation of non-linear observables are \emph{emulators} and the \emph{halo model} (HM). Emulators are machine learning inspired methods to interpolate accurately on large grids of simulations, and have been employed to great effect in a variety of cases in $\Lambda$CDM and beyond~\cite{heitmann_coyote_emulator,heitmann_miratitan_emulator,Rogers:2018smb,Pedersen:2020kaw}.

The HM~\cite{halo_model_cooray_sheth}, on the other hand, has the advantage of maintaining the speed of an emulator while being physics-inspired, and elements of the HM can be calibrated on a smaller number of simulations. \textsc{HMCode} is one example of the HM that competes with emulator accuracy on the power spectrum, $P(k)$, with a small number of parameters calibrated to simulation~\cite{HMCode_mead_2015}. As an example, \textsc{HMCode} has been successfully calibrated on fixed cosmology models of active galactic nucleus (AGN) feedback, and then used to predict the cosmological parameter dependence~\cite{mead_hmcode2020}. The method of accurate HM calibration is also used by the \emph{Euclid} consortium to model the non-linear power spectrum for cosmological parameter estimation from clusters~\cite{Euclid:2022dbc}. Being physics-inspired, the HM is thus highly suitable to apply to beyond $\Lambda$CDM models, where suitable simulations might be limited in dynamic range, sparse in parameter space, or non-existent.

In the following, we develop the mixed DM halo model for ULAs, greatly improving on the early work in Ref.~\cite{marsh_silk_cut_off_mass}. Our mixed HM draws inspiration from the treatment of neutrino clustering around halos~\cite{LoVerde:2013lta,massara_MDM_halo_model}, and the neutral hydrogen HM~\cite{Padmanabhan:2016odj,bauer_biased_tracer_H1}. Our model contains physically motivated elements that are suitable to calibrate against mixed DM simulations, when they reach the appropriate scales. 

Throughout this work we refer to the ultralight sub-component of DM as ``axions'', since the existence of such particles is motivated in the string theory landscape~\cite{Arvanitaki:2009fg,Mehta:2021pwf}. However, our treatment makes no assumption about the $CP$ properties of the ultralight component, so it is equally valid for scalars (e.g. dilaton-like fields~\cite{Hamaide:2022rwi}). Ultralight bosonic components with more degrees of freedom, such as a complex scalar~\cite{Li:2013nal} or vector~\cite{Gorghetto:2022sue}, are amenable to the same modelling (possessing a Jeans scales, and forming solitons) but differ in details that require treating separately (e.g. different linear transfer function and early Universe behaviour).

This paper is organised as follows. In Section~\ref{sec:ax_phy}, we briefly outline the relevant aspects of ULA cosmology. In Section~\ref{sec:HM} we develop the theory of the HM in the case of mixed DM. In Section~\ref{sec:results} we give the results of our model in terms of the non-linear power spectrum. We conclude in Section~\ref{sec:discussion}. Appendix~\ref{app:HMCode_Mead} discusses the modifications to the HM of \textsc{HMCode}, which we adopt as baseline. Appendix~\ref{app:neutrinos} shows the mixed DM halo model for massive neutrinos, following Ref.~\cite{massara_MDM_halo_model}, and assesses the effect of neutrino clustering in halos compared to the approximate treatment in \textsc{HMCode} (we find differences of order a few percent at $k\approx 1\, h\text{ Mpc}^{-1}$). Appendix~\ref{app:convergence} shows some convergence checks on our numerical implementation. We take baseline cosmological parameters as shown in Table~\ref{tab:param_cosmo}.
\begin{table}
    \centering
    \begin{tabular}{cc}
    \hline
    Parameter & Fiducial value \\
    \hline
    $h$ & 0.674 \\
    $\omega_{\mathrm{d}}$ & 0.12\\
    $\omega_{\mathrm{b}}$ & 0.02237 \\
    $f_{\mathrm{ax}}$ & 0.1 \\
    $N_{\textnormal{eff}}$ & 3.046 \\
    $A_s$ & $2.1 \times 10^{-9}$ \\
    $n_s$ & 0.97\\
    $k_{\textnormal{piv}}$\, [$\textnormal{Mpc}^{-1}$] & 0.05 \\
    \hline
    \end{tabular}
    \caption{Fiducial cosmological parameters and their values for our flat mixed dark mater (MDM) cosmology to compute the linear and non-linear power spectrum. $h$ is the Hubble parameter; $\omega_{\mathrm{d}}$ and $\omega_{\mathrm{b}}$ are the dark matter and baryon reduced density parameters, respectively, $f_{\mathrm{ax}} = \omega_{\mathrm{ax}} / \omega_{\mathrm{d}}$ the axion fraction , $N_{\textnormal{eff}}$ is the effective number of neutrinos, $A_s$ is the scalar amplitude, $n_s$ is the scalar spectral index and $k_{\textnormal{piv}}$ is the pivotal scale.}
    \label{tab:param_cosmo}
\end{table}

%%%%%%%%%%%%-----axion physics-----%%%%%%%%%%%%%%%%%
\section{Axion Physics}
\label{sec:ax_phy}
The halo model aims to predict the non-linear power spectrum given as input the linear power spectrum. We begin this section describing linear perturbation theory for ULAs, and key results (many more details can be found in e.g. Refs.~\cite{marsh_axion_cosmo,Hlozek:2014lca} and references therein). We then briefly describe the non-linear theory that underlies the mixed CDM-ULA simulations of Refs.~\cite{bodo_simulation,Lague_in_prep}, which gives the form of the halo density profiles we adopt.

\subsection{\label{subsec:ax_PT}Perturbation Theory}
Relic ULAs can be produced by the misalignment mechanism of a classical field~\cite{Preskill:1982cy,Abbott:1982af,Dine:1982ah}. The scalar field $\phi$ obeys the classical Klein-Gordon-Equation:
    \begin{equation}
        \Box\phi- m_{\textnormal{ax}}^2\phi = 0\,,
        \label{eqn:full_eom}
    \end{equation}
where $\Box$ is the D'Alembertian operator for spacetime metric $g_{\mu\nu}$, and we have assumed small displacements from the vacuum. Consider cosmological perturbation theory in the Newtonian gauge~\cite{Ma:1995ey}. At zeroth order in perturbations, the axion field obeys the ordinary differential equation:
    \begin{equation}
        \label{eq:klein_gordon_ax}
        \ddot{\phi}_0+3H\dot{\phi}_0+ m_{\textnormal{ax}}^2\phi_0 = 0\,,
    \end{equation}
where $\phi_0$ is a function only of time. Here $H$ denotes the Hubble parameter defined by $H = \dot a/a$, dots denote derivatives with respect to cosmic time, $t$, and $a$ is the cosmic scale factor.

Eq.~\eqref{eq:klein_gordon_ax} is the equation for a damped harmonic oscillator. As long as $m \ll H$ (early times) the field is ``frozen'', \ie the harmonic oscillator is overdamped. As the universe evolves $H \sim m_a$ and the field starts to oscillate. Since the axion field is overdamped to begin with, the initial conditions at time $t_i$ can be set to $\phi_0(t_i) = \phi_i$ and $\dot\phi(t_i) = 0$.

From the full Klein-Gordon-Equation, Eq.~\eqref{eqn:full_eom}, we can compute the equation of motion for the axion overdensity $\delta_{\textnormal{ax}}$ and thus find the (linear) matter power spectrum with ULAs. In the Newtonian gauge the perturbed field equation reads \cite{marsh_axion_cosmo}
    \begin{equation}
        \label{eq:pert_K_G_ax}
        \delta \phi'' + 2\mathcal{H}\delta \phi' + (k^2 + m_{\textnormal{ax}}^2 a^2)\delta\phi = (\Psi' + 3\Phi)\phi' - 2m_{\textnormal{ax}}^2 a^2\phi \Psi \,,
    \end{equation}
with the two Newtonian scalar potentials $\Phi$ and $\Psi$ and primes denotes derivative with respect to conformal time $\tau$ ($\textnormal{d}t = a(t) \textnormal{d}\tau$). We have assumed $\delta_{\textnormal{ax}}\ll 1$ and $\delta\phi\ll\phi_0$.

When the axion field $\phi$ starts to oscillate (when the damping term in Eq.~\ref{eq:klein_gordon_ax} becomes less than the mass, i.e. $H\lesssim m_{\rm ax}$) one can find with the WKB-Ansatz an expression for the effective axion sound speed \cite{sound_speed_ax}:
    \begin{equation}
        \label{eq:sound_speed_ax}
        c_{\textnormal{s}}^2 = \frac{\frac{k^2}{4m_{\textnormal{ax}}^2a^2}}{1+ \frac{k^2}{4m_{\textnormal{ax}}^2 a^2} } \,.
    \end{equation}
With the sound speed we obtain for the equation of motion for the axion overdensity in the Newtonian gauge \cite{marsh_axion_cosmo}
    \begin{equation}
        \label{eq:eom_ax}
        \delta ''+ \mathcal{H}\delta ' + c_{\textnormal{s}}^2k^2\delta - 3\mathcal{H}\Phi' +k^2\Psi - 3\Phi'' = 0 \, .
    \end{equation}

Compared with the equation of motion for the CDM overdensity, Eq.~\eqref{eq:eom_ax} has an extra term proportional to the sound speed. This term goes to zero as $k \to 0$ and thus ULAs behave like CDM on large scales. For small scales the sound speed is no longer negligible and the ULA overdensity oscillates instead of growing. This behaviour is different from CDM, since the CDM overdensity has a growing solution on all scales. The Jeans scale is the approximate scale where the transition between the two regimes takes place  \cite{Khlopov:1985jw}.

If $k \ll m_{\textnormal{ax}} a$ the sound speed reads:
    \begin{equation}
        \label{eq:sound_speed_approx_ax}
        c_{\textnormal{s}}^2 = \frac{k^2}{4m_{\textnormal{ax}}^2 a^2}  \,.
    \end{equation}
Then the Jeans scale reads \cite{marsh_axion_cosmo}
    \begin{equation}
        \label{eq:jeans_ax}
        k_{\textnormal{J}} = 66.5 a^{1/4} \left( \frac{\Omega_{\textnormal{ax}}h^2}{0.12} \right)^{1/4} \left( \frac{m_{\textnormal{ax}}}{10^{-22}\,\textnormal{eV}} \right)^{1/2}\,\textnormal{Mpc}^{-1}\,,
    \end{equation}
with $\bar \rho_{\textnormal{ax}} = \rho_{{\textnormal{ax}}, 0}a^{-3}$. Suppression of the ULA power spectrum relative to CDM begins at the Jeans scale at matter-radiation equality, $k_{\rm J,eq}$~\cite{halo_jeans_scale,bauer_biased_tracer_H1}.

With the above derivation we can summarise the effect on the matter power spectrum if the DM is a mixture of CDM and ULAs:
    \begin{enumerate}
        \item The axion overdensity behaves exactly as the CDM overdensity for large scales, \ie $k < k_{\textnormal{J}}$ and thus there is no change in the matter power spectrum for these scales.
        \item A suppression by a factor of $(1- \Omega_{\textnormal{ax}}/\Omega_{\textnormal{m}})^2$ for $k > k_{\textnormal{J}}$ because the axion oscillates on these scales and the overdensity is negligible compared to $\delta_{\textnormal{c}}$ and $\delta_{\textnormal{b}}$.
        \item An extra suppression again for $k > k_J$ due to the fact that the axion field suppresses the growing solution of the cold dark matter field, $\delta_{\textnormal{c}} \propto a^{\frac{-1 + \sqrt{25-24 \Omega_{\textnormal{ax}}/\Omega_{\textnormal{m}}}}{4}}$.
    \end{enumerate}
These effects lead to the presence of a step-like feature in the power spectrum, which begins near the Jeans scale at equality, and has an amplitude determined by the ULA fraction relative to CDM (see Refs.~\cite{Amendola:2005ad,Marsh:2010wq,Arvanitaki:2009fg}).

\begin{figure}[t!]
        \centering
        \includegraphics[width=\linewidth]{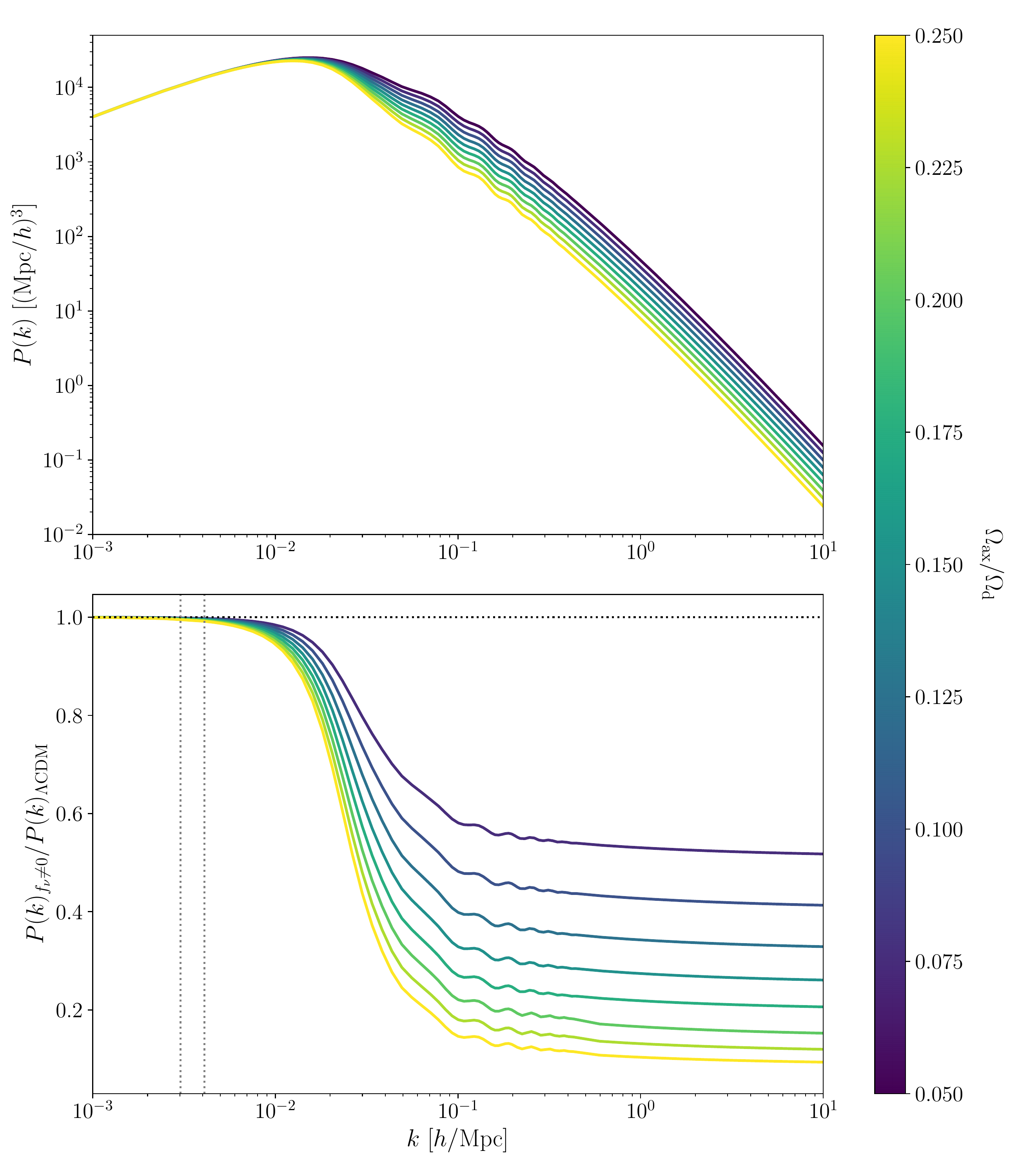}
        \caption{The linear matter power spectrum computed with \textsc{axionCAMB} with axions of a mass $m_{\textnormal{ax}} = 10^{-28}$\,eV and an axion fraction $0.05<\Omega_{\textnormal{ax}}/\Omega_{\textnormal{d}} \leq 0.25$  on the top. On the bottom the ratio between the different matter power spectra with axions to a $\Lambda$CDM power spectrum are shown. The dashed, vertical grey lines indicate $k_{\textnormal{J}}$ for the lowest axion fraction (right line) and highest fraction (left line) at $a_{\textnormal{eq}}$.}
        \label{fig:lin_PS_ax_and_ratio}
    \end{figure}

To illustrate the described behaviour, Fig.~\ref{fig:lin_PS_ax_and_ratio} shows the matter power spectrum with an axion mass $m_{\textnormal{ax}} = 10^{-28}$\,eV and an axion fraction $0.05<\Omega_{\textnormal{ax}}/\Omega_{\textnormal{d}} \leq 0.25$ (top) and the ratio to the power spectrum in a $\Lambda$CDM universe (bottom) computed with the Boltzmann code \textsc{axionCAMB}~\cite{axionCAMB}, which solves first order perturbation theory for ULAs in synchronous gauge coupled to all other $\Lambda$CDM components, taken here with adiabatic initial conditions in a radiation dominated Universe. 

\subsection{Non-Linear Regime}
\label{subsec:sims}

Taking Eq.~\eqref{eqn:full_eom}, we insert the WKB ansatz: 
\begin{equation}
    \phi = \frac{1}{\sqrt{2 m_{\rm ax}^2}}\left( \psi e^{im_{\rm ax}t}+ \psi^* e^{-im_{\rm ax}t}\right)\, ,
\end{equation}
working to linear order in perturbations of the metric, and taking the non-relativistic limit, we find the Schr\"{o}dinger-Poisson equations describing the complex field $\psi$:
\begin{align}
    i\frac{\partial}{\partial{t}} \psi&=-\frac{1}{2m}\nabla^2\psi + m_{\rm ax} \Phi\psi,
\\
\nabla^2\Phi &=4 \pi G \left[m_{\rm ax}\left(|\psi|^2-\langle|\psi|\rangle^2\right)+\delta\rho_{\rm f}\right],
\end{align}
where $G$ is Newton's constant, $\nabla$ is the flat space Laplacian, angle brackets denote spatial average, and $\delta\rho_{\rm f}$ denotes fluid density perturbations (e.g. CDM and baryons). This system of equations makes no assumptions about the smallness of density perturbations, and can be used to evolve ULAs into the non-linear regime while fully capturing wavelike dynamics~\cite{Widrow:1993qq,schive_soliton_1,soliton_velocity}.

The key features of ULAs in the non-linear regime are: the persistence of the Jeans scale, i.e. effective pressure opposing gravitational collapse, leading to the existence of stable solitons~\cite{schive_soliton_1}, condensation due to wave scattering~\cite{Levkov:2018kau}, relaxation~\cite{Mocz:2017wlg,Dalal:2022rmp}, and interference effects in the multi streaming regime~\cite{schive_soliton_1,Mocz:2019pyf,Gough:2022pof}. Many different numerical approaches have been adopted to solve the Schr\"{o}dinger-Poisson equations, with different methods being useful in different regimes of physical interest (e.g. Refs.~\cite{Nori:2018hud,Veltmaat:2016rxo,Chen:2020cef}). 

The code \textsc{axioNyx} (based on the \textsc{Nyx} code~\cite{Almgren:2013sz}) was presented and released in Ref.~\cite{bodo_simulation}, and has been used to simulate mixed ULA DM. These simualtions inform our MDM halo model. %In \textsc{axioNyx}, the ULA fluid is evolved on large scales using the Gaussian beams method~\cite{Schwabe:2021jne}, which tracks the phase of the wavefunction as evolved by quasi-particles. On increasing levels of refinement around halos, the full Schr\"{o}dinger-Poisson equations are evolved for the ULAs, which capture soliton formation and wave interference, while the CDM is evolved using an $N$-body solver (which in \textsc{Nyx} works on a grid). 
The MDM simulations of Ref.~\cite{bodo_simulation} studied spherical collapse, and noted that the ULAs follow CDM on large scales, while forming solitons on smaller scales. Ref.~\cite{Lague_in_prep} has extended the study of mixed DM to cosmological initial conditions, although box sizes limit the ability to measure the power spectrum and halo mass function. In order to properly evolve the ULA wavefunction, the simulation grid spacing must be smaller than half the de Broglie wavelength $\lambda_{\rm dB}\propto 1/m_{\rm ax}v$ of the ULAs~\cite{May_FDM_cosmo_sims}. In the presence of a dominant CDM component ($\Omega_{\textnormal{ax}}/\Omega_{\textnormal{d}}\lesssim 0.5$), the potential wells in which the wavefunction evolves become steeper, which increases the ULA velocity dispersion and decreases its wavelength. The mixed DM simulations thus require a higher resolution than their pure ULA counterparts. This makes running large cosmological volumes involving solving the full Schr\"odinger-Poisson with the MDM system difficult. Alternative MDM simulation algorithms based on a particle treatment of the ULAs (such as the one implemented in \textsc{ax-gadget}~\cite{Nori:2018hud}) could 
capture the large-scale dynamics of MDM and be a complement to the spectral and finite difference methods at a reduced computational cost.

%%%%%%%%%%%%%%%%%%%%%%%%%%%%%%%%%%%%%%%%%%%%%%%%%%%%%%%%%%%%%%
%%%%%%%%%%%%%%--------Theory Halo Model-------%%%%%%%%%%%%%%%%
%%%%%%%%%%%%%%%%%%%%%%%%%%%%%%%%%%%%%%%%%%%%%%%%%%%%%%%%%%%%%%
\section{\label{sec:HM}Halo Model Theory}
In this section we explore the halo model (HM)  as a theoretical approach to calculate the non-linear matter power spectrum. We first introduce the halo model in a standard $\Lambda$CDM cosmology, as reviewed by Ref.~\cite{halo_model_cooray_sheth}, in Section~\ref{subsec:LCDM_HM}. We then extend the HM to a mixed DM cosmology in Section~\ref{subsec:MDM_HM}, following a biased tracer treatment. The specific HM ingredients in the case of ULAs are shown in Section~\ref{subsec:ax_dens_profile}.

%%%%%%%%%%%%-----classical halo model-----%%%%%%%%%%%%%%%%%
\subsection{The \texorpdfstring{$\Lambda$}{L}CDM Halo Model}
\label{subsec:LCDM_HM}
%The halo model formulae from \cite{halo_model_cooray_sheth}
In a universe where all matter is assumed to be cold we can assume that all matter is contained in halos and thus the matter power spectrum is the sum of two terms, the one halo term $P^{\mathrm{1h}}$ (correlation in the same halo) and the two halo term $P^{\mathrm{2h}}$ (correlation between two different halos) \cite{halo_model_cooray_sheth}
    \begin{equation}
        \label{eq:non_lin_power_spec}
        P(k) = P^{\mathrm{1h}}(k) + P^{\mathrm{2h}}(k) \, ,
    \end{equation}
where the one halo and two halo terms have the following forms
    \begin{equation}
        \label{eq:one_halo_term}
        P^{\mathrm{1h}}(k) =\frac{1}{\Bar{\rho}^2} \int \mathrm{d}M M^2 n(M) \vert \Tilde{u}(k, M) \vert^2 \, ,
    \end{equation}
and
    \begin{equation}
        \label{eq:two_halo_term}
        P^{\mathrm{2h}}(k) = P^{\mathrm{L}}(k) \left[ \frac{1}{\Bar{\rho}} \int \mathrm{d}M M n(M) b(M) \vert \Tilde{u}(k, M) \vert \right]^2 \, .
    \end{equation}
The halo mass function (HMF) is $n(M)$, $b(M)$ is the halo bias, $\tilde{u}(k,M)$ is the Fourier transform of the halo density profile and $P^{\mathrm{L}}(k)$ is the linear matter power spectrum. These are defined in the following.

The improper integral in the two halo term, Eq.~\eqref{eq:two_halo_term}, should go to unity if $k \to 0$ because matter is unbiased with respect to itself~\cite{halo_model_cooray_sheth}. So, to ensure the correct behaviour for low $k$’s the mass interval should be chosen large enough. This numerical problem was studied in Ref.~\cite{correction_for_correct_two_halo_term} and was solved by adding some correction factors. The implications on the halo model are discussed in Appendix~\ref{app:convergence}.

The Fourier transform of a (normalised) radial density profile $u(r, M, z) = \rho(r, M, Z)/M$ is given by
    \begin{equation}
        \label{eq:fourier_tranform_dens_profile}
        \tilde u (k, M, z) = 4 \pi   \int_0^{r_{\textnormal{v}}}u(r, M, z) \frac{\sin(kr)}{kr}r^2 \dd r \, .
    \end{equation}
Here the profile is truncated at the virial radius $r_{\textnormal{v}}$, \ie we assume that the density profile is zero for $r > r_{\textnormal{v}}$ and that the mass of the halo is given by
    \begin{equation}
    \label{eq:rel_mass_vir_radius}
        M= \frac{4\pi}{3} \Bar{\rho} \Delta_{\textnormal{v}}(z) r_{\textnormal{v}}^3 \, ,
    \end{equation}
with $\Delta_{\textnormal{v}}$ the virial overdensity (see below). Still assuming a $\Lambda$CDM universe the density profile of a dark matter halo can be described by the Navarro-Frenk-White (NFW) profile~\cite{NFW_profile} 
   \begin{equation}
        \label{eq:NFW_profile}
        \rho_{\textnormal{NFW}}(r, M) = \frac{\rho_{\mathrm{char}}}{r/r_{\textnormal{s}} \left( 1 + r/r_{\textnormal{s}} \right)^2}\, ,
    \end{equation}
with $r_{\textnormal{s}}$ the scale radius and $\rho_{\textnormal{char}}$ the characteristic density of the profile which ensures that the integral over the NFW density profile gives the enclosed mass in Eq.~(\ref{eq:rel_mass_vir_radius}) and can be computed to be: 
    \begin{equation}
    \label{eq:rho_char_value}
        \rho_{\mathrm{char}} = \rho_{\textnormal{crit}} \frac{\Omega_{\textnormal{m}}(z) \Delta_{\textnormal{v}} c^3}{3  f(c)} \, ,
    \end{equation}
with $f(x) = -\frac{x}{1+x} + \ln(x+1)$ and $c = r_{\textnormal{v}}/r_{\textnormal{s}}$ is the halo concentration parameter.

Evaluating the Fourier transformation, Eq.~\eqref{eq:fourier_tranform_dens_profile}, with the NFW-profile gives~\cite{fourier_NFW_profile}:
    \begin{align}
        \Tilde{u}(k, M, z) &= \frac{1}{f(c)}  \bigg( \cos(b)(\mathrm{Ci}(b + kr_{\textnormal{v}}) - \mathrm{Ci}(b))  \nonumber \\
        \label{eq:dens_profile_kspace_computed}
        & \hspace{0.1cm} + \sin (b)(\mathrm{Si}(b+kr_{\textnormal{v}})  \left. - \mathrm{Si}(b)) - \frac{\sin(k r_{\textnormal{v}})}{b + k r_{\textnormal{v}}} \right) \,.
    \end{align}
Here $b = k r_{\textnormal{v}} /c$ and Si$(x)$ and Ci$(x)$ are the sine and cosine integrals. To calculate the halo mass function (HMF) $n(M)$ and the halo bias $b(M)$ we need the variance of the linear power spectrum
    \begin{align}
        \label{eq:def_sigma}
        &\sigma(R)^2 = \frac{1}{2 \pi^2} \int_0^\infty P^{\mathrm{L}}(k, z) \tilde{W}(Rk)^2 k^2 \dd k \\
        & \hspace{1cm} \mathrm{with} \quad \tilde{W}(x) = \frac{3}{x^3} (\sin\, x - x \cos\, x) . \nonumber
    \end{align}
Here we assumed a spherical top hat window function, $W$, in real space. The variance above can be transformed to a function of the halo mass by $M = 4/3 \pi \bar{\rho} R^3$ .
The halo mass function, $n(M)$, is given by~\cite{Press:1973iz}:
    \begin{equation}
        \label{eq:HMF}
         n(M, z ) = \frac{1}{M} \frac{\dd \tilde{n}}{\textnormal{dln}M} = \frac{1}{2} \frac{\Bar{\rho}(z)}{M^2} f(\nu) \left \vert \frac{\mathrm{dln}\sigma^2}{\mathrm{dln} M} \right \vert \,.
    \end{equation}
%
%Some authors define $\tilde{n}$ from Eq.~\ref{eq:HMF} as the halo mass function and not $\dd \tilde{n}/ \dd M$ as we do.
where $\tilde{n}$ is the halo number density, $\nu=\delta_{\rm crit}(z)/\sigma(M, z)$, with $\delta_{\rm crit}$ the critical linear density threshold for halo collapse, and $f(\nu)$ is the multiplicity function, which for ellipsoidal collapse is given by the Sheth-Tormen (ST) multiplicity function~\cite{sheth_tormen_multiplicity_function}
    \begin{equation}
    \label{eq:ST_muliplicity}
        f_{\textnormal{ST}}(\nu) = A \sqrt{\frac{2}{\pi}} \sqrt{q}\nu (1 + (\sqrt{q}\nu)^{-2p})e^{-\frac{q\nu^2}{2}}\, ,
    \end{equation}
with $A = 0.3222,\ p = 0.3,\ q = 0.707$. The last term in Eq.~\eqref{eq:HMF} can be calculated by inserting the definition of the variance $\sigma(M, z)$ and using $x \mathrm{d}x = \mathrm{dln}x$:
    \begin{align}
        \frac{\mathrm{d\, ln}\sigma^2}{\mathrm{d\, ln} M}&= \frac{3}{\sigma^2 R^4 \pi^2} \int_0^\infty \mathrm{d}k \frac{P^{\mathrm{L}}(k)}{k^2} \Tilde{I}(k, R) \\
        \Tilde{I}(k, R) &= (\sin(kR) - kR\cos (kR)) \nonumber \\
        & \hspace{0.1cm}\left[ \sin(kR) \left(1-\frac{3}{(kr)^2}\right) + \frac{3}{kR} \cos(kR) \right ].
    \end{align}

The halo bias can computed with the theory of \cite{sheth_tormen_multiplicity_function} to be
    \begin{equation}
        \label{eq:halo bias}
        b(m, z) = 1+ \frac{1}{\delta_{\rm crit}} \left ( q\nu^2 - 1 + \frac{2p}{1+(\sqrt{q}\nu )^{2p}} \right)\,.
    \end{equation}
It remains to specify two quantities: the virial overdensity $\Delta_{\mathrm{v}}$, and the concentration parameter, $c(M)$, of the NFW profile. The first one is found for a $\Lambda$CDM cosmology by simulations and a fitting formula was constructed which reads \cite{virial_overdensity_formula}
    \begin{equation}
        \label{eq:Delta_vir_LCDM}
        \Delta_{\textnormal{v}}(z) =\frac{18\pi^2+82x - 39x^2}{\Omega_{\textnormal{m}}(z)}  \,,
    \end{equation}
with $x = \Omega_{\textnormal{m}}(z) - 1$.

To find a functional form of the concentration parameter we follow Ref.~\cite{concentration_parameter} and assign for each halo of mass $M$ at redshift $z$ a formation redshift $z_{\textnormal{f}}$ by the equation
    \begin{equation}
        \label{eq:def_z_form}
        M_*(z_{\textnormal{f}}) = 0.01M \, ,
    \end{equation}
where $M_*(z)$ is the collapsing mass defined by $\sigma(M_*(z)) = \delta_{\rm crit}(z)$. Here we assumed that $\delta_{\rm crit}$ is constant and is given by $\delta_{\rm crit} = 1.686$. An equivalent definition for the formation redshift is given by:
    \begin{equation}
        \label{eq:def_z_form_2}
       \frac{D(z_{\textnormal{f}})}{D(z)} \sigma(0.01M, z) = \delta_{\rm crit}(z) \, ,
    \end{equation}
here $D(z)$ is the linear growth factor $D(z)/D(0) = \delta_L(\mathbf{x}, z)/\delta_L(\mathbf{x}, 0)$. With the equations above and fits to simulations the concentration parameter is \cite{concentration_parameter}
    \begin{equation}
        \label{eq:concen_param}
        c(M, z) = 4 \left( \frac{1+z_{\textnormal{f}}(M, z)}{1+z} \right)\, .
    \end{equation}
Note that the defining equation for the formation redshift Eq.~\eqref{eq:def_z_form_2} can give $z_{\textnormal{f}} < z$. In this case the authors of \cite{concentration_parameter} forced $z_{\textnormal{f}} = z$ and the concentration parameter is $c(M, z) = 4$. Thus, the prefactor in Eq.~\eqref{eq:concen_param} is called the minimum halo concentration.

%%%%%%%%%%%%-----MDM halo model-----%%%%%%%%%%%%%%%%%
\subsection{\label{subsec:MDM_HM}Mixed Dark Matter Halo Model}

In any MDM model where the clustering properties of the components differ, the HM is more complex. As we saw in Section~\ref{subsec:ax_PT} ULAs cannot cluster on small scales, \ie smaller than the Jeans scale and thus the assumption that all matter is contained in halos is no longer valid. We also expect that the Schrodinger-Poisson equation will cause the internal structure of ULA density profiles to depart from the CDM profile on small scales. We consider MDM models where the ULA component is a sub-dominant component of the DM. We follow closely the model for massive neutrinos plus CDM (mixed hot and cold DM) of  Ref.~\cite{massara_MDM_halo_model}, which we reproduce in more detail in Appendix~\ref{app:neutrinos} (the authors of this paper developed a general approach of a MDM halo model with a sub-domiant component as a biased tracer of the CDM and thus their method is applicable in our analysis).

One of the main assumption in Ref.~\cite{massara_MDM_halo_model} is that the non-cold component of the DM, \ie massive neutrinos, clusters in the potential wells of the cold matter halos, see \eg Refs.~\cite{neutrino_clustering_around_chalos_1,neutrino_clustering_around_chalos_2,LoVerde:2013lta}. Thus the non-cold component is treated as a biased tracer of the cold matter (for simplicity we neglect the non-trivial clustering of baryons on the scales of interest). A similar biased tracer technique is used in the halo model of neutral hydrogen, \eg see Refs.~\cite{villaescusa_biased_tracer_HI,bauer_biased_tracer_H1}, which can match well to simulations. ULAs are expected to behave in the same way: tracing the dominant cold matter in halos above the Jeans scale with some characteristic internal density profile. 

Since the matter power spectrum is proportional to the matter overdensity squared, $P(k) \propto \delta_{\textnormal{m}}^2$, the total matter overdensity in a MDM cosmology is a sum of the cold matter $\delta_{\textnormal{c}}$ and ULAs $\delta_{\textnormal{a}}$:
    \begin{equation}
        \label{eq:matter_mix_delta}
        \delta_{\textnormal{m}} = \frac{\Omega_{\textnormal{c}}}{\Omega_{\textnormal{m}}} \delta_{\textnormal{c}} + \frac{\Omega_{\textnormal{a}}}{\Omega_{\textnormal{m}}} \delta_{\textnormal{a}}\,,
    \end{equation}
with $\Omega_{\textnormal{c}}$ the weighted sum of the density parameters of CDM and baryons and $\Omega_{\textnormal{a}}$ the density parameter of axions. Then the power spectrum reads:~\footnote{Note that we consider only adiabatic perturbations, where the cross correlation $P_{\rm c,a}\propto \delta_{\rm c}\delta_{\rm a}$.}
    \begin{equation}
        \label{eq:PS_MDM}
        P_{\textnormal{m}}(k) = \left(\frac{\Omega_{\textnormal{c}}}{\Omega_{\textnormal{m}}} \right)^2 P_{\textnormal{c}}(k) + \frac{2\Omega_{\textnormal{c}} \Omega_{\textnormal{a}}}{\Omega^2_{\textnormal{m}}} P_{\textnormal{c,a}}(k) + \left( \frac{\Omega_{\textnormal{a}}}{\Omega_{\textnormal{m}}} \right)^2 
        P_{\textnormal{a}}(k)\,,
    \end{equation}
where $P_{\textnormal{c}}(k)$, $P_{\textnormal{c,a}}(k) $ and $P_{\textnormal{a}}(k)$ are the cold matter, cross and ULA power spectrum respectively. We can already see from the prefactors of the above equation that the main contribution to the non-linear total matter power spectrum comes from the cold part, because we assume $\Omega_{\textnormal{a}} \ll \Omega_{\textnormal{c}}$.

For the cold matter power spectrum we can use the standard halo model described in Section~\ref{subsec:LCDM_HM} and thus $P_{\textnormal{c}}(k)$ is given by Equation~\eqref{eq:non_lin_power_spec}, since for cold matter we still assume that all (cold) matter is bound into halos.

Next we have to find expressions for the cross and non-cold non-linear power spectrum. As explained above ULAs have a component that cannot cluster and thus evolve approximately linearly. Thus, the axion overdensity has a component in halos and a linear component, and can be written as:
    \begin{equation}
        \label{eq:delta_non_cold}
        \delta_{\textnormal{a}} = F_\textnormal{h} \delta_{\textnormal{h}} + (1-F_\textnormal{h}) \delta_{\textnormal{L}}\,.
    \end{equation}
Here $\delta_{\textnormal{h}}$ and $\delta_{\textnormal{L}}$ are the halo and linear overdensities, respectively, and $F_\textnormal{h}$ is the ULA fraction in halos, \ie $F_\textnormal{h} \in [0, 1]$. With this overdensity the cross and ULA power spectra read~\cite{massara_MDM_halo_model}
    \begin{align}
        \label{eq:cross_PS}
        &P_{\textnormal{c, a}}(k) = F_\textnormal{h} P_\textnormal{c, a}^{\textnormal{h}}(k) + (1- F_\textnormal{h}) \sqrt{P_\textnormal{c}(k) P_\textnormal{a}^\textnormal{L} (k) }\,, \\
        \label{eq:non-cold_PS}
        &P_{\textnormal{a}}(k) = F_\textnormal{h}^2 P_\textnormal{a}^{\textnormal{h}}(k) + 2F_\textnormal{h}(1- F_\textnormal{h}) \sqrt{P_\textnormal{a}^\textnormal{h}(k) P_\textnormal{a}^\textnormal{L}(k)} \nonumber \\
        & \hspace{1.cm}+ (1-F_\textnormal{h})^2 P_\textnormal{a}^\textnormal{L}(k)\,,
    \end{align}
where $P_\textnormal{c, a}^{\textnormal{h}}$ and $P_\textnormal{a}^{\textnormal{h}}$ are the cross and axion non-linear power spectra, respectively, and $P_\textnormal{c,a}^{\textnormal{L}}$ and $P_\textnormal{a}^{\textnormal{L}}$ the corresponding linear power spectra. The linear power spectra can be computed directly with \textsc{axionCAMB} \cite{axionCAMB} by using the transfer functions and the primordial power spectrum, whereas for the non-linear part we use the HM.

To find the expression for the cross and axion non-linear power spectrum the biased tracer technique plays an important role. As mentioned above this method assumes that axion halos only form around cold matter halos (c-halos) and thus the halo mass function for ULAs is the same as for the cold field, $n(M_{\rm ax}) \textnormal{d}M_{\rm ax} = n(M_\textnormal{c})\dd M_\textnormal{c}$, and the linear axion halo bias corresponds to the c-halo bias, $b(M_{\rm ax}) = b(M_\textnormal{c})$ (\ie the axion halo mass is itself a function of the c-halo mass).

Another quantity which also will change is the exact form of the halo mass function in Equation~\ref{eq:HMF} which can be rewritten in the following form
    \begin{equation}
        \label{eq:HMF_cold}
        n_{\textnormal{c}}\dd M_{\textnormal{c}} = \frac{\bar{\rho}_{\textnormal{tot}}}{M} f(\nu) \frac{1}{\nu} \dd \nu\,,
    \end{equation}
with $M = M_{\textnormal{c}} +  M_{\textnormal{a}}$ and $\nu = \delta_{\textnormal{c}}/\sigma(M)$. In the case of a MDM with massive neutrinos N-body simulations showed that the HMF is fitted much better if only the cold matter field is used in the peak height variable $\nu$ \cite{HMF_halobias_fct_of_Mc_sim_1}, \cite{HMF_halobias_fct_of_Mc_sim_2}, \cite{HMF_halobias_fct_of_Mc_sim_3}. The authors of this paper series also made predictions for the halo bias and showed that also for the bias only the cold matter field should be used. We assume that this continues to be true for other biased tracers such as ULAs. 

Computing the cross and axion non-linear power spectra with the halo model with the biased tracer technique as well as the cold matter description gives for the one and two halo term of the cross power 
    \begin{align}
        \label{eq:cross_PS_1h}
        P_\textnormal{c,a}^{\textnormal{1h}}(k) &= \frac{1}{ F_{\textnormal{h}} \bar \rho_{\textnormal{c}}\bar \rho_{\textnormal{a}}}\int_{M_{\textnormal{cut}}}^\infty \dd M_{\textnormal{c}} n(M_{\textnormal{c}}) M_{\textnormal{c}} M_{\textnormal{a}}(M_{\textnormal{c}}) \nonumber \\
        & \hspace{2.5cm} \tilde u_{\textnormal{c}}(k, M_{\textnormal{c}}) \tilde u_{\textnormal{a}} (k, M_{\textnormal{c}})\,, \\
        \label{eq:cross_PS_2h}
        P_\textnormal{c, a}^{\textnormal{2h}}(k) &= \frac{1}{ F_{\textnormal{h}} \bar \rho_{\textnormal{c}}\bar \rho_{\textnormal{a}}}\int_0^\infty \dd M_{\textnormal{c}} n(M_{\textnormal{c}}) b(M_{\textnormal{c}}) M_{\textnormal{c}} \tilde u_{\textnormal{c}}(k, M_{\textnormal{c}}) \nonumber \\
        &  \times \int_{M_{\textnormal{cut}}}^\infty \dd M_{\textnormal{c}} n(M_{\textnormal{c}}) b(M_{\textnormal{c}}) M_{\textnormal{a}}(M_{\textnormal{c}}) \tilde u_{\textnormal{a}}(k, M_{\textnormal{c}}) P_{\textnormal{c}}^{\textnormal{L}}(k)\,
    \end{align}
and for the one and two halo of the ULA power:
    \begin{align}
        \label{eq:non-cold_PS_1h}
        P_\textnormal{a}^{\textnormal{1h}}(k) &= \frac{1}{ (F_{\textnormal{h}}\bar \rho_{\textnormal{a}})^2}\int_{M_{\textnormal{cut}}}^\infty \dd M_{\textnormal{c}} n(M_{\textnormal{c}}) M_{\textnormal{a}}(M_{\textnormal{c}})^2 \tilde u_{\textnormal{a}} (k, M_{\textnormal{c}})^2\,, \\
        \label{eq:non-cold_PS_2h}
        P_\textnormal{a}^{\textnormal{2h}}(k) &= \left[ \frac{1}{ F_{\textnormal{h}} \bar \rho_{\textnormal{a}}} \int_{M_{\textnormal{cut}}}^\infty \dd M_{\textnormal{c}} n(M_{\textnormal{c}}) b(M_{\textnormal{c}}) M_{\textnormal{a}}(M_{\textnormal{c}})\times \right.\nonumber \\ %\times \nonumber \\
        & \hspace{0.5cm} \left. \tilde u_{\textnormal{a}}(k, M_{\textnormal{c}}) \right ]^2 P_{\textnormal{c}}^{\textnormal{L}}(k)\,.
    \end{align}
Here we have introduced $M_{\textnormal{cut}}$: the cut-off mass below which the axions can no longer cluster. The cut-off mass is also involved in the clustered fraction $F_\textnormal{h}$ which is defined via \cite{massara_MDM_halo_model}
    \begin{equation}
        \label{eq:clustered_frac}
        F_\textnormal{h} = \frac{1}{ \bar\rho_{\textnormal{a}}} \int_{M_{\textnormal{cut}}}^\infty \dd M_{\textnormal{c}} n(M_{\textnormal{c}}) b(M_{\textnormal{c}}) M_{\textnormal{a}}(M_{\textnormal{c}}) \,.
    \end{equation}
This means the three new quantities we have to specify to complete the MDM HM are: the cut-off mass, $M_{\textnormal{cut}}$, the ULA halo mass relation, \ie the ULA halo mass as a function of the c-halo mass, and the ULA halo density profile.

%%%%%%%%%%%%-------axion density profile---%%%%%%%%%%%%%%%%%
\subsection{\label{subsec:ax_dens_profile}Cut-off Mass, Axion Halo Mass Relations and Axion Halo Density Profile}
\textbf{Cut-off Mass}
To find an expression for the cut-off mass we will follow Ref.~\cite{marsh_silk_cut_off_mass}, where we proposed that in a pure ULA cosmology no halo will form if the halo Jeans scale is larger than the virial radius. The halo Jeans scale is similar to the linear Jeans scale, but depends on the halo profile of the c-halo. The halo Jeans scale is given by~\cite{halo_jeans_scale}:
    \begin{align}
        \label{eq:halo_jeans_ax}
        k_{\textnormal{hJ}} = &66.5 (1+z)^{-1/4} \left( \frac{\Omega_{\textnormal{m}}h^2}{0.12} \right)^{1/4} \left( \frac{m_a}{10^{-22}\,\textnormal{eV}} \right)^{1/2} \nonumber \\
        &\left( \frac{\rho_{\textnormal{NFW}}(r_{\textnormal{hJ}})}{\bar{\rho}_{m}} \right)^{1/4}\,\textnormal{Mpc}^{-1}\,.
    \end{align}
Here $r_{\textnormal{hJ}}$ is the halo Jeans length calculated by converting $r = \pi/k$. Note that here the radius is set to half the wavelength as in Ref.~\cite{marsh_silk_cut_off_mass} instead of $r = 2\pi / k$ as in Ref.~\cite{halo_jeans_scale}. 

\begin{figure}
        \centering
        \includegraphics[width=\linewidth]{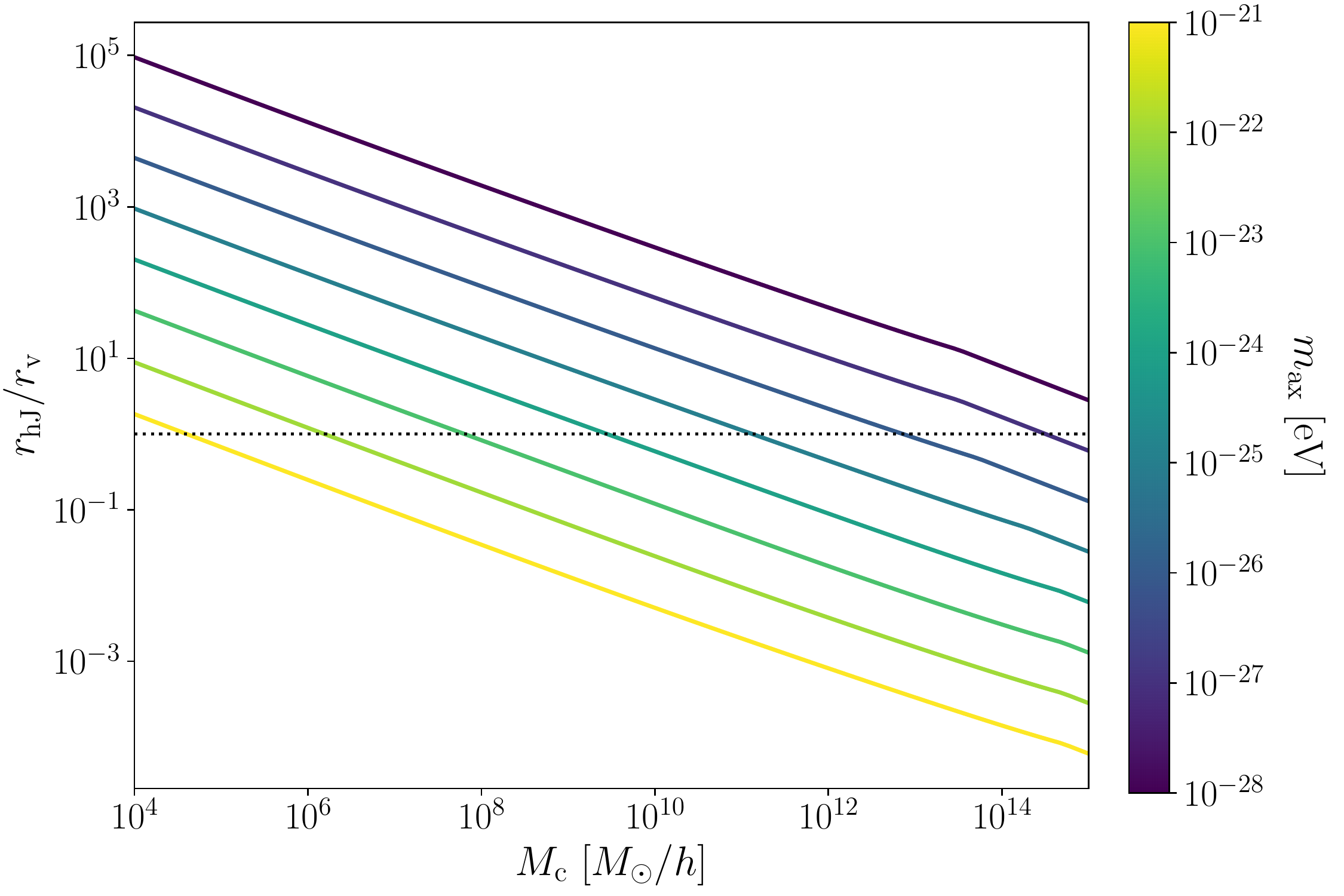}
        \caption{The ratio of the halo Jeans scale $r_{\textnormal{hJ}}$ to the virial radius $r_{\textnormal{v}}$ as a function of the c-halo mass for different axion masses colour coded as indicated in the colourbar. The horizontal dashed line shows the point for which the two quantities are equal and determines the cut-off mass.}
        \label{fig:ratio_r_hJ_r_v}
    \end{figure}

Since we assume that $r_{\textnormal{hJ}} \leq r_{\textnormal{v}}$ the NFW profile can be approximated as%
   \begin{equation}
        \label{eq:NFW_profile_aprox}
        \rho_{\textnormal{NFW}}(r_{\textnormal{hJ}}) \approx \frac{\bar \rho_{\mathrm{m}} \Delta_{\textnormal{v}} c^2 r_{\textnormal{v}}}{3 f(c)r_{\textnormal{hJ}}}\,
    \end{equation}
with $f(x)$ the function as in Eq.~\eqref{eq:rho_char_value}. Inserting this approximation back into Equation~\eqref{eq:halo_jeans_ax} and using $k_{\textnormal{hJ}} = \pi/r_{\textnormal{hJ}}$, the halo Jeans length becomes
    \begin{align}
        \label{eq:halo_jeans_ax_final}
        r_{\textnormal{hJ}} \approx &2.2 a^{-1/3} \left( \frac{m_a}{10^{-22}\,\textnormal{eV}} \right)^{-2/3} \left (  \frac{100 f(c)}{f(10) c^2}\right)^{1/3} \nonumber \\
        &\left( \frac{M}{10^{10}M_\odot} \right )^{-1/9}  \left( \frac{\Omega_{\textnormal{m}}h^2}{0.12} \right)^{-2/9}  \,\textnormal{kpc}\,.
    \end{align}
Fig.~\ref{fig:ratio_r_hJ_r_v} shows the ratio of the halo Jeans length to the virial radius as a function of the dark matter halo mass for different axion masses. The horizontal dashed line in the figure indicates the point where the ratio of the two length is equal to one and thus gives the cut-off mass $M_{\textnormal{cut}}$ for the ULA halo.

\begin{figure}
        \centering
        \includegraphics[width=\linewidth]{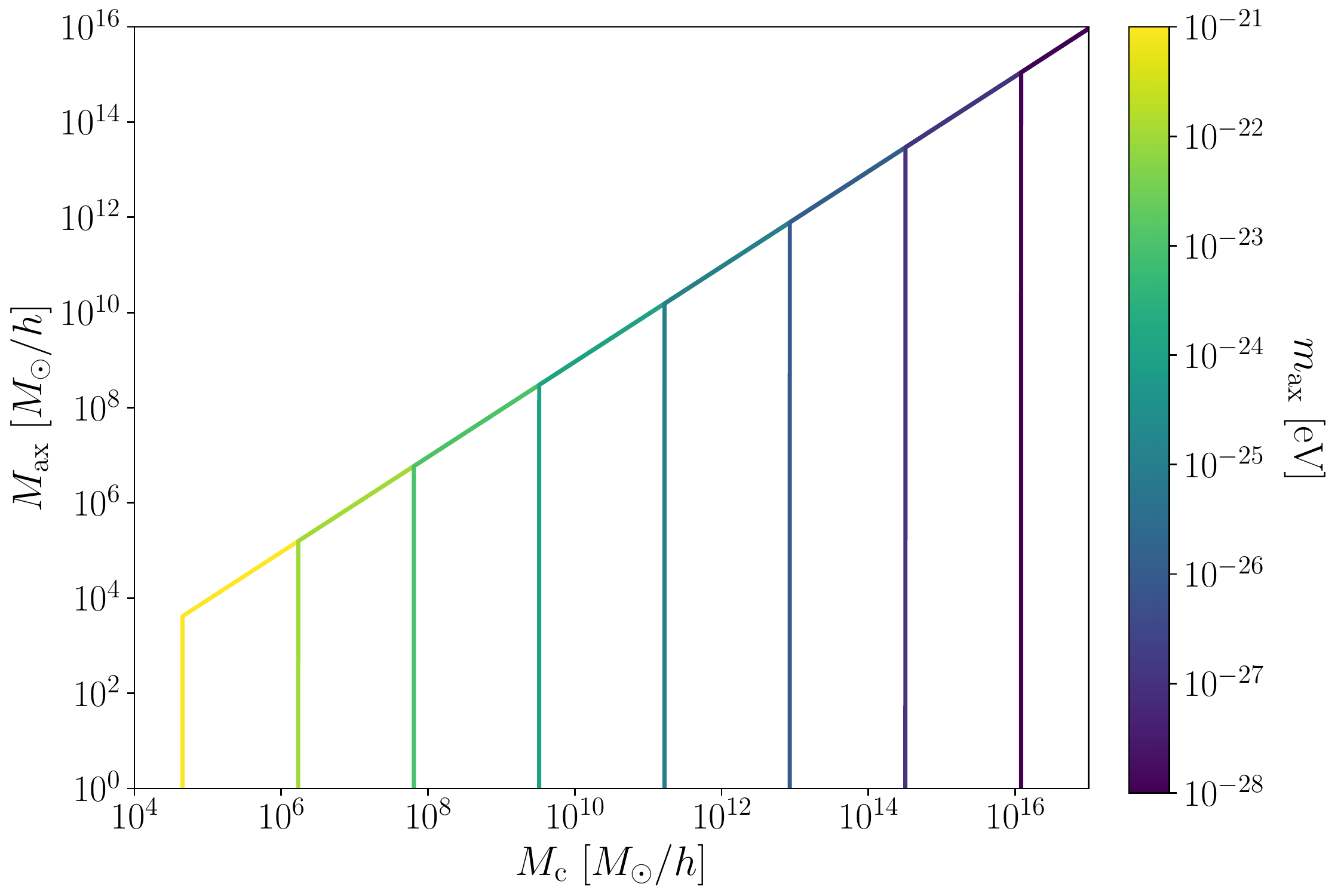}
        \caption{The mass of the axion halo as a function of the corresponding c-halo mass for axions in the mass range $10^{-28}\,\textnormal{eV} \leq m_{\textnormal{ax}} \leq 10^{-21}\,$eV as colour coded in the colourbar and an axion fraction of $f_{\textnormal{ax}} = 0.1$. Below the cut-off mass for each axion mass the axion halo mass is assumed to be zero.}
        \label{fig:ax_halo_mass_relation}
    \end{figure}

\textbf{ULA Halo Mass Relation} In this paper we assume, that for a ULA halo which is located around a cold matter halo with mass $M_{\textnormal{c}}$ the mass is given by the cosmic abundance, \ie the ULA halo mass relation is $M_{\textnormal{ax}} = \frac{\Omega_{\textnormal{ax}}}{\Omega_{\textnormal{c}}} M_{\textnormal{c}}$. Since ULAs cannot cluster into halos on small scales, this relation is only given for cold matter halos above the cut-off mass $M_{\textnormal{cut}}$ and below $M_{\textnormal{cut}}$ the ULA halo mass is assumed to be zero. In Fig.~\ref{fig:ax_halo_mass_relation} the halo mass relation for axion halos is plotted for axions in the mass range of $10^{-28}\,\textnormal{eV} \leq m_{\textnormal{ax}} \leq 10^{-21}\,$eV and an axion fraction of 0.1. In the next section we will see how the axion halo mass relation helps to find the axion halo density profile.

\textbf{Axion Halo Density Profile} In a cosmology with ULAs we have no fitting function of the axion density profile. There are simulations with mixture of ULAs and cold dark matter, though, that tell us something about the shape of the profile \cite{bodo_simulation}. However, in the case of a pure axion DM cosmology a density profile is found by simulations and a fitting formula was determined in \cite{schive_soliton_1}. The high resolution simulations showed that the core of the axion density profile is given by a soliton whereas the outer regions follow a CDM NFW-profile as in Eq.~\eqref{eq:NFW_profile}. Ref.~\cite{schive_soliton_1} found that the soliton in a pure ULA cosmology is well fitted by:
    \begin{align}
        \label{eq:soliton_profile_pure_ax}
        \rho_{\textnormal{c}}(r) = &\frac{1.9 (1+z)}{(1+9.1\times 10^{-2} (r/r_{\textnormal{c}})^2)^8} \left( \frac{r_{\textnormal{c}}}{\textnormal{kpc}} \right)^{-4} \nonumber \\
        &\left( \frac{m_{\textnormal{ax}}}{10^{-23}\,\textnormal{eV}} \right)^{-2}  \,M_\odot \textnormal{pc}^{-3}\,,
    \end{align}
with $r_{\textnormal{c}}$ the core radius where the density drops to one half of the central density. Further, Ref.~\cite{schive_soliton_2} determined this core radius to be:
    \begin{align}
        \label{eq:soliton_core_radius_pure_ax}
        r_{\textnormal{c}} = &1.6 (1+z)^{-1/2} \left( \frac{m_{\textnormal{ax}}}{10^{-22}\,\textnormal{eV}} \right)^{-1} \left( \frac{\Delta_{\textnormal{v}}(z)}{\Delta_{\textnormal{v}}(0)} \right)^{-1/6} \nonumber \\
        &\left( \frac{M_{\textnormal{h}}}{10^9\,M_\odot} \right)^{-1/3} \,\textnormal{kpc}\,.
    \end{align}
Here $M_{\textnormal{h}}$ is the mass of the ULA halo. 

Ref.~\cite{bodo_simulation} have simulated spherical collapse of halos in mixed CDM-ULA models, which showed that also in this case the ULA halo density profile is given by a soliton core and an NFW profile in the outer regions. The soliton core forms only as long as $f_{\textnormal{ax}} \geq 0.1$. For lower axion fractions the simulations showed that strong fluctuations in the central density profile do not allow a fit to the soliton profile \cite{bodo_simulation}. Therefore, we restricted the axion fraction to the range $f_{\textnormal{ax}} \in [0.1, 0.5]$ (the upper bound is given by the biased tracer approach), and consider only halos that host soliton cores. We are conservative with the lower bound, but new high resolution simulations of mixed CDM-ULA model could show that we can remove the lower bound and that we only have to set the upper bound of the axion fraction, \ie  $f_{\textnormal{ax}} \leq 0.5$. The NFW profile for the axion halo in the outer regions was found to be the same as the surrounding c-halo scaled by the cosmic abundance $\Omega_{\textnormal{ax}}/\Omega_{\textnormal{c}}$. Since the axion halos form in the potential wells of the c-halos the cold matter influenced the ground state solution of the ULA and the solution given by the soliton above has to be modified. We keep the same shape, Eq.~\eqref{eq:soliton_profile_pure_ax}, but determine the core radius from the CDM virial velocity, and rescale the central soliton density. \\
To construct a new soliton radius we used the radius defined by the characteristic velocity, $v_\textnormal{c}$, which scales like the de Broglie radius. Fits to simulations~\cite{soliton_velocity} give
    \begin{equation}
        \label{eq:soliton_velocity}
        r_{\textnormal{c}} = \frac{2\pi}{7.5} \frac{\hbar}{m_{\textnormal{ax}}v_{\textnormal{c}}}\,,
    \end{equation}
if the core is in equilibrium with its host halo. If we take the characteristic velocity equal to the virial velocity~\cite{soliton_velocity_virial_vel_1,soliton_velocity_virial_vel_2}, $v_{\textnormal{v}} = (GM_{\textnormal{h}}/r_{\textnormal{v}})^{1/2}$, then the soliton radius becomes:
    \begin{align}
        \label{eq:soliton_radius_MDM}
        r_{\textnormal{c}} = &1.2 (1+z)^{-1/2} \left( \frac{m_{\textnormal{ax}}}{10^{-22}\,\textnormal{eV}} \right)^{-1} \left( \frac{\Delta_{\textnormal{v}}(z)}{\Delta_{\textnormal{v}}(0)} \right)^{-1/6} \nonumber \\
        &\left( \frac{M_{\textnormal{h}}}{10^9\,M_\odot} \right)^{-1/6} \left( \frac{\Omega_{\textnormal{m}}h^2}{0.12} \right)^{-1/6} \,\textnormal{kpc}\,.
    \end{align}
In addition to the modified core radius we have to change the central density of the soliton such that the ULA halo has the correct mass given by the ULA halo mass relation described above. We thus rescale the soliton density with a factor $A$, which is set by fixing:
    \begin{equation}
        \label{eq:profile_ax_halo}
        M_{\textnormal{ax}} = 4 \pi \int_0^{r_{\textnormal{v}}} \dd r (\Theta(r_\textnormal{i} - r)A\rho_\textnormal{c}(r) + \Theta(r-r_\textnormal{i})\rho_\textnormal{NFW}(r) ) r^2 \,,
    \end{equation}
with $\Theta$ the Heaviside step function and $r_\textnormal{i}$ the radius where the two profiles cross. 
    \begin{figure}
        \centering
        \includegraphics[width=\linewidth]{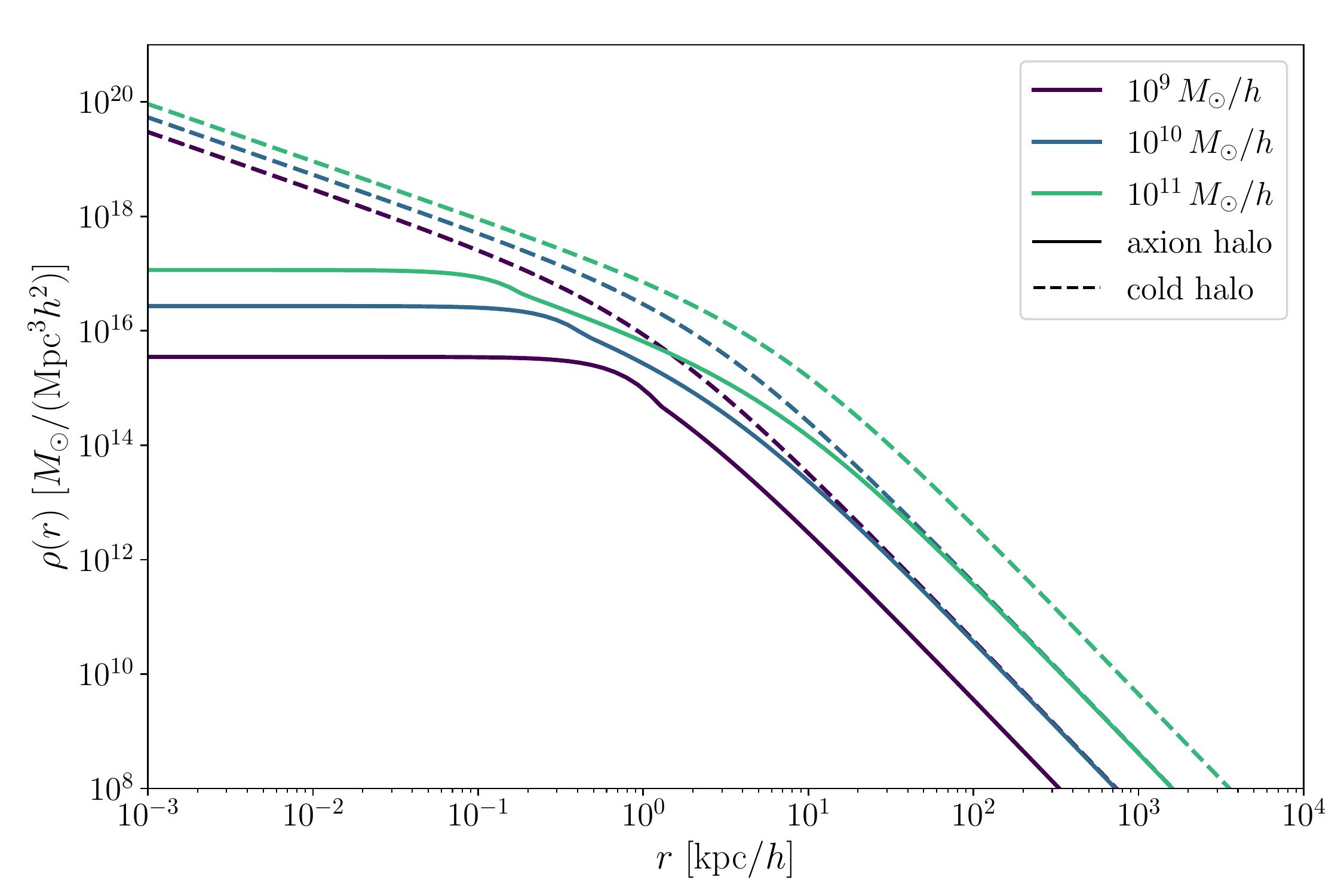}
        \caption{The axion halo density profiles for three different c-halos for $m_{\textnormal{a}}= 10^{-22}\,$eV with $f_{\textnormal{ax}} = 0.1$ as solid lines. The profile has a soliton core, see Eq.~\eqref{eq:soliton_profile_pure_ax}, with a soliton radius as in Eq.~\eqref{eq:soliton_radius_MDM} and an NFW profile on the outer region (see the text for the exact construction). For comparison also the corresponding NFW profiles of the c-halos are plotted in dashed lines.}
        \label{fig:ax_profile}
    \end{figure} 

With this at hand we can now determine the ULA halo density profile. By computing the central density scaling, $A$, we found that Eq.~\eqref{eq:profile_ax_halo} has no solution if the axion halo mass is very close to the cut-off mass. This is because the virial radius of a halo with a mass equal to the cut-off mass is very similar to the soliton radius of Eq.~\eqref{eq:soliton_radius_MDM} where the soliton density falls rapidly with increasing radius. Therefore, we decide to set the new cut-off mass to a little higher value where no solution is found for the central density of the soliton profile. The axion halo profile is shown in Figure~\ref{fig:ax_profile} for three different halo masses and an axion mass of $m_{\textnormal{a}} = 10^{-22}\,\textnormal{eV}$ with $f_{\textnormal{ax}}= 0.1$. The profiles show the soliton core and an NFW profile for large $r$. For comparison also the NFW profiles of the corresponding c-halo are shown in dashed lines. We can see that the axion profile is less massive in the core and shows a flat core rather than a cusp like the NFW profile. Our constructed profiles resemble closely the simulated profiles of Ref.~\cite{bodo_simulation}.

%%%%%%%%%%%%%%%%%%%%%%%%%%%%%%%%%%%%%%%%%%%%%%%%%%%%%%%%%%%%%%
%%%%%%%%%%%%%%%%%%%--------Results-------%%%%%%%%%%%%%%%%%%%%%
%%%%%%%%%%%%%%%%%%%%%%%%%%%%%%%%%%%%%%%%%%%%%%%%%%%%%%%%%%%%%%
\section{\label{sec:results}Results}

    \begin{figure}
        \centering
        \includegraphics[width=\linewidth]{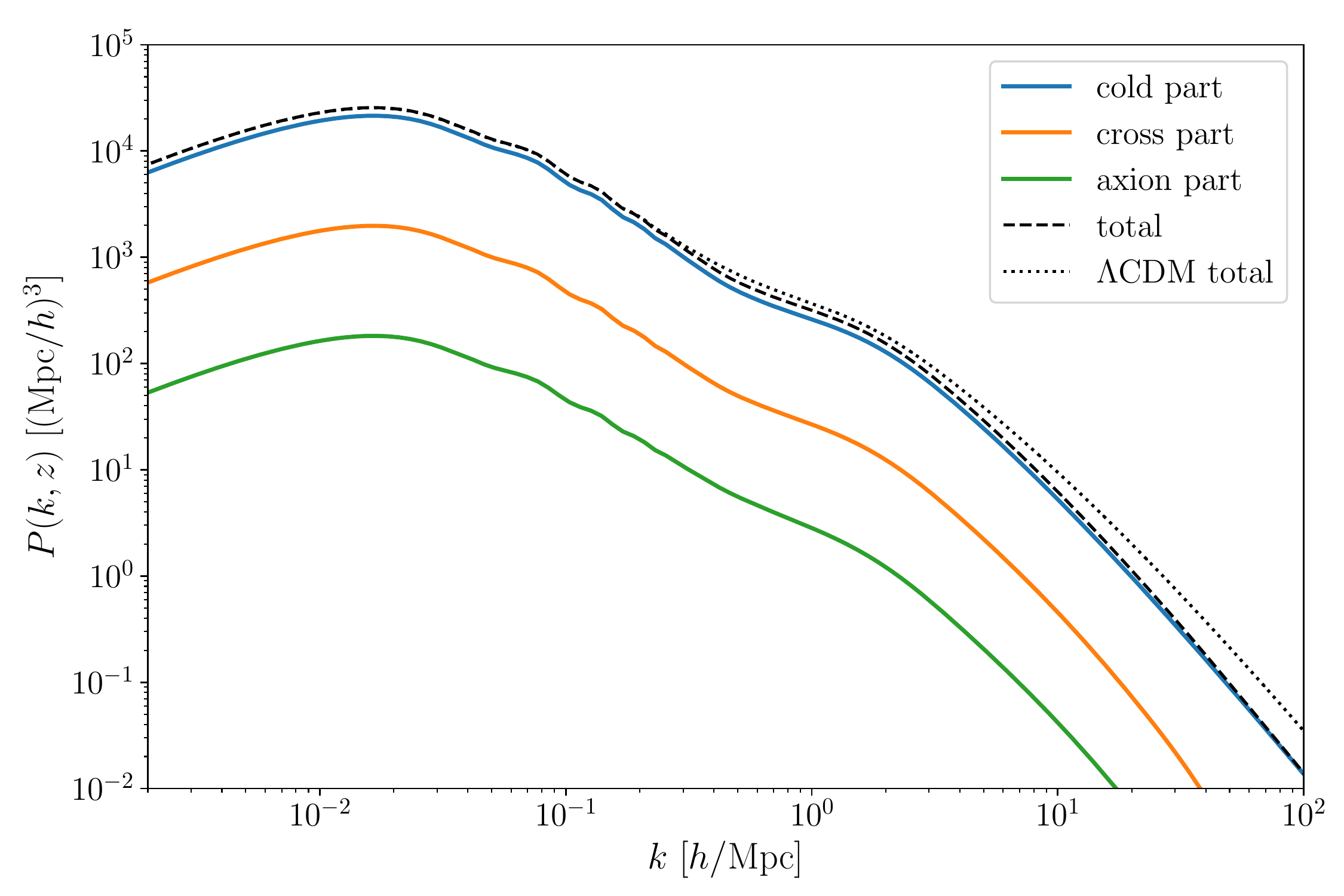}
        \caption{The non-linear power spectrum from the halo model in a MDM cosmology with axions of $m_{\textnormal{ax}}= 10^{-25}\,$eV and $f_{\textnormal{ax}} = 0.1$. The three terms of the power spectrum as in Eq.~\eqref{eq:PS_MDM} are shown as well as the resulting total non-linear matter power spectrum in dashed black. For comparison also the total matter power spectrum in a $\Lambda$CDM cosmology is represented as a black dotted line.}
        \label{fig:PS_with_ax}
    \end{figure} 

\subsection{Power Spectrum from MDM Halo Model}
\label{subsec:comp_LCDM}
\begin{figure}
        \centering
        \includegraphics[width=\linewidth]{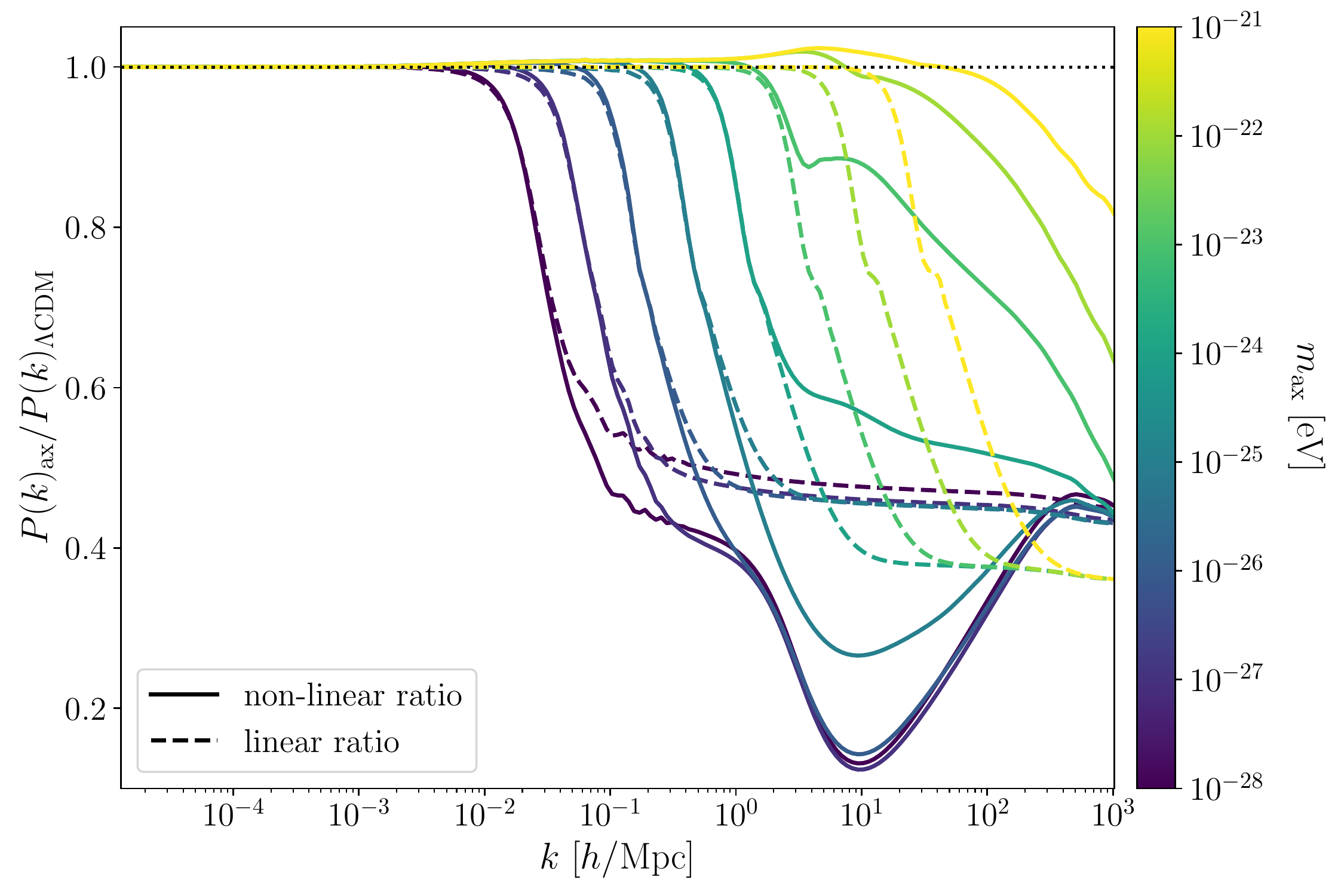}
        \caption{The ratio of the non-linear (solid lines) and linear (dashed lines) power spectrum in a MDM cosmology with axions in the mass range $10^{-28}\,\textnormal{eV} \leq m_{\textnormal{ax}} \leq 10^{-21}\,$eV and 10\,\% axions to the $\Lambda$CDM case. The non-linear power spectra are calculated with the halo model described in this work and the linear ones are calculated with \textsc{axionCAMB}. }
        \label{fig:ratio_axHM_LCDM}
    \end{figure}

The non-linear power spectrum with the extended halo model described in the previous Section is shown in Fig.~\ref{fig:PS_with_ax} for a MDM cosmology with $m_{\textnormal{ax}}= 10^{-25}\,$eV and $f_{\textnormal{ax}} = 0.1$. To understand the influence of ULAs on the power spectrum in more detail, we compare the MDM halo model with the $\Lambda$CDM halo model in Fig.~\ref{fig:ratio_axHM_LCDM} for a ULA fraction of 0.1 and mass range $10^{-28}\,\textnormal{eV} \leq m_{\textnormal{a}} \leq 10^{-21}\,$eV. As expected the non-linear power spectrum shows a suppression on large wavenumbers compared to pure CDM, asymptoting to a constant step-size. The size of the step is fixed by the relative abundance, $\Omega_{\rm ax}/\Omega_{\rm d}$, and transitions with $m_{\rm ax}$ as $k_{\rm J,eq}$ crosses through $1\, h\text{ Mpc}^{-1}$.

The suppression scale in the non-linear power can start at very different wavenumbers compared to the linear theory, with the difference depending on the ULA mass (a similar effect has also been seen before in Refs.~\cite{Hlozek:2016lzm,Marsh:2016vgj}). In linear theory, suppression relative to CDM comes from the mass dependence of the axion Jeans scale. But why does the suppression wavenumber in the non-linear power spectrum change? This can be understood when we look at the formula for the non-linear power spectrum which is given by a one and two halo term, see Eq.~\eqref{eq:non_lin_power_spec}, and the transition between these two terms is around $k_{\textnormal{t}} \sim 1\, h\textnormal{Mpc}^{-1}$. Furthermore, the two halo term is proportional to the linear power spectrum and thus, as long as the suppression of the linear power spectrum starts below the transition wavenumber, \ie $k_{\textnormal{J,eq}} < k_{\textnormal{t}}$, the suppression of the non-linear power spectrum starts at the same scale as the linear one. This occurs for masses $m_{\rm ax}\lesssim 10^{-23}\text{ eV}$. 

For wavenumbers higher than the transition wavenumber the one halo term starts to dominate and the non-linear power spectrum departs strongly from the linear one. Hence, the location of the suppression no longer depends simply on the linear Jeans scale. For higher mass ULAs with linear Jeans scale $k_{\textnormal{J,eq}} > k_{\textnormal{t}}$, the difference to the pure CDM case is driven by the one-halo term, which in turn is dominated by the cold halos themselves. The power spectrum drops when halos contributing to it become less massive than the cut-off mass. Additional suppression is driven ``passively'' by the effect of ULAs in reducing the clustering of CDM on small scales, which reduces the variance of CDM density fluctuations.

Another feature which can be seen in Fig.~\ref{fig:ratio_axHM_LCDM} is that the lowest masses considered, $m_{\textnormal{ax}} \leq 10^{-27}\,$eV, the ratio has a spoon-like shape. A similar shape was found if one compares the non-linear power spectrum with massive neutrinos to the power spectrum of a $\Lambda$CDM cosmology computed with the halo model Refs.~\cite{massara_MDM_halo_model,spoon_neutrino_halo_model}, or also from simulation, see \eg Refs.~\cite{spoon_neutrino_sim_1,spoon_neutrino_sim_2}. The appearance of a similar feature for ULAs is thus not unexpected.

\begin{figure*}
        \centering
        \includegraphics[width=\textwidth]{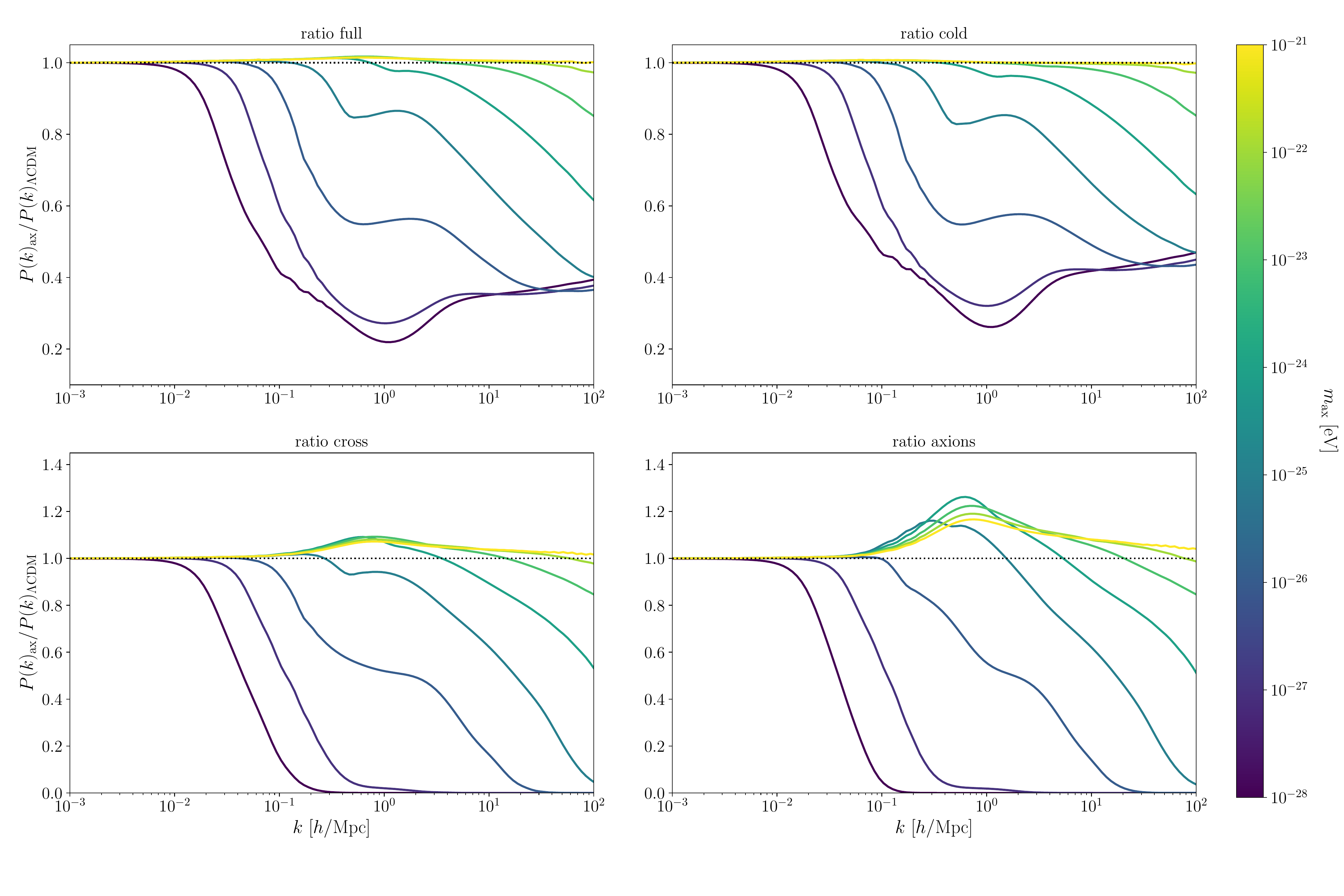}
        \caption{The ratio of the halo model in a MDM cosmology with axions (top left) and all three terms as in Eq.~\eqref{eq:non_lin_power_spec} (top right: cold part, bottom left: cross part and bottom right: axion part) to the $\Lambda$CDM halo model. Here the axions have a mass of $10^{-28}\,\textnormal{eV} \leq m_{\textnormal{ax}} \leq 10^{-21}\,$eV and 10\,\% of the DM are axions. }
        \label{fig:ratio_axHM_LCDM_all_parts}
    \end{figure*}

The last feature we can see in Fig.~\ref{fig:ratio_axHM_LCDM} is an \emph{enhancement} in power for the MDM model compared to pure CDM around $k \sim 1\, h \textnormal{Mpc}^{-1}$ for the higher ULA masses. We can understand this when we look at the ratio of the three different parts of the power spectrum in Eq.~\eqref{eq:non_lin_power_spec} to the $\Lambda$CDM halo model, as shown in Fig.~\ref{fig:ratio_axHM_LCDM_all_parts}. In the ratios of the cross and axion parts (bottom left and right panels) we see a strong enhancement at the scales mentioned above and this comes from the shape of the ULA halo density profile which is different from the cold one, i.e. this is caused by the coherence of the soliton, which increases the correlation function of the ULA field on small scales. The enhancement is not present for all ULA masses, since it requires a conincidence between the one-to-two halo transition in $P(k)$, and the size of the soliton in the halo mass dominating the power at this scale, which occurs for $m_{\rm ax}\approx 10^{-22}\text{ eV}$. This prediction of our model is in complete agreement with the simulations of Ref.~\cite{Nori:2018pka}, who observed a small increase in the power for $m_{\rm ax}\approx 10^{-22}\text{ eV}$ in pure ULA cosmologies only after accounting for the effect of the ``quantum pressure'' terms in the effective fluid description of the Schr\"{o}dinger-Poisson equation. 

In Fig.~\ref{fig:ratio_axHM_LCDM} we have shown for illustration the relative power over a wide $k$ range, and in particular for very large wavenumbers. This means, however, that the halo model is evaluated up to very small lengths where internal properties of individual halos have to be taken into account, \eg baryonic feedback from star formation or active galactic nuclei \cite{baryonic_feedback}. Thus our model is idealised on these scales, and should not be considered realistic. See Refs.~\cite{halo_model_max_k_1,mead_hmcode2020,massara_MDM_halo_model} for more discussion. We expect our model to be relatively accurate up to approximately $k = 10\,h\textnormal{Mpc}^{-1}$.
\\

\subsection{DE-like Axions}
\begin{figure}
        \centering
        \includegraphics[width=\linewidth]{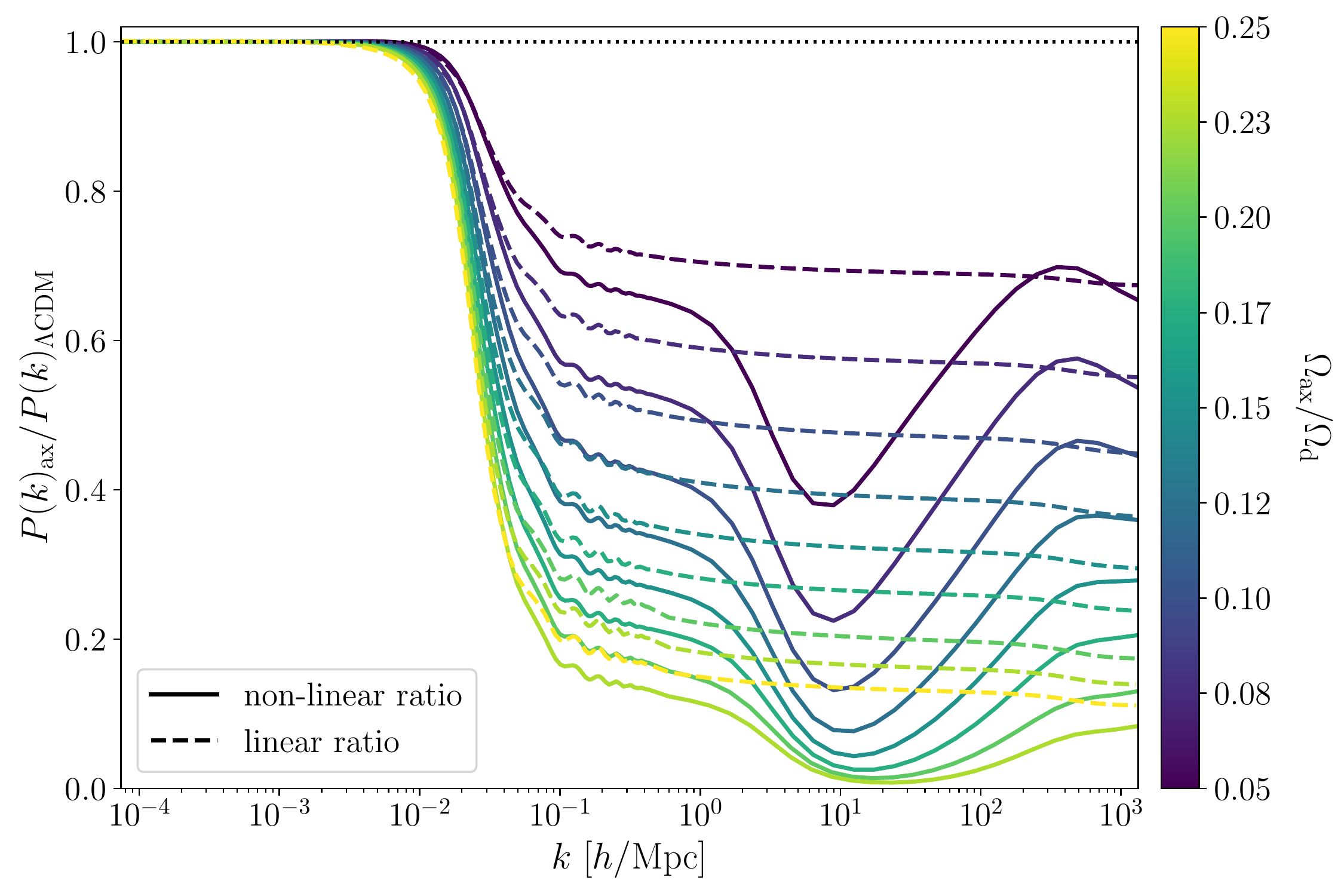}
        \caption{The ratio of the non-linear (solid lines) and linear (dashed lines) power spectrum in a MDM cosmology with axions of a mass $10^{-28}\,$eV and for an axion fraction in the range $0.05 \leq f_{\mathrm{ax}} \leq 0.25$ to the $\Lambda$CDM case.}
        \label{fig:ratio_axHM_LCDM_m_e-28}
    \end{figure}

Our halo model should work extremely well for dark energy (DE) like ULAs, as defined by Ref.~\cite{Hlozek:2014lca} with $m_{\rm ax}\leq 10^{-28}\text{ eV}$, where no simulations are available at the moment. 
As mentioned above in the discussion of Fig.~\ref{fig:ratio_axHM_LCDM} a spoon like shape is seen for very light ULAs, \ie DE-like axions. We want investigate this feature further comparing to the $\Lambda$CDM power spectrum for different ULA fractions, \ie $0.05 \leq f_{\mathrm{ax}} \leq 0.25$, and an axion mass of $m_{\textnormal{ax}} = 10^{-28}\,$eV, shown in Fig.~\ref{fig:ratio_axHM_LCDM_m_e-28}.
Here the we use our mixed Halo model for axion fractions below the discussed lower bound of $f_{\mathrm{ax}}$ in Sec.~\ref{subsec:ax_dens_profile}. But at this low fractions and small ULA mass the effect of the axions on the non-linear power spectrum is very small and thus we decided to extend the mixed halo model for lower ULA mass also to smaller axion fractions.
We observe that the spoon like shape is more dominant for smaller axion fraction and faded away when the fraction is raised. 

For lower masses still, $m_{\rm ax}<10^{-28}\text{ eV}$, we see from Fig.~\ref{fig:ax_halo_mass_relation} that we do not expect such ULAs to reside in any cosmologically known halos $M_h\lesssim 10^{15}M_\odot$. This justifies the approximation taken in Ref.~\cite{bauer_biased_tracer_H1} to remove ULAs entirely from the HM at low masses, and has the same effect as the removal of neutrinos from halos in the case of \textsc{HMCode} (the effect of this approximation compared to the full mixed halo model of neutrinos is discussed in Appendix~\ref{app:neutrinos}). This suggests that for $m_{\rm ax}<10^{-28}\text{ eV}$ one can leverage the accuracy of \textsc{HMCode} for ULAs at any density fraction allowed by current constraints from linear scales~\cite{Hlozek:2017zzf}, although Lagrangian perturbation theory~\cite{Lague:2021frh} will also be accurate in this regime. This is due to the fact that the axion perturbations $\delta_{\rm ax}$ have a scale-dependent growth which is suppressed on small scales. In the case of DE-like axions, the perturbations on scales $k\lesssim 0.1\, h\text{ Mpc}^{-1}$ will not grow until the present day and remain in the linear regime. 

%

%%%%%%%%%%%%%%%%%%%%%%%%%%%%%%%%%%%%%%%%%%%%%%%%%%%%%%%%%%%%%%
%%%%%%%%%%%%%%%%%%%--------Discussion-------%%%%%%%%%%%%%%%%%%
%%%%%%%%%%%%%%%%%%%%%%%%%%%%%%%%%%%%%%%%%%%%%%%%%%%%%%%%%%%%%%
\section{\label{sec:discussion}Discussion and Outlook}

We presented in this paper an improved halo model for a cosmology composed of CDM and ULAs. The standard pure CDM halo model assumes that all matter is bound in halos and that the two-point correlation between the matter is given by the correlations inside one halo and between two separate halos. However, in a mixed dark matter cosmology with a sub-dominant ULA component, because of the ULA Jeans scale we can no longer assume that all matter is bound into halos. More generally, due to their different clustering properties CDM and ULAs must be treated differently. In the power spectrum we have a cold-cold part, a cross correlation between the cold and the non-cold matter, and an ULA self-clustering part. The ULA power spectrum has a ``smooth'' component from the matter that is not contained in halos and a component that can cluster inside halos on large enough scales

Our model accounts for all of these effects. Our model assumes (in a manner that is consistent with observations) that ULAs make up only a sub-dominant component of the total matter, and thus we treat ULAs as a biased tracer of the cold matter. In the spirit of the biased tracer model (which has been applied successfully  to neutrinos and neutral hydrogen), we have proposed a density profile for ULAs inside halos, as well as a relationship between the ULA and CDM halo masses, including a cut-off mass below which ULAs do not cluster inside halos. For the ULA density profile in a mixed halo, no fitting formulae are available in the literature, and we proposed a model based on Ref.~\cite{schive_soliton_1} fitting formula in a pure ULA cosmology, and observations from simulations in mixed ULA-CDM spherical collapse of Ref.~\cite{bodo_simulation}. 

For ULAs in the mass range $10^{-28}\text{ eV}\lesssim m_{\rm ax}\lesssim 10^{-23}\text{ eV}$ our model predicts that the suppression scale between the MDM power spectrum and pure CDM occurs near the linear theory Jeans scale, with an additional spoon-like suppression similar to the case of neutrinos, before asymptoting to a constant fixed by the relative DM density fractions. At larger ULA masses, the suppression scale is controlled by the one-halo term, and moves out to larger wavenumbers. For ULAs with $m_{\rm ax}\approx 10^{-22}\text{ eV}$ we observe a small enhancement in the power relative to CDM on intermediate scales, which we attribute to the role of solitons in the power spectrum. This prediction of our model is in qualitative agreement with the simulations of Ref.~\cite{Nori:2018pka}. Finally, we also considered DE-like ULAs with $m_{\rm ax}\lesssim 10^{-28}\text{ eV}$ and also found a spoon-like feature in the power spectrum. 

Our model is inspired by the mixed ULA-CDM simulations of Ref.~\cite{bodo_simulation}, who propose the density profile we use based on the Schro\"{o}dinger-Poisson equation, and also many simulations that observe a minimum halo mass in pure ULA cosmologies in N-body (e.g. Ref.~\cite{Schive:2015kza}). However, we have not been able to calibrate our model on cosmological mixed ULA-CDM simulations. Some such simulations are in preparation~\cite{Lague_in_prep}. However, the box size of these simulations is too small to have a large number of halos. Thus we cannot test the cut-off mass in our model. The relative power spectrum is in principle well predicted in relatively small boxes~\cite{Corasaniti:2016epp}, however this is only true on scales where there are many halos contributing to the power such that the one-halo term is well sampled. We have also found that mixed DM simulations are limited in resolution of the density profile~\cite{Lague_in_prep}. Nonetheless, this shows us the way forward to calibrating our halo model in future.

An interesting extension of our work would be to include multiple ULA sub-components, as one might expect in an ``axiverse''~\cite{Arvanitaki:2009fg}. The principles of the biased tracer model could be adopted for each component, and one might expect e.g. ``nested solitons'' in the density profile~\cite{Luu:2018afg}. Indeed, mixed ULA simulations were recently reported~\cite{Gosenca:2023yjc}. The linear transfer function in such a model is expected to demonstrate multiple step features, and one would expect this to be reflected in our model also in the halo mass function and non-linear power. An accurate linear transfer function has, however, not been computed in such a case. Extending e.g. \textsc{axionCAMB} to multiple fields would be evidently worthwhile.

We expect our halo model to be extremely useful in future analyses of cosmological data. We demonstrated recently that the halo model for pure ULAs can be used in analysis of Dark Energy Survey data~\cite{Dentler:2021zij}. Ref.~\cite{Dentler:2021zij} was only able to constrain the ULA mass, and could not vary the fraction at low masses due to the lack of an appropriate halo model. The model presented here fulfils that purpose and will allow for a combined CMB+DES analysis covering all masses and fractions across the ULA parameter space. Such an analysis will plug an important gap in current ULA constraints between $10^{-25}\text{ eV}$ and $10^{-23}\text{ eV}$. This has the potential to probe new parameter space of string theory models~\cite{Arvanitaki:2009fg,Demirtas:2021gsq,Mehta:2021pwf}, some of which now make specific predictions in this region~\cite{Cicoli:2021gss}, and has wider implications for the understanding of DM at the low-mass frontier~\cite{Grin:2019mub,AlvesBatista:2021gzc}. Our model will continue to be useful for constraining ULAs with next generation cosmological data such as Simons Observatory~\cite{SimonsObservatory:2018koc}, CMB-S4~\cite{Dvorkin:2022bsc}, and \emph{Euclid}~\cite{Amendola:2016saw}.

%%%%%%%%%%%%%%%%%%%%%%%%%%%%%%%%%%%%%%%%%%%%%%%%%%%%%%%%%%%%%%
%%%%%%%%%%%%%%%--------acknowledgements-------%%%%%%%%%%%%%%%%
%%%%%%%%%%%%%%%%%%%%%%%%%%%%%%%%%%%%%%%%%%%%%%%%%%%%%%%%%%%%%%
\begin{acknowledgments}
We acknowledge useful discussions with Mona Dentler, Jens Niemeyer, and Bodo Schwabe. We acknowledge Jurek Bauer for collaboration in an early stage of this project. DJEM is supported by an Ernest Rutherford Fellowship from the Science and Technologies Facilities Council (UK). This work made use of the open-source libraries \textsc{matplotlib}~\cite{matplotlib}, \textsc{numpy}~\cite{numpy}, \textsc{scipy}~\cite{scipy}, \textsc{astropy}~\cite{astropy}.

\end{acknowledgments}

%%%%%%%%%%%%%%%%%%%%%%%%%%%%%%%%%%%%%%%%%%%%%%%%%%%%%%%%%%%%%%
%%%%%%%%%%%%%%%--------Appendix-------%%%%%%%%%%%%%%%%
%%%%%%%%%%%%%%%%%%%%%%%%%%%%%%%%%%%%%%%%%%%%%%%%%%%%%%%%%%%%%%
\appendix

\section{\textsc{HMcode} Parameters and Comparison}
\label{app:HMCode_Mead}
The halo model is a very good and quick model to find the non-linear power spectrum and is used in a lot of different codes. One of the most frequently used codes. the \textsc{HMcode}, is provided by Ref.~\cite{HMCode_mead_2020}, with previous versions Refs.~\cite{HMCode_mead_2015,HMCode_mead_2016}. \textsc{HMCode} introduces a number of parameters to improve the model in its fit to simulations over the standard HM. The parameters were fit to simulations of Ref.~\cite{heitmann_coyote_emulator} such that the model accurately matches these simulations as well as the simulations from Ref.~\cite{heitmann_miratitan_emulator}. The newest code \textsc{HMCode} shows excellent agreement to simulations for $k \lesssim 10\,h\textnormal{Mpc}^{-1}$ and $z<2$ with a root mean square of at most 2.5\% \cite{HMCode_mead_2020}.

Since \textsc{HMCode} is a frequently used halo model code and shows excellent results,  we decide to implement the parameter in our halo model code, called \textsc{axionHMcode}. When cosmological mixed DM ULA simulations become available we can compare them with our MDM halo model with the parameters from \textsc{HMCode} to see if these parameters also improve the HM with ULAs. In total there are six new parameter and we will discuss them in this Appendix.

\begin{figure}
        \centering
        \includegraphics[width=\linewidth]{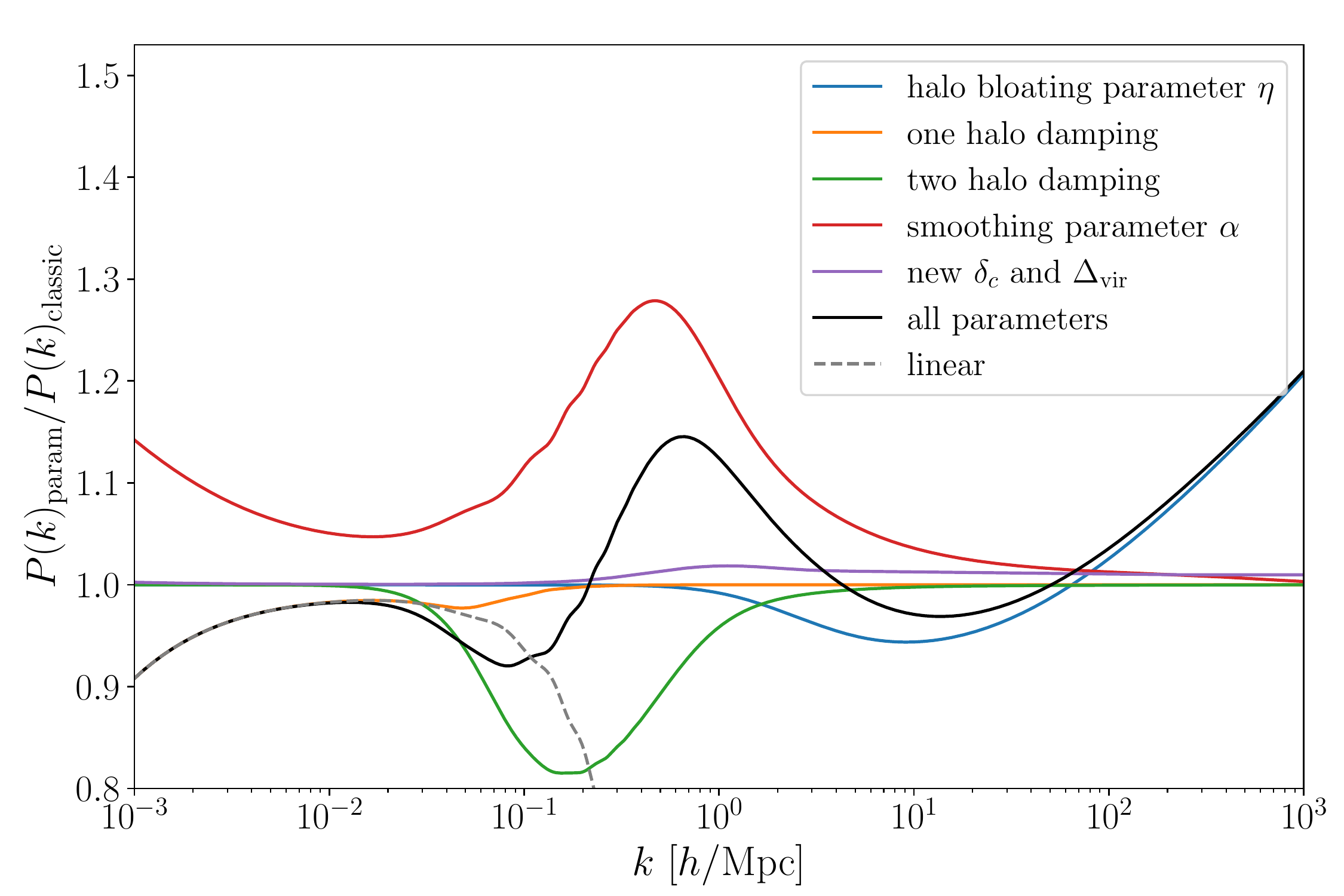}
        \caption{Ratio between the halo model with one of the parameters in \textsc{HMCode} and the standard halo model. In the ratio with the two halo damping three parameters are involved (see the text for more details). Each parameter shows a different effect on the non-linear power spectrum. The black line shows the ratio when all parameters and a non-$\Lambda$CDM approach for the overdensities are used.}
        \label{fig:ratio_parmeter_HMcode}
    \end{figure}

\noindent \textbf{Halo bloating term}: The NFW profile from Eq.~\eqref{eq:NFW_profile} is modified in \textsc{HMcode} by the parameter $\eta$ such that
   \begin{equation}
        \label{eq:NFW_profile_modified}
        \rho_{\textnormal{NFW}}(r, M, \eta) = \frac{\rho_{\mathrm{char}}}{r/(\nu^\eta r_s) \left( 1 + r/(\nu^\eta r_s) \right)^2}, \qquad r \leq r_{\textnormal{v}}\nu^\eta \, ,
    \end{equation}
with
    \begin{equation}
        \label{eq:def_eta}
        \eta(z) = 0.1281 \sigma_{8,\,\textnormal{cc}}^{-0.3644}(z) > 0 \, ,
    \end{equation}
where $\sigma_{8,\,\textnormal{cc}}$ refers to the variance of the cold matter linear power spectrum for $R  = 8\,h^{-1}\,\mathrm{Mpc}$ in Eq.~\eqref{eq:def_sigma}. It can be shown that the halo bloating term influences the Fourier transformation of the NFW profile, Eq.~\eqref{eq:dens_profile_kspace_computed}, by scaling the wavenumber $k$ in the following way \cite{halo_bloating_term_kspace_prof}:
    \begin{equation}
        \label{eq:eta_fourie_NFW}
        \Tilde{u}(k, M, z) \to  \Tilde{u}(\nu^\eta k, M, z)\,.
    \end{equation}
Since $\eta > 0$ the $k$-space halo density profile is pushed to higher $k$'s for small halos and the profile for massive halos is transformed to smaller wavenumbers. The exact shape of $\Tilde{u}(k, M, z)$ is very important for the one halo term and smaller halos play an increasingly large role for high $k$'s. This means that at a scale around $k \sim 1\,h\textnormal{Mpc}^{-1}$, where the one halo term becomes the dominant component, the halo bloating term shows an effect on the non-linear power spectrum. Since on larger scales the contribution from halos with high masses dominates we expect a suppression in the power spectrum, because $\nu^\eta < 1$. Then for higher and higher $k$'s the shapes of smaller halos are more important and the $\eta$ parameter will give an enhancement in the non-linear power spectrum ($\nu^\eta > 1$). This described effect can be seen in Fig.~\ref{fig:ratio_parmeter_HMcode}, where the ratio between the halo model to the halo bloating term and the standard model is shown in blue. 

\textbf{One halo term damping}: In the standard halo model approach the one halo term is constant on large scales. However, this is not the correct behaviour due to mass and momentum conservation. It was shown in \cite{one_halo_term_damping_reason} that the one halo term should grow like $P^{1\textnormal{h}}(k) \propto k^4$ at small $k$ (i.e. damp away compared to constant at small $k$). In \textsc{HMCode} this is implemented trough the modification
    \begin{equation}
        \label{eq:one_halo_damping}
        P^{1\textnormal{h}}(k)  \to P^{1\textnormal{h}}(k) \frac{(k/k_*)^4}{1+(k/k_*)^4}\,.
    \end{equation}
The modified one halo term grows as expected and is suppressed on large scales. This also ensures that on large scales the non-linear power spectrum is given by the two halo term and hence is equal to the linear power spectrum. The suppression depends on the free parameter $k_*$ which is fitted from simulations to be \cite{HMCode_mead_2020} 
    \begin{equation}
        \label{eq:k_star_fit}
        k_* = 0.05618\sigma_{8,\,\textnormal{cc}}^{-1.013}(z)\,h\textnormal{Mpc}^{-1}\,.
    \end{equation}
Since the one halo term is suppressed on large scales, the total non-linear power is also suppressed on large scales if the one halo term is modified as above. This can be seen in Fig.~\ref{fig:ratio_parmeter_HMcode} in orange. 

\textbf{Two halo term damping}: 
Like the one halo term also the two halo term is damped on some scales. These lengths are scales larger than a particular wavenumber $k_d$. The damping takes the form
    \begin{equation}
        \label{eq:two_halo_damping}
        P^{2\textnormal{h}}(k)  \to P^{2\textnormal{h}}(k) \left( 1-f \frac{(k/k_{\textnormal{d}})^{n_{\textnormal{d}}}}{1+(k/k_{\textnormal{d}})^{n_{\textnormal{d}}}} \right)\,,
    \end{equation}
where $n_{\textnormal{d}},\ k_{\textnormal{d}}$ and $f$ are fitting parameters. The three parameters in Eq.~\eqref{eq:two_halo_damping} are fitted and given by \cite{HMCode_mead_2020}
    \begin{align}
        \label{eq:k_d_fit}
        k_{\textnormal{d}} &= 0.05699\sigma_{8,\,\textnormal{cc}}^{-1.089}(z)\,h\textnormal{Mpc}^{-1}\,, \\
        \label{eq:f_fit}
        f &= 0.2696\sigma_{8,\,\textnormal{cc}}^{0.9403}(z)\,, \\
        \label{eq:n_d_fit}
        n_{\textnormal{d}} &= 2.853\,.
    \end{align}
We see that the damping power $n_{\textnormal{d}} > 0$ and hence there is a damping for $k \gg k_{\textnormal{d}}$ as long as the two halo term dominates in the non-linear power spectrum. The explained behaviour can be seen in the Fig.~\ref{fig:ratio_parmeter_HMcode} in green. 

\textbf{Smoothing parameter}: 
In the standard halo model the power spectrum is the sum of the one halo and two halo term, see Eq.~\eqref{eq:non_lin_power_spec}. However, if the two halo term is of comparable size to the one halo term (transition region), the assumption of a purely additive behaviour is too simple. The scale of the transition region is also known as the quasi-linear regime and Ref.~\cite{HMCode_mead_2015} found that the halo model performed quite poorly at the quasi-linear regime. Hence the \textsc{HMcode} introduces a transition parameter by modelling:
    \begin{equation}
        \label{eq:alpha_param}
       P(k) =  (P^{\mathrm{1h}}(k)^{\alpha} + P^{\mathrm{2h}}(k)^{\alpha})^{1/\alpha} \,,
    \end{equation}
where $\alpha$ is the parameter that shapes the transition. If $\alpha < 1$ the transition is smoothed whereas for $\alpha > 1$ the transition is sharper. The general form of the smoothing parameter is assumed to be
    \begin{equation}
        \label{eq:alpha_general}
        \alpha = a b^{n_{\textnormal{eff}}(z)}\,.
    \end{equation}
Here $a$ and $b$ are the fitting parameters and $n_{\textnormal{eff}}(z)$ is the effective spectral index at the non-linear length $R_{\textnormal{nl}}$, where $\sigma_{\textnormal{cc}}(R_{\textnormal{nl}}, z) = \delta_{\textnormal{c}}$:
    \begin{equation}
        \label{eq:neff_def}
        n_{\textnormal{eff}}(z) = - \frac{\dd\,\textnormal{ln} \sigma_{\textnormal{cc}}^2(R, z)}{\dd\,\textnormal{ln} R} \bigg|_{\sigma = \delta_{\rm crit}} -3\,.
    \end{equation}
In the newest version \textsc{HMCode} the smoothing parameter is fitted to be \cite{HMCode_mead_2020}
    \begin{equation}
        \label{eq:alpha_fit}
        \alpha = 1.876 \times 1.603^{n_{\textnormal{eff}}(z)}\,.
    \end{equation}
The effect of the smoothing parameter can be seen in Fig.~\ref{fig:ratio_parmeter_HMcode} in red and a clear enhancement around the transition region is visible since $\alpha < 1$ in the analysed cosmology. Note that the halo model with the smoothing parameter is not equal to the standard halo model on large scales because only the smoothing parameter is used and the one halo term is not damped in this plot and thus the constant one halo term influences the power spectrum on large scales. 

In Figure~\ref{fig:ratio_parmeter_HMcode} also the ratio with all parameters of the \textsc{HMCode} is shown by a solid black line and the ratio to the linear power spectrum is shown in dotted grey. On large scales the non-linear power spectrum with all parameters described above and the linear power spectrum coincide and the same is true if only the one halo damping is used. Therefore, we use only the one halo damping in \textsc{axionHMcode} as default to ensure the correct behaviour on large scales. The other parameters are optional in our halo model, and free parameters in the code. 

\section{Massive Neutrinos}\label{app:neutrinos}

The full HM for massive neutrinos was developed similarly to our model in Ref.~\cite{massara_MDM_halo_model}, and calibrated to a small range of simulations. Crucially, the simulations in this case include the distinct dynamics of massive neutrinos compared to CDM, and allow them to non-linearly cluster around halos. 

In contrast, the \textsc{MiraTitan}~\cite{heitmann_miratitan_emulator} emulator simulates only the \emph{linear} evolution of neutrinos and does not allow them to cluster inside halos. Thus, the massive neutrino model in \textsc{HMCode}, which is calibrated to \textsc{MiraTitan}, takes an approximate treatment of massive neutrinos that removes them from halos. The accuracy of this treatment, as applied to \textsc{MiraTitan}, is discussed in depth in Ref.~\cite{HMCode_mead_2020}.
We developed our own implementation of the massive neutrino halo model of Ref.~\cite{massara_MDM_halo_model}~\footnote{Available at \url{https://github.com/SophieMLV/nuHMcode}}. Comparing this treatment to the \textsc{HMCode} approximation can be used as an estimate of the theoretical error in \textsc{HMCode}, and consequently in \textsc{MiraTitan}, of neglecting full neutrino dynamics. We briefly present these results here. A more complete comparison, using e.g. larger neutrino simulations such as Ref.~\cite{Emberson:2016ecv}, would be interesting, but is beyond the scope of the present work.

\textbf{Neutrino Halo model ingredients:} The halo model with massive neutrinos uses the same formulae as the model with ULAs described in Section~\ref{subsec:MDM_HM}. The difference is that the massive neutrino halos have a different shape, halo mass relation and the cut-off mass. The advantage for the model with massive neutrinos is that we have fitting functions for the neutrino halo ($\nu$-halo) profiles for $\sum m_\nu = 0.3\,$eV and $ \sum m_\nu = 0.6\,$eV from simulations in Ref.~\cite{neutrino_halo_profile}. The authors found the following fitting function
    \begin{equation}
        \label{eq:nu_halo_fitting}
        \rho_\nu (r, M_{\textnormal{c}}) = \frac{\delta_{\textnormal{core}}}{1+ \left(r/r_{\textnormal{core}}\right)^\alpha} \bar{\rho}_\nu\,.
    \end{equation}
Here $\delta_{\textnormal{core}}$, $r_{\textnormal{core}}$ and $\alpha$ are fitting parameters and depend on the mass of the corresponding c-halo $M_{\textnormal{c}}$. 

However, the resolution of the N-body simulations in Ref.~\cite{massara_MDM_halo_model} was not high enough to resolve the core of the neutrino halos for c-halos of mass below $M_{\textnormal{c}} \sim 10^{13.5}\,h^{-1}\,M_\odot$. Hence, the profile in this case is chosen to behave like
    \begin{equation}
        \label{eq:nu_halo_fitting_2}
        \rho_\nu (r) = \frac{\kappa}{r^\beta} \bar{\rho}_\nu\,
    \end{equation}
and to reproduce the outer region of the neutrino density profile as in Eq.~\eqref{eq:nu_halo_fitting}. Here $\kappa$ and $\beta$ are again determine by fitting to the simulation results. For a total neutrino masses of $\sum m_\nu = 0.3\,$eV and $ \sum m_\nu = 0.6\,$eV the parameters are given in Table~\ref{tab:nu_halo_params}.
    \begin{table*}
        \centering
        \begin{tabular}{ccc}
        \hline
        parameter & $\sum m_\nu$ = 0.3\,eV & $\sum m_\nu$ = 0.6\,eV \\
        \hline
        $\delta_{\textnormal{core}}$ & $6.056\times 10^{-8} M_{\textnormal{c}}^{0.58}$ & $3.7478\times 10^{-8} M_{\textnormal{c}}^{0.64}$\\
        $r_{\textnormal{core}}$ [$h^{-1}$kpc] & $4.029\times 10^{-8} M_{\textnormal{c}}^{0.68}$ & $2.046\times 10^{-4}M_{\textnormal{c}}^{0.43}$ \\
        $\alpha$ & $-6.69+0.24\log(M_{\textnormal{c}})$ & $-4.62 + 0.19 \log(M_{\textnormal{c}})$\\
        $\beta$ &  $-2.06+0.09\log(M_{\textnormal{c}})$ & $-3.64 + 0.15 \log(M_{\textnormal{c}})$\\
        $\kappa$ [$(h^{-1}\textnormal{kpc})^{-\beta}$] & $0.19+3.242\times 10^{-19} M_{\textnormal{c}}^{1.5}$ & $0.24 + 1.144\times 10^{-20} M_{\textnormal{c}}^{1.7}$\\
        \hline
        \end{tabular}
        \caption{Fitting functions for the parameters $\delta_{\textnormal{core}}$, $r_{\textnormal{core}}$, $\alpha$, $\beta$ and $\kappa$ in Eq.~\eqref{eq:nu_halo_fitting} and Eq.~\eqref{eq:nu_halo_fitting_2} as in figure~10 of \cite{neutrino_halo_profile}.}
        \label{tab:nu_halo_params}
    \end{table*}

\begin{figure}
        \centering
        \includegraphics[width=\linewidth]{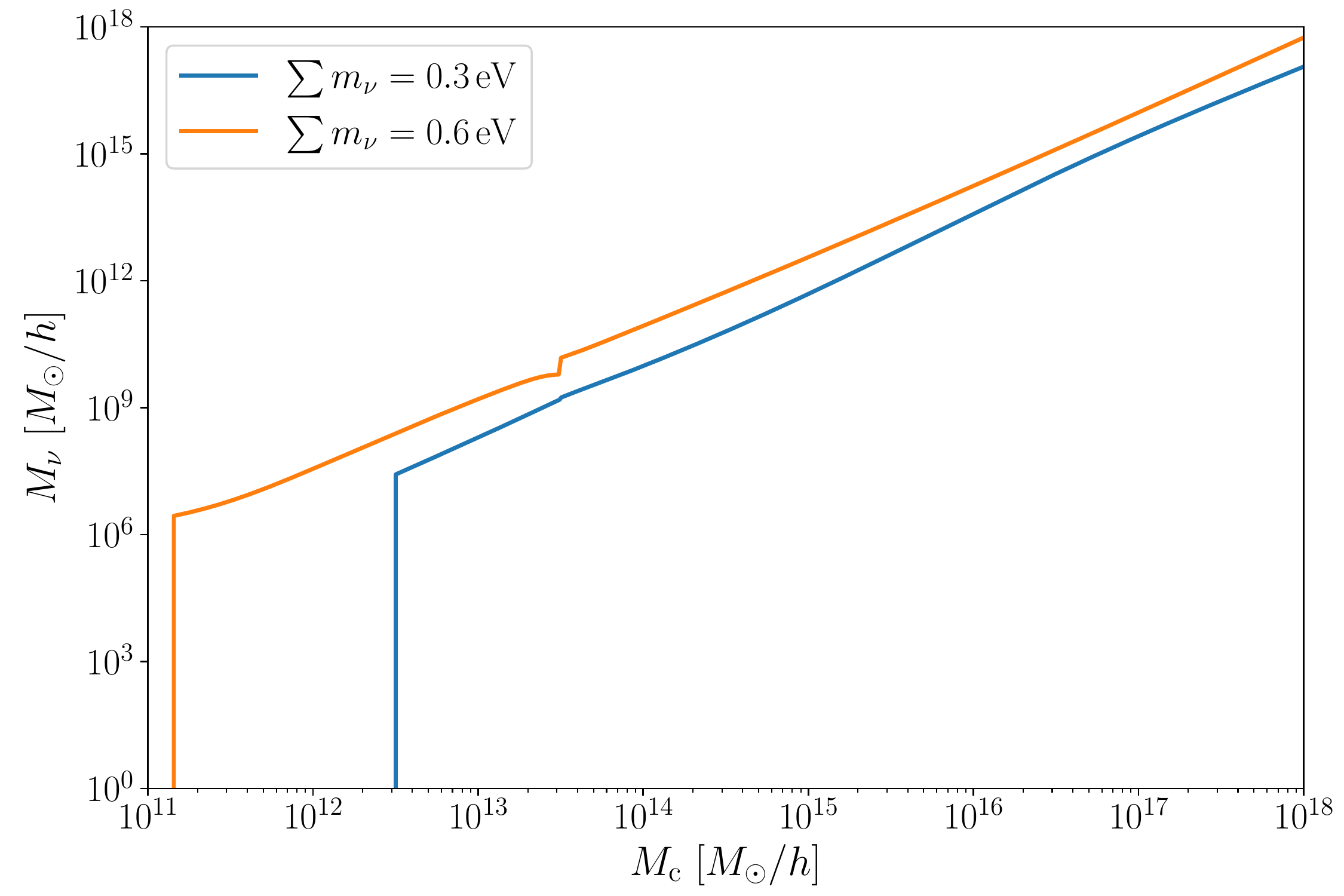}
        \caption{The mass of the neutrino halo as a function of the corresponding c-halo for a total neutrino mass of $\sum m_\nu = 0.3\,$eV in blue and for $\sum m_\nu = 0.6\,$eV in orange. At the cut-off mass the halo mass relation drops to zero, \ie there is no $\nu$-halo for a c-halo with mass below the cut-off mass. The jump in the halo mass relation around $M_{\textnormal{c}} = 10^{13.5}\,h^{-1}\,M_\odot$ comes from the change in the density profile, Eq.~\eqref{eq:nu_halo_fitting} and \eqref{eq:nu_halo_fitting_2}. Note that these lines do not have a slope equal to the cosmic abundance $\Omega_\nu / \Omega_{\textnormal{c}}$.}
        \label{fig:nu_halo_mass_relation}
    \end{figure}

Different approaches can be made to find a cut-off mass. We continue to follow Ref.~\cite{massara_MDM_halo_model} who defined $M_{\textnormal{cut}}$ as the c-halo for which the corresponding $\nu$-halo has a mass of at least 10\,\% of the background neutrino density enclosed in the same radius:
    \begin{equation}
        \label{eq:cut-off_mass_nu}
        M_\nu(M_{\textnormal{cut}}) = 0.1 \frac{4\pi \bar \rho_\nu}{3} r_{\textnormal{v}}^3(M_{\textnormal{cut}})\,.
    \end{equation}
with the cut-off mass, $M_{\textnormal{cut}}$, the $\nu$-halo density profile, $\rho_\nu$, and the halo mass relation for the $\nu$-halo, $M_\nu(M_{\textnormal{c}})$. Fig.~\ref{fig:nu_halo_mass_relation} shows the mass of the $\nu$-halo as a function of the corresponding cold matter halo. In the figure we also see the transition between the two neutrino halo profiles around $M_{\textnormal{c}} = 10^{13.5}\,h^{-1}\,M_\odot$ as a small discontinuity in $M_\nu(M_{\textnormal{c}})$.
\\
\\
\textbf{Comparison to \textsc{HMCode}:} We want to compare the results from our halo model code with massive neutrinos, \textsc{$\nu$HMCode}, with the \textsc{HMCode} predictions. The massive neutrinos are implemented such that they evolve linearly and do not cluster in halos at all. This means Ref.~\cite{HMCode_mead_2020} used the normal $\Lambda$CDM halo model as in Section~\ref{sec:HM}, but removed the massive neutrino from the halo density profile, Eq.~\eqref{eq:dens_profile_kspace_computed}, by transforming the profile as
    \begin{equation}
        \label{eq:dens_profile_kspace_cold_transform_nu_!}
        \Tilde{u}(k, M) \to  (1-f_\nu)\Tilde{u}(k, M) \,.
    \end{equation}

The modifications of the halo model for \textsc{HMCode} in Appendix~\ref{app:HMCode_Mead} can also be adopted to our \textsc{$\nu$HMCode}. Note that for the smoothing parameter $\alpha$ the smoothing is only applied to the cold matter part of the fully extended halo model. We decided to smooth only the cold part because the expression for the cold part is given by the standard halo model, Eq.~\eqref{eq:non_lin_power_spec}, by using the cold matter quantities rather than the total matter terms and thus the expression is very similar to the one where the smoothing parameter is used in \textsc{HMcode}.

\begin{figure}
        \centering
        \includegraphics[width=\linewidth]{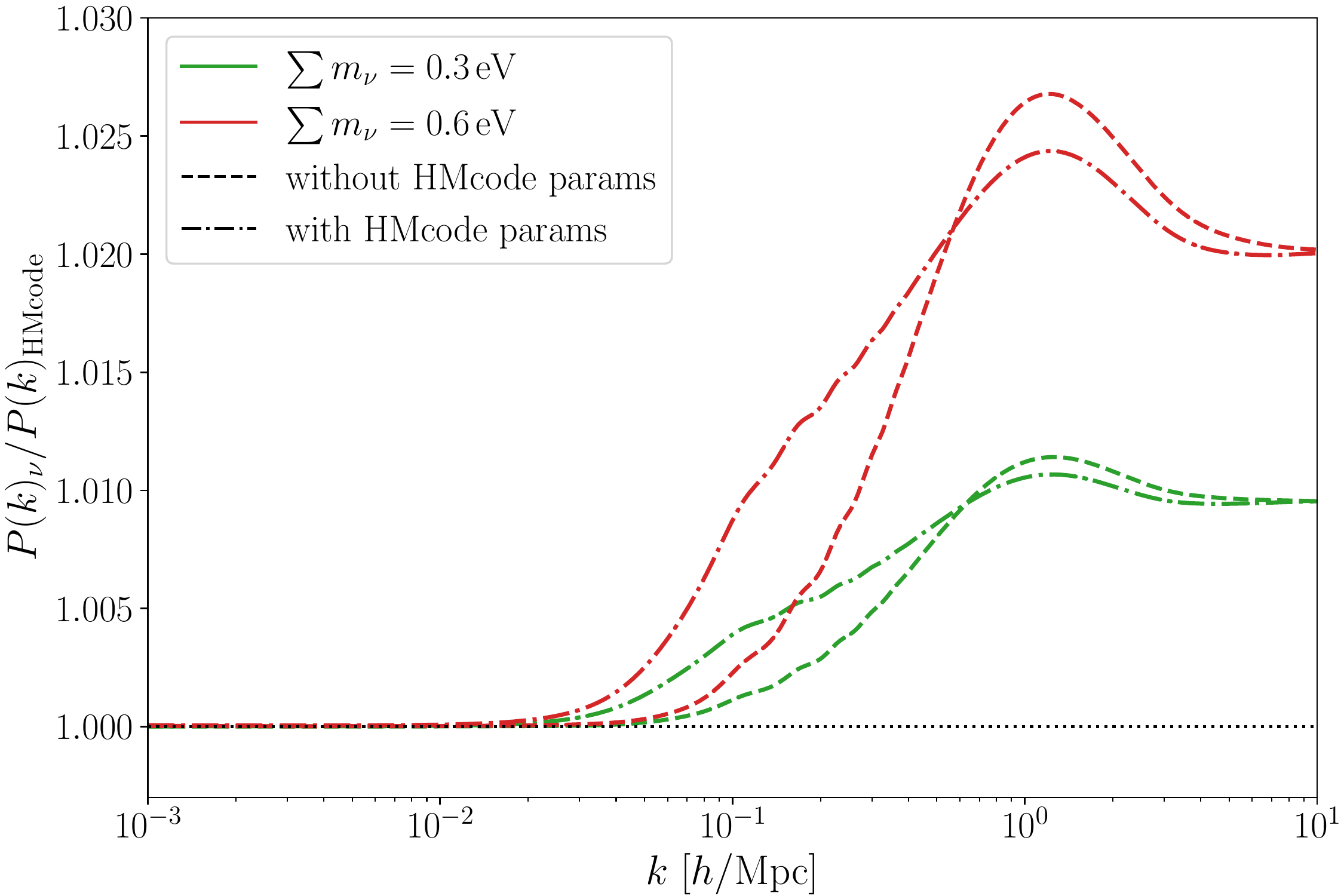}
        \caption{The ratio of the \textsc{$\nu$HMcode} without the \textsc{HMCode} parameters (dashed lines) and with the \textsc{HMCode} parameters (dashed-dotted lines) to the corresponding approximation of the power spectrum with massive neutrinos as in \textsc{HMCode} for $\sum m_\nu = 0.3$\,eV in green and $\sum m_\nu = 0.6$\,eV in red.}
        \label{fig:ratio_nuHm_HMCode}
    \end{figure}

So, we compare here the following models first our full \textsc{$\nu$HMcode} without the parameters with the \textsc{HMcode} massive neutrino approximation also with no parameters and second the \textsc{$\nu$HMcode} with all parameters with the \textsc{HMCode} approximation also with all parameters. The difference between these models can be understood as the effect of the clustered treatment of the massive neutrinos on the non-linear power spectrum.
%since my model takes non-linearities of massive neutrinos into account whereas \cite{mead_hmcode-2020_2021} models the neutrinos completely unclustered, \ie linearly.\\
The ratios of the models for both massive neutrino masses can be seen in Figure~\ref{fig:ratio_nuHm_HMCode} and we see that the full treatment in \textsc{$\nu$HMcode} has an enhancement for large wavenumbers in both configurations and for both masses. The extra power comes from the additional clustering of massive neutrinos inside halos, which were taken into account in our halo model. The difference for a total neutrino mass $\sum m_\nu = 0.3$\,eV is not larger than 1\,\% and for $\sum m_\nu = 0.3$\,eV the discrepancy never reaches 3\,\% for wavenumbers below $k \sim 10\,h\textnormal{Mpc}^{-1}$.  Thus the advantage of using the full treatment of massive neutrinos is very small and the question is whether it is worthwhile to use the more computationally intensive but more accurate model or to work with the simplified model which achieves comparable results in much less time. Moreover, the difference of the two models as shown in Fig.~\ref{fig:ratio_nuHm_HMCode} can be used for an approximation of the error of simulations which treat massive neutrinos only linearly. 
%

%%%%%%%%%%%%-------Convergence Checks---%%%%%%%%%%%%%%%%%
\section{Convergence Checks}
\label{app:convergence}
In Section~\ref{sec:HM} we mentioned that we have to check if the integral involved in the two halo term, see Eq.~\eqref{eq:two_halo_term}, converges. In theory this is an improper integral, but in the numerical computation the integral has to be evaluated on a finite interval $[M_{\textnormal{min}}, M_{\textnormal{max}}]$. With the finite interval we can no longer ensure that to $k \to 0$ the integral converges to
    \begin{equation}
        \label{eq:integral_two_halo_k_0}
        \frac{1}{\Bar{\rho}} \int_0^\infty \mathrm{d}M M n(M) b(M) = 1 \,,
    \end{equation}
which implies that on large scales the non-linear power spectrum is equal to the linear one (this makes sense because we live in an isotropic and homogeneous universe). So we have to be careful when choosing the boundaries for the integral of the two halo term. This problem was discussed in the Appendix A of Ref.~\cite{correction_for_correct_two_halo_term} and the authors showed that a relatively simple modification of the two halo term integral solves this problem. 

Note that the problem of the finite interval is not the upper bound, since the HMF $n(M)$ has an exponential cut-off approximately around $10^{16}M_\odot\,h^{-1}$ and thus if we take $M_{\textnormal{max}}$ sufficiently large it can be understood as infinite. Instead, the lower bound is a problem due to the constantly rising HMF which states that a large amount of the matter is contained in low mass halos \cite{correction_for_correct_two_halo_term} and thus the correction of the two halo term integral depends on the lower bound $M_{\textnormal{min}}$.

So the solution of Ref.~\cite{correction_for_correct_two_halo_term} is defining a new function
    \begin{equation}
        \label{eq:correction_factor_2HT}
        A(M_{\textnormal{min}}) = 1 - \frac{1}{\Bar{\rho}} \int_{M_{\textnormal{min}}}^\infty \mathrm{d}M M n(M) b(M) \,,
    \end{equation}
which gives the missing part of the integral below $M_{\textnormal{min}}$. With this at hand we can define a new halo mass function which takes care of the missing part from halos with masses below $M_{\textnormal{min}}$ \cite{correction_HMF}. The new HMF is given by the transformation
    \begin{equation}
        \label{eq:new_HMF}
        n(M) \to n(M) + \frac{A(M_{\textnormal{min}}) \delta(M -M_{\textnormal{min}}) }{b(M_{\textnormal{min}}) M_{\textnormal{min}}/\bar \rho} \,,
    \end{equation}
where $\delta$ is the Dirac delta distribution. This new HMF gives a correction to the two halo term integral:
    \begin{equation}
        \label{eq:correction_2HT_integral}
         \frac{1}{\Bar{\rho}} \int_{ M_{\textnormal{min}}}^\infty \mathrm{d}M M n(M) b(M) \vert \Tilde{u}(k, M) \vert + \frac{A(M_{\textnormal{min}}) \tilde{u}(k, M_{\textnormal{min}}) }{M_{\textnormal{min}}/\bar \rho} \,.
    \end{equation}
So as $k \to 0$ the first term goes to $1/\bar{\rho} \int_{M_{\textnormal{min}}}^\infty \mathrm{d}M M n(M)$ and the second term $\tilde{u}(k,  M_{\textnormal{min}}) \to M_{\textnormal{min}}/\bar \rho$. Using Eq.~\eqref{eq:correction_factor_2HT} we see that the new two halo term integral goes to one. So, we always use this new two halo term integral when we compute the two halo term in our code.

For the one halo term we do not need this correction since in the one halo term low mass halos contribute very little \cite{correction_for_correct_two_halo_term}.

\bibliography{apssamp}% Produces the bibliography via BibTeX.

%merlin.mbs apsrev4-1.bst 2010-07-25 4.21a (PWD, AO, DPC) hacked
%Control: key (0)
%Control: author (72) initials jnrlst
%Control: editor formatted (1) identically to author
%Control: production of article title (-1) disabled
%Control: page (0) single
%Control: year (1) truncated
%Control: production of eprint (0) enabled
\begin{thebibliography}{104}%
\makeatletter
\providecommand \@ifxundefined [1]{%
 \@ifx{#1\undefined}
}%
\providecommand \@ifnum [1]{%
 \ifnum #1\expandafter \@firstoftwo
 \else \expandafter \@secondoftwo
 \fi
}%
\providecommand \@ifx [1]{%
 \ifx #1\expandafter \@firstoftwo
 \else \expandafter \@secondoftwo
 \fi
}%
\providecommand \natexlab [1]{#1}%
\providecommand \enquote  [1]{``#1''}%
\providecommand \bibnamefont  [1]{#1}%
\providecommand \bibfnamefont [1]{#1}%
\providecommand \citenamefont [1]{#1}%
\providecommand \href@noop [0]{\@secondoftwo}%
\providecommand \href [0]{\begingroup \@sanitize@url \@href}%
\providecommand \@href[1]{\@@startlink{#1}\@@href}%
\providecommand \@@href[1]{\endgroup#1\@@endlink}%
\providecommand \@sanitize@url [0]{\catcode `\\12\catcode `\$12\catcode
  `\&12\catcode `\#12\catcode `\^12\catcode `\_12\catcode `\%12\relax}%
\providecommand \@@startlink[1]{}%
\providecommand \@@endlink[0]{}%
\providecommand \url  [0]{\begingroup\@sanitize@url \@url }%
\providecommand \@url [1]{\endgroup\@href {#1}{\urlprefix }}%
\providecommand \urlprefix  [0]{URL }%
\providecommand \Eprint [0]{\href }%
\providecommand \doibase [0]{http://dx.doi.org/}%
\providecommand \selectlanguage [0]{\@gobble}%
\providecommand \bibinfo  [0]{\@secondoftwo}%
\providecommand \bibfield  [0]{\@secondoftwo}%
\providecommand \translation [1]{[#1]}%
\providecommand \BibitemOpen [0]{}%
\providecommand \bibitemStop [0]{}%
\providecommand \bibitemNoStop [0]{.\EOS\space}%
\providecommand \EOS [0]{\spacefactor3000\relax}%
\providecommand \BibitemShut  [1]{\csname bibitem#1\endcsname}%
\let\auto@bib@innerbib\@empty
%</preamble>
\bibitem [{\citenamefont {Aghanim}\ \emph
  {et~al.}(2020{\natexlab{a}})\citenamefont {Aghanim} \emph
  {et~al.}}]{Planck:2018nkj}%
  \BibitemOpen
  \bibfield  {author} {\bibinfo {author} {\bibfnamefont {N.}~\bibnamefont
  {Aghanim}} \emph {et~al.} (\bibinfo {collaboration} {Planck}),\ }\href
  {\doibase 10.1051/0004-6361/201833880} {\bibfield  {journal} {\bibinfo
  {journal} {Astron. Astrophys.}\ }\textbf {\bibinfo {volume} {641}},\ \bibinfo
  {pages} {A1} (\bibinfo {year} {2020}{\natexlab{a}})},\ \Eprint
  {http://arxiv.org/abs/1807.06205} {arXiv:1807.06205 [astro-ph.CO]}
  \BibitemShut {NoStop}%
\bibitem [{\citenamefont {Aghanim}\ \emph
  {et~al.}(2020{\natexlab{b}})\citenamefont {Aghanim} \emph
  {et~al.}}]{Planck:2018vyg}%
  \BibitemOpen
  \bibfield  {author} {\bibinfo {author} {\bibfnamefont {N.}~\bibnamefont
  {Aghanim}} \emph {et~al.} (\bibinfo {collaboration} {Planck}),\ }\href
  {\doibase 10.1051/0004-6361/201833910} {\bibfield  {journal} {\bibinfo
  {journal} {Astron. Astrophys.}\ }\textbf {\bibinfo {volume} {641}},\ \bibinfo
  {pages} {A6} (\bibinfo {year} {2020}{\natexlab{b}})},\ \bibinfo {note}
  {[Erratum: Astron.Astrophys. 652, C4 (2021)]},\ \Eprint
  {http://arxiv.org/abs/1807.06209} {arXiv:1807.06209 [astro-ph.CO]}
  \BibitemShut {NoStop}%
\bibitem [{\citenamefont {Alves~Batista}\ \emph {et~al.}(2021)\citenamefont
  {Alves~Batista} \emph {et~al.}}]{AlvesBatista:2021gzc}%
  \BibitemOpen
  \bibfield  {author} {\bibinfo {author} {\bibfnamefont {R.}~\bibnamefont
  {Alves~Batista}} \emph {et~al.},\ }\href@noop {} {\  (\bibinfo {year}
  {2021})},\ \Eprint {http://arxiv.org/abs/2110.10074} {arXiv:2110.10074
  [astro-ph.HE]} \BibitemShut {NoStop}%
\bibitem [{\citenamefont {Hlozek}\ \emph
  {et~al.}(2015{\natexlab{a}})\citenamefont {Hlozek}, \citenamefont {Grin},
  \citenamefont {Marsh},\ and\ \citenamefont {Ferreira}}]{Hlozek:2014lca}%
  \BibitemOpen
  \bibfield  {author} {\bibinfo {author} {\bibfnamefont {R.}~\bibnamefont
  {Hlozek}}, \bibinfo {author} {\bibfnamefont {D.}~\bibnamefont {Grin}},
  \bibinfo {author} {\bibfnamefont {D.~J.~E.}\ \bibnamefont {Marsh}}, \ and\
  \bibinfo {author} {\bibfnamefont {P.~G.}\ \bibnamefont {Ferreira}},\ }\href
  {\doibase 10.1103/PhysRevD.91.103512} {\bibfield  {journal} {\bibinfo
  {journal} {Phys. Rev. D}\ }\textbf {\bibinfo {volume} {91}},\ \bibinfo
  {pages} {103512} (\bibinfo {year} {2015}{\natexlab{a}})},\ \Eprint
  {http://arxiv.org/abs/1410.2896} {arXiv:1410.2896 [astro-ph.CO]} \BibitemShut
  {NoStop}%
\bibitem [{\citenamefont {Hlozek}\ \emph {et~al.}(2018)\citenamefont {Hlozek},
  \citenamefont {Marsh},\ and\ \citenamefont {Grin}}]{Hlozek:2017zzf}%
  \BibitemOpen
  \bibfield  {author} {\bibinfo {author} {\bibfnamefont {R.}~\bibnamefont
  {Hlozek}}, \bibinfo {author} {\bibfnamefont {D.~J.~E.}\ \bibnamefont
  {Marsh}}, \ and\ \bibinfo {author} {\bibfnamefont {D.}~\bibnamefont {Grin}},\
  }\href {\doibase 10.1093/mnras/sty271} {\bibfield  {journal} {\bibinfo
  {journal} {Mon. Not. Roy. Astron. Soc.}\ }\textbf {\bibinfo {volume} {476}},\
  \bibinfo {pages} {3063} (\bibinfo {year} {2018})},\ \Eprint
  {http://arxiv.org/abs/1708.05681} {arXiv:1708.05681 [astro-ph.CO]}
  \BibitemShut {NoStop}%
\bibitem [{\citenamefont {Kobayashi}\ \emph {et~al.}(2017)\citenamefont
  {Kobayashi}, \citenamefont {Murgia}, \citenamefont {De~Simone}, \citenamefont
  {Ir\v{s}i\v{c}},\ and\ \citenamefont {Viel}}]{Kobayashi:2017jcf}%
  \BibitemOpen
  \bibfield  {author} {\bibinfo {author} {\bibfnamefont {T.}~\bibnamefont
  {Kobayashi}}, \bibinfo {author} {\bibfnamefont {R.}~\bibnamefont {Murgia}},
  \bibinfo {author} {\bibfnamefont {A.}~\bibnamefont {De~Simone}}, \bibinfo
  {author} {\bibfnamefont {V.}~\bibnamefont {Ir\v{s}i\v{c}}}, \ and\ \bibinfo
  {author} {\bibfnamefont {M.}~\bibnamefont {Viel}},\ }\href {\doibase
  10.1103/PhysRevD.96.123514} {\bibfield  {journal} {\bibinfo  {journal} {Phys.
  Rev. D}\ }\textbf {\bibinfo {volume} {96}},\ \bibinfo {pages} {123514}
  (\bibinfo {year} {2017})},\ \Eprint {http://arxiv.org/abs/1708.00015}
  {arXiv:1708.00015 [astro-ph.CO]} \BibitemShut {NoStop}%
\bibitem [{\citenamefont {Lagu\"e}\ \emph {et~al.}(2022)\citenamefont
  {Lagu\"e}, \citenamefont {Bond}, \citenamefont {Hlo\v{z}ek}, \citenamefont
  {Rogers}, \citenamefont {Marsh},\ and\ \citenamefont {Grin}}]{Lague:2021frh}%
  \BibitemOpen
  \bibfield  {author} {\bibinfo {author} {\bibfnamefont {A.}~\bibnamefont
  {Lagu\"e}}, \bibinfo {author} {\bibfnamefont {J.~R.}\ \bibnamefont {Bond}},
  \bibinfo {author} {\bibfnamefont {R.}~\bibnamefont {Hlo\v{z}ek}}, \bibinfo
  {author} {\bibfnamefont {K.~K.}\ \bibnamefont {Rogers}}, \bibinfo {author}
  {\bibfnamefont {D.~J.~E.}\ \bibnamefont {Marsh}}, \ and\ \bibinfo {author}
  {\bibfnamefont {D.}~\bibnamefont {Grin}},\ }\href {\doibase
  10.1088/1475-7516/2022/01/049} {\bibfield  {journal} {\bibinfo  {journal}
  {JCAP}\ }\textbf {\bibinfo {volume} {01}},\ \bibinfo {pages} {049} (\bibinfo
  {year} {2022})},\ \Eprint {http://arxiv.org/abs/2104.07802} {arXiv:2104.07802
  [astro-ph.CO]} \BibitemShut {NoStop}%
\bibitem [{\citenamefont {Witten}(1984)}]{Witten:1984dg}%
  \BibitemOpen
  \bibfield  {author} {\bibinfo {author} {\bibfnamefont {E.}~\bibnamefont
  {Witten}},\ }\href {\doibase 10.1016/0370-2693(84)90422-2} {\bibfield
  {journal} {\bibinfo  {journal} {Phys. Lett. B}\ }\textbf {\bibinfo {volume}
  {149}},\ \bibinfo {pages} {351} (\bibinfo {year} {1984})}\BibitemShut
  {NoStop}%
\bibitem [{\citenamefont {Svrcek}\ and\ \citenamefont
  {Witten}(2006)}]{Svrcek:2006yi}%
  \BibitemOpen
  \bibfield  {author} {\bibinfo {author} {\bibfnamefont {P.}~\bibnamefont
  {Svrcek}}\ and\ \bibinfo {author} {\bibfnamefont {E.}~\bibnamefont
  {Witten}},\ }\href {\doibase 10.1088/1126-6708/2006/06/051} {\bibfield
  {journal} {\bibinfo  {journal} {JHEP}\ }\textbf {\bibinfo {volume} {06}},\
  \bibinfo {pages} {051} (\bibinfo {year} {2006})},\ \Eprint
  {http://arxiv.org/abs/hep-th/0605206} {arXiv:hep-th/0605206} \BibitemShut
  {NoStop}%
\bibitem [{\citenamefont {Conlon}(2006)}]{Conlon:2006tq}%
  \BibitemOpen
  \bibfield  {author} {\bibinfo {author} {\bibfnamefont {J.~P.}\ \bibnamefont
  {Conlon}},\ }\href {\doibase 10.1088/1126-6708/2006/05/078} {\bibfield
  {journal} {\bibinfo  {journal} {JHEP}\ }\textbf {\bibinfo {volume} {05}},\
  \bibinfo {pages} {078} (\bibinfo {year} {2006})},\ \Eprint
  {http://arxiv.org/abs/hep-th/0602233} {arXiv:hep-th/0602233} \BibitemShut
  {NoStop}%
\bibitem [{\citenamefont {Arvanitaki}\ \emph {et~al.}(2010)\citenamefont
  {Arvanitaki}, \citenamefont {Dimopoulos}, \citenamefont {Dubovsky},
  \citenamefont {Kaloper},\ and\ \citenamefont
  {March-Russell}}]{Arvanitaki:2009fg}%
  \BibitemOpen
  \bibfield  {author} {\bibinfo {author} {\bibfnamefont {A.}~\bibnamefont
  {Arvanitaki}}, \bibinfo {author} {\bibfnamefont {S.}~\bibnamefont
  {Dimopoulos}}, \bibinfo {author} {\bibfnamefont {S.}~\bibnamefont
  {Dubovsky}}, \bibinfo {author} {\bibfnamefont {N.}~\bibnamefont {Kaloper}}, \
  and\ \bibinfo {author} {\bibfnamefont {J.}~\bibnamefont {March-Russell}},\
  }\href {\doibase 10.1103/PhysRevD.81.123530} {\bibfield  {journal} {\bibinfo
  {journal} {Phys. Rev. D}\ }\textbf {\bibinfo {volume} {81}},\ \bibinfo
  {pages} {123530} (\bibinfo {year} {2010})},\ \Eprint
  {http://arxiv.org/abs/0905.4720} {arXiv:0905.4720 [hep-th]} \BibitemShut
  {NoStop}%
\bibitem [{\citenamefont {Mehta}\ \emph {et~al.}(2021)\citenamefont {Mehta},
  \citenamefont {Demirtas}, \citenamefont {Long}, \citenamefont {Marsh},
  \citenamefont {McAllister},\ and\ \citenamefont {Stott}}]{Mehta:2021pwf}%
  \BibitemOpen
  \bibfield  {author} {\bibinfo {author} {\bibfnamefont {V.~M.}\ \bibnamefont
  {Mehta}}, \bibinfo {author} {\bibfnamefont {M.}~\bibnamefont {Demirtas}},
  \bibinfo {author} {\bibfnamefont {C.}~\bibnamefont {Long}}, \bibinfo {author}
  {\bibfnamefont {D.~J.~E.}\ \bibnamefont {Marsh}}, \bibinfo {author}
  {\bibfnamefont {L.}~\bibnamefont {McAllister}}, \ and\ \bibinfo {author}
  {\bibfnamefont {M.~J.}\ \bibnamefont {Stott}},\ }\href {\doibase
  10.1088/1475-7516/2021/07/033} {\bibfield  {journal} {\bibinfo  {journal}
  {JCAP}\ }\textbf {\bibinfo {volume} {07}},\ \bibinfo {pages} {033} (\bibinfo
  {year} {2021})},\ \Eprint {http://arxiv.org/abs/2103.06812} {arXiv:2103.06812
  [hep-th]} \BibitemShut {NoStop}%
\bibitem [{\citenamefont {Demirtas}\ \emph {et~al.}(2021)\citenamefont
  {Demirtas}, \citenamefont {Gendler}, \citenamefont {Long}, \citenamefont
  {McAllister},\ and\ \citenamefont {Moritz}}]{Demirtas:2021gsq}%
  \BibitemOpen
  \bibfield  {author} {\bibinfo {author} {\bibfnamefont {M.}~\bibnamefont
  {Demirtas}}, \bibinfo {author} {\bibfnamefont {N.}~\bibnamefont {Gendler}},
  \bibinfo {author} {\bibfnamefont {C.}~\bibnamefont {Long}}, \bibinfo {author}
  {\bibfnamefont {L.}~\bibnamefont {McAllister}}, \ and\ \bibinfo {author}
  {\bibfnamefont {J.}~\bibnamefont {Moritz}},\ }\href@noop {} {\  (\bibinfo
  {year} {2021})},\ \Eprint {http://arxiv.org/abs/2112.04503} {arXiv:2112.04503
  [hep-th]} \BibitemShut {NoStop}%
\bibitem [{\citenamefont {Cicoli}\ \emph {et~al.}(2022)\citenamefont {Cicoli},
  \citenamefont {Guidetti}, \citenamefont {Righi},\ and\ \citenamefont
  {Westphal}}]{Cicoli:2021gss}%
  \BibitemOpen
  \bibfield  {author} {\bibinfo {author} {\bibfnamefont {M.}~\bibnamefont
  {Cicoli}}, \bibinfo {author} {\bibfnamefont {V.}~\bibnamefont {Guidetti}},
  \bibinfo {author} {\bibfnamefont {N.}~\bibnamefont {Righi}}, \ and\ \bibinfo
  {author} {\bibfnamefont {A.}~\bibnamefont {Westphal}},\ }\href {\doibase
  10.1007/JHEP05(2022)107} {\bibfield  {journal} {\bibinfo  {journal} {JHEP}\
  }\textbf {\bibinfo {volume} {05}},\ \bibinfo {pages} {107} (\bibinfo {year}
  {2022})},\ \Eprint {http://arxiv.org/abs/2110.02964} {arXiv:2110.02964
  [hep-th]} \BibitemShut {NoStop}%
\bibitem [{\citenamefont {Hlo\v{z}ek}\ \emph {et~al.}(2017)\citenamefont
  {Hlo\v{z}ek}, \citenamefont {Marsh}, \citenamefont {Grin}, \citenamefont
  {Allison}, \citenamefont {Dunkley},\ and\ \citenamefont
  {Calabrese}}]{Hlozek:2016lzm}%
  \BibitemOpen
  \bibfield  {author} {\bibinfo {author} {\bibfnamefont {R.}~\bibnamefont
  {Hlo\v{z}ek}}, \bibinfo {author} {\bibfnamefont {D.~J.~E.}\ \bibnamefont
  {Marsh}}, \bibinfo {author} {\bibfnamefont {D.}~\bibnamefont {Grin}},
  \bibinfo {author} {\bibfnamefont {R.}~\bibnamefont {Allison}}, \bibinfo
  {author} {\bibfnamefont {J.}~\bibnamefont {Dunkley}}, \ and\ \bibinfo
  {author} {\bibfnamefont {E.}~\bibnamefont {Calabrese}},\ }\href {\doibase
  10.1103/PhysRevD.95.123511} {\bibfield  {journal} {\bibinfo  {journal} {Phys.
  Rev. D}\ }\textbf {\bibinfo {volume} {95}},\ \bibinfo {pages} {123511}
  (\bibinfo {year} {2017})},\ \Eprint {http://arxiv.org/abs/1607.08208}
  {arXiv:1607.08208 [astro-ph.CO]} \BibitemShut {NoStop}%
\bibitem [{\citenamefont {Bauer}\ \emph {et~al.}(2020)\citenamefont {Bauer},
  \citenamefont {Marsh}, \citenamefont {Hložek}, \citenamefont {Padmanabhan},\
  and\ \citenamefont {Laguë}}]{bauer_biased_tracer_H1}%
  \BibitemOpen
  \bibfield  {author} {\bibinfo {author} {\bibfnamefont {J.~B.}\ \bibnamefont
  {Bauer}}, \bibinfo {author} {\bibfnamefont {D.~J.~E.}\ \bibnamefont {Marsh}},
  \bibinfo {author} {\bibfnamefont {R.}~\bibnamefont {Hložek}}, \bibinfo
  {author} {\bibfnamefont {H.}~\bibnamefont {Padmanabhan}}, \ and\ \bibinfo
  {author} {\bibfnamefont {A.}~\bibnamefont {Laguë}},\ }\href {\doibase
  10.1093/mnras/staa3300} {\bibfield  {journal} {\bibinfo  {journal} {Monthly
  Notices of the Royal Astronomical Society}\ }\textbf {\bibinfo {volume}
  {500}},\ \bibinfo {pages} {3162} (\bibinfo {year} {2020})},\ \bibinfo {note}
  {arXiv:2003.09655}\BibitemShut {NoStop}%
\bibitem [{\citenamefont {Farren}\ \emph {et~al.}(2022)\citenamefont {Farren},
  \citenamefont {Grin}, \citenamefont {Jaffe}, \citenamefont {Hlo\v{z}ek},\
  and\ \citenamefont {Marsh}}]{Farren:2021jcd}%
  \BibitemOpen
  \bibfield  {author} {\bibinfo {author} {\bibfnamefont {G.~S.}\ \bibnamefont
  {Farren}}, \bibinfo {author} {\bibfnamefont {D.}~\bibnamefont {Grin}},
  \bibinfo {author} {\bibfnamefont {A.~H.}\ \bibnamefont {Jaffe}}, \bibinfo
  {author} {\bibfnamefont {R.}~\bibnamefont {Hlo\v{z}ek}}, \ and\ \bibinfo
  {author} {\bibfnamefont {D.~J.~E.}\ \bibnamefont {Marsh}},\ }\href {\doibase
  10.1103/PhysRevD.105.063513} {\bibfield  {journal} {\bibinfo  {journal}
  {Phys. Rev. D}\ }\textbf {\bibinfo {volume} {105}},\ \bibinfo {pages}
  {063513} (\bibinfo {year} {2022})},\ \Eprint
  {http://arxiv.org/abs/2109.13268} {arXiv:2109.13268 [astro-ph.CO]}
  \BibitemShut {NoStop}%
\bibitem [{\citenamefont {Abazajian}\ \emph {et~al.}(2016)\citenamefont
  {Abazajian} \emph {et~al.}}]{CMB-S4:2016ple}%
  \BibitemOpen
  \bibfield  {author} {\bibinfo {author} {\bibfnamefont {K.~N.}\ \bibnamefont
  {Abazajian}} \emph {et~al.} (\bibinfo {collaboration} {CMB-S4}),\ }\href@noop
  {} {\  (\bibinfo {year} {2016})},\ \Eprint {http://arxiv.org/abs/1610.02743}
  {arXiv:1610.02743 [astro-ph.CO]} \BibitemShut {NoStop}%
\bibitem [{\citenamefont {Dvorkin}\ \emph {et~al.}(2022)\citenamefont {Dvorkin}
  \emph {et~al.}}]{Dvorkin:2022bsc}%
  \BibitemOpen
  \bibfield  {author} {\bibinfo {author} {\bibfnamefont {C.}~\bibnamefont
  {Dvorkin}} \emph {et~al.},\ }in\ \href@noop {} {\emph {\bibinfo {booktitle}
  {{2022 Snowmass Summer Study}}}}\ (\bibinfo {year} {2022})\ \Eprint
  {http://arxiv.org/abs/2203.07064} {arXiv:2203.07064 [hep-ph]} \BibitemShut
  {NoStop}%
\bibitem [{\citenamefont {Flitter}\ and\ \citenamefont
  {Kovetz}(2022)}]{Flitter:2022pzf}%
  \BibitemOpen
  \bibfield  {author} {\bibinfo {author} {\bibfnamefont {J.}~\bibnamefont
  {Flitter}}\ and\ \bibinfo {author} {\bibfnamefont {E.~D.}\ \bibnamefont
  {Kovetz}},\ }\href@noop {} {\  (\bibinfo {year} {2022})},\ \Eprint
  {http://arxiv.org/abs/2207.05083} {arXiv:2207.05083 [astro-ph.CO]}
  \BibitemShut {NoStop}%
\bibitem [{\citenamefont {Vogelsberger}\ \emph {et~al.}(2020)\citenamefont
  {Vogelsberger}, \citenamefont {Marinacci}, \citenamefont {Torrey},\ and\
  \citenamefont {Puchwein}}]{sims_review}%
  \BibitemOpen
  \bibfield  {author} {\bibinfo {author} {\bibfnamefont {M.}~\bibnamefont
  {Vogelsberger}}, \bibinfo {author} {\bibfnamefont {F.}~\bibnamefont
  {Marinacci}}, \bibinfo {author} {\bibfnamefont {P.}~\bibnamefont {Torrey}}, \
  and\ \bibinfo {author} {\bibfnamefont {E.}~\bibnamefont {Puchwein}},\ }\href
  {\doibase 10.1038/s42254-019-0127-2} {\bibfield  {journal} {\bibinfo
  {journal} {Nature Rev. Phys.}\ }\textbf {\bibinfo {volume} {2}},\ \bibinfo
  {pages} {42} (\bibinfo {year} {2020})},\ \Eprint
  {http://arxiv.org/abs/1909.07976} {arXiv:1909.07976 [astro-ph.GA]}
  \BibitemShut {NoStop}%
\bibitem [{\citenamefont {Heitmann}\ \emph {et~al.}(2013)\citenamefont
  {Heitmann}, \citenamefont {Lawrence}, \citenamefont {Kwan}, \citenamefont
  {Habib},\ and\ \citenamefont {Higdon}}]{heitmann_coyote_emulator}%
  \BibitemOpen
  \bibfield  {author} {\bibinfo {author} {\bibfnamefont {K.}~\bibnamefont
  {Heitmann}}, \bibinfo {author} {\bibfnamefont {E.}~\bibnamefont {Lawrence}},
  \bibinfo {author} {\bibfnamefont {J.}~\bibnamefont {Kwan}}, \bibinfo {author}
  {\bibfnamefont {S.}~\bibnamefont {Habib}}, \ and\ \bibinfo {author}
  {\bibfnamefont {D.}~\bibnamefont {Higdon}},\ }\href {\doibase
  10.1088/0004-637X/780/1/111} {\bibfield  {journal} {\bibinfo  {journal} {The
  Astrophysical Journal}\ }\textbf {\bibinfo {volume} {780}},\ \bibinfo {pages}
  {111} (\bibinfo {year} {2013})}\BibitemShut {NoStop}%
\bibitem [{\citenamefont {Heitmann}\ \emph {et~al.}(2016)\citenamefont
  {Heitmann}, \citenamefont {Bingham}, \citenamefont {Lawrence}, \citenamefont
  {Bergner}, \citenamefont {Habib}, \citenamefont {Higdon}, \citenamefont
  {Pope}, \citenamefont {Biswas}, \citenamefont {Finkel}, \citenamefont
  {Frontiere},\ and\ \citenamefont
  {Bhattacharya}}]{heitmann_miratitan_emulator}%
  \BibitemOpen
  \bibfield  {author} {\bibinfo {author} {\bibfnamefont {K.}~\bibnamefont
  {Heitmann}}, \bibinfo {author} {\bibfnamefont {D.}~\bibnamefont {Bingham}},
  \bibinfo {author} {\bibfnamefont {E.}~\bibnamefont {Lawrence}}, \bibinfo
  {author} {\bibfnamefont {S.}~\bibnamefont {Bergner}}, \bibinfo {author}
  {\bibfnamefont {S.}~\bibnamefont {Habib}}, \bibinfo {author} {\bibfnamefont
  {D.}~\bibnamefont {Higdon}}, \bibinfo {author} {\bibfnamefont
  {A.}~\bibnamefont {Pope}}, \bibinfo {author} {\bibfnamefont {R.}~\bibnamefont
  {Biswas}}, \bibinfo {author} {\bibfnamefont {H.}~\bibnamefont {Finkel}},
  \bibinfo {author} {\bibfnamefont {N.}~\bibnamefont {Frontiere}}, \ and\
  \bibinfo {author} {\bibfnamefont {S.}~\bibnamefont {Bhattacharya}},\ }\href
  {\doibase 10.3847/0004-637X/820/2/108} {\bibfield  {journal} {\bibinfo
  {journal} {The Astrophysical Journal}\ }\textbf {\bibinfo {volume} {820}},\
  \bibinfo {pages} {108} (\bibinfo {year} {2016})}\BibitemShut {NoStop}%
\bibitem [{\citenamefont {Rogers}\ \emph {et~al.}(2019)\citenamefont {Rogers},
  \citenamefont {Peiris}, \citenamefont {Pontzen}, \citenamefont {Bird},
  \citenamefont {Verde},\ and\ \citenamefont {Font-Ribera}}]{Rogers:2018smb}%
  \BibitemOpen
  \bibfield  {author} {\bibinfo {author} {\bibfnamefont {K.~K.}\ \bibnamefont
  {Rogers}}, \bibinfo {author} {\bibfnamefont {H.~V.}\ \bibnamefont {Peiris}},
  \bibinfo {author} {\bibfnamefont {A.}~\bibnamefont {Pontzen}}, \bibinfo
  {author} {\bibfnamefont {S.}~\bibnamefont {Bird}}, \bibinfo {author}
  {\bibfnamefont {L.}~\bibnamefont {Verde}}, \ and\ \bibinfo {author}
  {\bibfnamefont {A.}~\bibnamefont {Font-Ribera}},\ }\href {\doibase
  10.1088/1475-7516/2019/02/031} {\bibfield  {journal} {\bibinfo  {journal}
  {JCAP}\ }\textbf {\bibinfo {volume} {02}},\ \bibinfo {pages} {031} (\bibinfo
  {year} {2019})},\ \Eprint {http://arxiv.org/abs/1812.04631} {arXiv:1812.04631
  [astro-ph.CO]} \BibitemShut {NoStop}%
\bibitem [{\citenamefont {Pedersen}\ \emph {et~al.}(2021)\citenamefont
  {Pedersen}, \citenamefont {Font-Ribera}, \citenamefont {Rogers},
  \citenamefont {McDonald}, \citenamefont {Peiris}, \citenamefont {Pontzen},\
  and\ \citenamefont {Slosar}}]{Pedersen:2020kaw}%
  \BibitemOpen
  \bibfield  {author} {\bibinfo {author} {\bibfnamefont {C.}~\bibnamefont
  {Pedersen}}, \bibinfo {author} {\bibfnamefont {A.}~\bibnamefont
  {Font-Ribera}}, \bibinfo {author} {\bibfnamefont {K.~K.}\ \bibnamefont
  {Rogers}}, \bibinfo {author} {\bibfnamefont {P.}~\bibnamefont {McDonald}},
  \bibinfo {author} {\bibfnamefont {H.~V.}\ \bibnamefont {Peiris}}, \bibinfo
  {author} {\bibfnamefont {A.}~\bibnamefont {Pontzen}}, \ and\ \bibinfo
  {author} {\bibfnamefont {A.}~\bibnamefont {Slosar}},\ }\href {\doibase
  10.1088/1475-7516/2021/05/033} {\bibfield  {journal} {\bibinfo  {journal}
  {JCAP}\ }\textbf {\bibinfo {volume} {05}},\ \bibinfo {pages} {033} (\bibinfo
  {year} {2021})},\ \Eprint {http://arxiv.org/abs/2011.15127} {arXiv:2011.15127
  [astro-ph.CO]} \BibitemShut {NoStop}%
\bibitem [{\citenamefont {Cooray}\ and\ \citenamefont
  {Sheth}(2002)}]{halo_model_cooray_sheth}%
  \BibitemOpen
  \bibfield  {author} {\bibinfo {author} {\bibfnamefont {A.}~\bibnamefont
  {Cooray}}\ and\ \bibinfo {author} {\bibfnamefont {R.}~\bibnamefont {Sheth}},\
  }\href {\doibase https://doi.org/10.1016/S0370-1573(02)00276-4} {\bibfield
  {journal} {\bibinfo  {journal} {Physics Reports}\ }\textbf {\bibinfo {volume}
  {372}},\ \bibinfo {pages} {1} (\bibinfo {year} {2002})},\ \bibinfo {note}
  {arXiv: astro-ph/0206508}\BibitemShut {NoStop}%
\bibitem [{\citenamefont {Mead}\ \emph
  {et~al.}(2015{\natexlab{a}})\citenamefont {Mead}, \citenamefont {Peacock},
  \citenamefont {Heymans}, \citenamefont {Joudaki},\ and\ \citenamefont
  {Heavens}}]{HMCode_mead_2015}%
  \BibitemOpen
  \bibfield  {author} {\bibinfo {author} {\bibfnamefont {A.~J.}\ \bibnamefont
  {Mead}}, \bibinfo {author} {\bibfnamefont {J.~A.}\ \bibnamefont {Peacock}},
  \bibinfo {author} {\bibfnamefont {C.}~\bibnamefont {Heymans}}, \bibinfo
  {author} {\bibfnamefont {S.}~\bibnamefont {Joudaki}}, \ and\ \bibinfo
  {author} {\bibfnamefont {A.~F.}\ \bibnamefont {Heavens}},\ }\href {\doibase
  10.1093/mnras/stv2036} {\bibfield  {journal} {\bibinfo  {journal} {Monthly
  Notices of the Royal Astronomical Society}\ }\textbf {\bibinfo {volume}
  {454}},\ \bibinfo {pages} {1958} (\bibinfo {year}
  {2015}{\natexlab{a}})}\BibitemShut {NoStop}%
\bibitem [{\citenamefont {Mead}\ \emph
  {et~al.}(2021{\natexlab{a}})\citenamefont {Mead}, \citenamefont {Brieden},
  \citenamefont {Tröster},\ and\ \citenamefont {Heymans}}]{mead_hmcode2020}%
  \BibitemOpen
  \bibfield  {author} {\bibinfo {author} {\bibfnamefont {A.}~\bibnamefont
  {Mead}}, \bibinfo {author} {\bibfnamefont {S.}~\bibnamefont {Brieden}},
  \bibinfo {author} {\bibfnamefont {T.}~\bibnamefont {Tröster}}, \ and\
  \bibinfo {author} {\bibfnamefont {C.}~\bibnamefont {Heymans}},\ }\href
  {\doibase 10.1093/mnras/stab082} {\bibfield  {journal} {\bibinfo  {journal}
  {Monthly Notices of the Royal Astronomical Society}\ }\textbf {\bibinfo
  {volume} {502}},\ \bibinfo {pages} {1401} (\bibinfo {year}
  {2021}{\natexlab{a}})},\ \bibinfo {note} {arXiv: 2009.01858}\BibitemShut
  {NoStop}%
\bibitem [{\citenamefont {Castro}\ \emph {et~al.}(2022)\citenamefont {Castro}
  \emph {et~al.}}]{Euclid:2022dbc}%
  \BibitemOpen
  \bibfield  {author} {\bibinfo {author} {\bibfnamefont {T.}~\bibnamefont
  {Castro}} \emph {et~al.} (\bibinfo {collaboration} {Euclid}),\ }\href@noop {}
  {\  (\bibinfo {year} {2022})},\ \Eprint {http://arxiv.org/abs/2208.02174}
  {arXiv:2208.02174 [astro-ph.CO]} \BibitemShut {NoStop}%
\bibitem [{\citenamefont {Marsh}\ and\ \citenamefont
  {Silk}(2014)}]{marsh_silk_cut_off_mass}%
  \BibitemOpen
  \bibfield  {author} {\bibinfo {author} {\bibfnamefont {D.~J.~E.}\
  \bibnamefont {Marsh}}\ and\ \bibinfo {author} {\bibfnamefont
  {J.}~\bibnamefont {Silk}},\ }\href {\doibase 10.1093/mnras/stt2079}
  {\bibfield  {journal} {\bibinfo  {journal} {Monthly Notices of the Royal
  Astronomical Society}\ }\textbf {\bibinfo {volume} {437}},\ \bibinfo {pages}
  {2652} (\bibinfo {year} {2014})},\ \bibinfo {note}
  {arXiv:1307.1705}\BibitemShut {NoStop}%
\bibitem [{\citenamefont {LoVerde}\ and\ \citenamefont
  {Zaldarriaga}(2014)}]{LoVerde:2013lta}%
  \BibitemOpen
  \bibfield  {author} {\bibinfo {author} {\bibfnamefont {M.}~\bibnamefont
  {LoVerde}}\ and\ \bibinfo {author} {\bibfnamefont {M.}~\bibnamefont
  {Zaldarriaga}},\ }\href {\doibase 10.1103/PhysRevD.89.063502} {\bibfield
  {journal} {\bibinfo  {journal} {Phys. Rev. D}\ }\textbf {\bibinfo {volume}
  {89}},\ \bibinfo {pages} {063502} (\bibinfo {year} {2014})},\ \Eprint
  {http://arxiv.org/abs/1310.6459} {arXiv:1310.6459 [astro-ph.CO]} \BibitemShut
  {NoStop}%
\bibitem [{\citenamefont {Massara}\ \emph {et~al.}(2014)\citenamefont
  {Massara}, \citenamefont {Villaescusa-Navarro},\ and\ \citenamefont
  {Viel}}]{massara_MDM_halo_model}%
  \BibitemOpen
  \bibfield  {author} {\bibinfo {author} {\bibfnamefont {E.}~\bibnamefont
  {Massara}}, \bibinfo {author} {\bibfnamefont {F.}~\bibnamefont
  {Villaescusa-Navarro}}, \ and\ \bibinfo {author} {\bibfnamefont
  {M.}~\bibnamefont {Viel}},\ }\href {\doibase 10.1088/1475-7516/2014/12/053}
  {\bibfield  {journal} {\bibinfo  {journal} {Journal of Cosmology and
  Astroparticle Physics}\ }\textbf {\bibinfo {volume} {2014}},\ \bibinfo
  {pages} {053} (\bibinfo {year} {2014})},\ \bibinfo {note}
  {arXiv:1410.6813}\BibitemShut {NoStop}%
\bibitem [{\citenamefont {Padmanabhan}\ and\ \citenamefont
  {Refregier}(2017)}]{Padmanabhan:2016odj}%
  \BibitemOpen
  \bibfield  {author} {\bibinfo {author} {\bibfnamefont {H.}~\bibnamefont
  {Padmanabhan}}\ and\ \bibinfo {author} {\bibfnamefont {A.}~\bibnamefont
  {Refregier}},\ }\href {\doibase 10.1093/mnras/stw2706} {\bibfield  {journal}
  {\bibinfo  {journal} {Mon. Not. Roy. Astron. Soc.}\ }\textbf {\bibinfo
  {volume} {464}},\ \bibinfo {pages} {4008} (\bibinfo {year} {2017})},\ \Eprint
  {http://arxiv.org/abs/1607.01021} {arXiv:1607.01021 [astro-ph.CO]}
  \BibitemShut {NoStop}%
\bibitem [{\citenamefont {Hamaide}\ \emph {et~al.}(2022)\citenamefont
  {Hamaide}, \citenamefont {M\"uller},\ and\ \citenamefont
  {Marsh}}]{Hamaide:2022rwi}%
  \BibitemOpen
  \bibfield  {author} {\bibinfo {author} {\bibfnamefont {L.}~\bibnamefont
  {Hamaide}}, \bibinfo {author} {\bibfnamefont {H.}~\bibnamefont {M\"uller}}, \
  and\ \bibinfo {author} {\bibfnamefont {D.~J.~E.}\ \bibnamefont {Marsh}},\
  }\href {\doibase 10.1103/PhysRevD.106.123509} {\bibfield  {journal} {\bibinfo
   {journal} {Phys. Rev. D}\ }\textbf {\bibinfo {volume} {106}},\ \bibinfo
  {pages} {123509} (\bibinfo {year} {2022})},\ \Eprint
  {http://arxiv.org/abs/2210.03705} {arXiv:2210.03705 [astro-ph.CO]}
  \BibitemShut {NoStop}%
\bibitem [{\citenamefont {Li}\ \emph {et~al.}(2014)\citenamefont {Li},
  \citenamefont {Rindler-Daller},\ and\ \citenamefont {Shapiro}}]{Li:2013nal}%
  \BibitemOpen
  \bibfield  {author} {\bibinfo {author} {\bibfnamefont {B.}~\bibnamefont
  {Li}}, \bibinfo {author} {\bibfnamefont {T.}~\bibnamefont {Rindler-Daller}},
  \ and\ \bibinfo {author} {\bibfnamefont {P.~R.}\ \bibnamefont {Shapiro}},\
  }\href {\doibase 10.1103/PhysRevD.89.083536} {\bibfield  {journal} {\bibinfo
  {journal} {Phys. Rev. D}\ }\textbf {\bibinfo {volume} {89}},\ \bibinfo
  {pages} {083536} (\bibinfo {year} {2014})},\ \Eprint
  {http://arxiv.org/abs/1310.6061} {arXiv:1310.6061 [astro-ph.CO]} \BibitemShut
  {NoStop}%
\bibitem [{\citenamefont {Gorghetto}\ \emph {et~al.}(2022)\citenamefont
  {Gorghetto}, \citenamefont {Hardy}, \citenamefont {March-Russell},
  \citenamefont {Song},\ and\ \citenamefont {West}}]{Gorghetto:2022sue}%
  \BibitemOpen
  \bibfield  {author} {\bibinfo {author} {\bibfnamefont {M.}~\bibnamefont
  {Gorghetto}}, \bibinfo {author} {\bibfnamefont {E.}~\bibnamefont {Hardy}},
  \bibinfo {author} {\bibfnamefont {J.}~\bibnamefont {March-Russell}}, \bibinfo
  {author} {\bibfnamefont {N.}~\bibnamefont {Song}}, \ and\ \bibinfo {author}
  {\bibfnamefont {S.~M.}\ \bibnamefont {West}},\ }\href {\doibase
  10.1088/1475-7516/2022/08/018} {\bibfield  {journal} {\bibinfo  {journal}
  {JCAP}\ }\textbf {\bibinfo {volume} {08}},\ \bibinfo {pages} {018} (\bibinfo
  {year} {2022})},\ \Eprint {http://arxiv.org/abs/2203.10100} {arXiv:2203.10100
  [hep-ph]} \BibitemShut {NoStop}%
\bibitem [{\citenamefont {Marsh}(2016{\natexlab{a}})}]{marsh_axion_cosmo}%
  \BibitemOpen
  \bibfield  {author} {\bibinfo {author} {\bibfnamefont {D.~J.~E.}\
  \bibnamefont {Marsh}},\ }\href {\doibase 10.1016/j.physrep.2016.06.005}
  {\bibfield  {journal} {\bibinfo  {journal} {Physics Reports}\ }\textbf
  {\bibinfo {volume} {643}},\ \bibinfo {pages} {1} (\bibinfo {year}
  {2016}{\natexlab{a}})},\ \bibinfo {note} {arXiv:1510.07633}\BibitemShut
  {NoStop}%
\bibitem [{\citenamefont {Schwabe}\ \emph {et~al.}(2020)\citenamefont
  {Schwabe}, \citenamefont {Gosenca}, \citenamefont {Behrens}, \citenamefont
  {Niemeyer},\ and\ \citenamefont {Easther}}]{bodo_simulation}%
  \BibitemOpen
  \bibfield  {author} {\bibinfo {author} {\bibfnamefont {B.}~\bibnamefont
  {Schwabe}}, \bibinfo {author} {\bibfnamefont {M.}~\bibnamefont {Gosenca}},
  \bibinfo {author} {\bibfnamefont {C.}~\bibnamefont {Behrens}}, \bibinfo
  {author} {\bibfnamefont {J.~C.}\ \bibnamefont {Niemeyer}}, \ and\ \bibinfo
  {author} {\bibfnamefont {R.}~\bibnamefont {Easther}},\ }\href {\doibase
  10.1103/PhysRevD.102.083518} {\bibfield  {journal} {\bibinfo  {journal}
  {Physical Review D}\ }\textbf {\bibinfo {volume} {102}},\ \bibinfo {pages}
  {083518} (\bibinfo {year} {2020})},\ \bibinfo {note}
  {arXiv:2007.08256}\BibitemShut {NoStop}%
\bibitem [{\citenamefont {Lague}\ \emph {et~al.}(2022)\citenamefont {Lague}
  \emph {et~al.}}]{Lague_in_prep}%
  \BibitemOpen
  \bibfield  {author} {\bibinfo {author} {\bibfnamefont {A.}~\bibnamefont
  {Lague}} \emph {et~al.},\ }\href@noop {} {\bibfield  {journal} {\bibinfo
  {journal} {In preparation}\ } (\bibinfo {year} {2022})}\BibitemShut {NoStop}%
\bibitem [{\citenamefont {Preskill}\ \emph {et~al.}(1983)\citenamefont
  {Preskill}, \citenamefont {Wise},\ and\ \citenamefont
  {Wilczek}}]{Preskill:1982cy}%
  \BibitemOpen
  \bibfield  {author} {\bibinfo {author} {\bibfnamefont {J.}~\bibnamefont
  {Preskill}}, \bibinfo {author} {\bibfnamefont {M.~B.}\ \bibnamefont {Wise}},
  \ and\ \bibinfo {author} {\bibfnamefont {F.}~\bibnamefont {Wilczek}},\ }\href
  {\doibase 10.1016/0370-2693(83)90637-8} {\bibfield  {journal} {\bibinfo
  {journal} {Phys. Lett. B}\ }\textbf {\bibinfo {volume} {120}},\ \bibinfo
  {pages} {127} (\bibinfo {year} {1983})}\BibitemShut {NoStop}%
\bibitem [{\citenamefont {Abbott}\ and\ \citenamefont
  {Sikivie}(1983)}]{Abbott:1982af}%
  \BibitemOpen
  \bibfield  {author} {\bibinfo {author} {\bibfnamefont {L.}~\bibnamefont
  {Abbott}}\ and\ \bibinfo {author} {\bibfnamefont {P.}~\bibnamefont
  {Sikivie}},\ }\href {\doibase 10.1016/0370-2693(83)90638-X} {\bibfield
  {journal} {\bibinfo  {journal} {Phys. Lett. B}\ }\textbf {\bibinfo {volume}
  {120}},\ \bibinfo {pages} {133} (\bibinfo {year} {1983})}\BibitemShut
  {NoStop}%
\bibitem [{\citenamefont {Dine}\ and\ \citenamefont
  {Fischler}(1983)}]{Dine:1982ah}%
  \BibitemOpen
  \bibfield  {author} {\bibinfo {author} {\bibfnamefont {M.}~\bibnamefont
  {Dine}}\ and\ \bibinfo {author} {\bibfnamefont {W.}~\bibnamefont
  {Fischler}},\ }\href {\doibase 10.1016/0370-2693(83)90639-1} {\bibfield
  {journal} {\bibinfo  {journal} {Phys. Lett. B}\ }\textbf {\bibinfo {volume}
  {120}},\ \bibinfo {pages} {137} (\bibinfo {year} {1983})}\BibitemShut
  {NoStop}%
\bibitem [{\citenamefont {Ma}\ and\ \citenamefont
  {Bertschinger}(1995)}]{Ma:1995ey}%
  \BibitemOpen
  \bibfield  {author} {\bibinfo {author} {\bibfnamefont {C.-P.}\ \bibnamefont
  {Ma}}\ and\ \bibinfo {author} {\bibfnamefont {E.}~\bibnamefont
  {Bertschinger}},\ }\href {\doibase 10.1086/176550} {\bibfield  {journal}
  {\bibinfo  {journal} {Astrophys. J.}\ }\textbf {\bibinfo {volume} {455}},\
  \bibinfo {pages} {7} (\bibinfo {year} {1995})},\ \Eprint
  {http://arxiv.org/abs/astro-ph/9506072} {arXiv:astro-ph/9506072} \BibitemShut
  {NoStop}%
\bibitem [{\citenamefont {chan Hwang}\ and\ \citenamefont
  {Noh}(2009)}]{sound_speed_ax}%
  \BibitemOpen
  \bibfield  {author} {\bibinfo {author} {\bibfnamefont {J.}~\bibnamefont {chan
  Hwang}}\ and\ \bibinfo {author} {\bibfnamefont {H.}~\bibnamefont {Noh}},\
  }\href {\doibase https://doi.org/10.1016/j.physletb.2009.08.031} {\bibfield
  {journal} {\bibinfo  {journal} {Physics Letters B}\ }\textbf {\bibinfo
  {volume} {680}},\ \bibinfo {pages} {1} (\bibinfo {year} {2009})}\BibitemShut
  {NoStop}%
\bibitem [{\citenamefont {Khlopov}\ \emph {et~al.}(1985)\citenamefont
  {Khlopov}, \citenamefont {Malomed},\ and\ \citenamefont
  {Zeldovich}}]{Khlopov:1985jw}%
  \BibitemOpen
  \bibfield  {author} {\bibinfo {author} {\bibfnamefont {M.}~\bibnamefont
  {Khlopov}}, \bibinfo {author} {\bibfnamefont {B.~A.}\ \bibnamefont
  {Malomed}}, \ and\ \bibinfo {author} {\bibfnamefont {I.~B.}\ \bibnamefont
  {Zeldovich}},\ }\href@noop {} {\bibfield  {journal} {\bibinfo  {journal}
  {Mon. Not. Roy. Astron. Soc.}\ }\textbf {\bibinfo {volume} {215}},\ \bibinfo
  {pages} {575} (\bibinfo {year} {1985})}\BibitemShut {NoStop}%
\bibitem [{\citenamefont {Hu}\ \emph {et~al.}(2000)\citenamefont {Hu},
  \citenamefont {Barkana},\ and\ \citenamefont {Gruzinov}}]{halo_jeans_scale}%
  \BibitemOpen
  \bibfield  {author} {\bibinfo {author} {\bibfnamefont {W.}~\bibnamefont
  {Hu}}, \bibinfo {author} {\bibfnamefont {R.}~\bibnamefont {Barkana}}, \ and\
  \bibinfo {author} {\bibfnamefont {A.}~\bibnamefont {Gruzinov}},\ }\href
  {\doibase 10.1103/PhysRevLett.85.1158} {\bibfield  {journal} {\bibinfo
  {journal} {Physical Review Letters}\ }\textbf {\bibinfo {volume} {85}},\
  \bibinfo {pages} {1158} (\bibinfo {year} {2000})},\ \bibinfo {note}
  {arXiv:astro-ph/0003365}\BibitemShut {NoStop}%
\bibitem [{\citenamefont {Amendola}\ and\ \citenamefont
  {Barbieri}(2006)}]{Amendola:2005ad}%
  \BibitemOpen
  \bibfield  {author} {\bibinfo {author} {\bibfnamefont {L.}~\bibnamefont
  {Amendola}}\ and\ \bibinfo {author} {\bibfnamefont {R.}~\bibnamefont
  {Barbieri}},\ }\href {\doibase 10.1016/j.physletb.2006.08.069} {\bibfield
  {journal} {\bibinfo  {journal} {Phys. Lett. B}\ }\textbf {\bibinfo {volume}
  {642}},\ \bibinfo {pages} {192} (\bibinfo {year} {2006})},\ \Eprint
  {http://arxiv.org/abs/hep-ph/0509257} {arXiv:hep-ph/0509257} \BibitemShut
  {NoStop}%
\bibitem [{\citenamefont {Marsh}\ and\ \citenamefont
  {Ferreira}(2010)}]{Marsh:2010wq}%
  \BibitemOpen
  \bibfield  {author} {\bibinfo {author} {\bibfnamefont {D.~J.~E.}\
  \bibnamefont {Marsh}}\ and\ \bibinfo {author} {\bibfnamefont {P.~G.}\
  \bibnamefont {Ferreira}},\ }\href {\doibase 10.1103/PhysRevD.82.103528}
  {\bibfield  {journal} {\bibinfo  {journal} {Phys. Rev. D}\ }\textbf {\bibinfo
  {volume} {82}},\ \bibinfo {pages} {103528} (\bibinfo {year} {2010})},\
  \Eprint {http://arxiv.org/abs/1009.3501} {arXiv:1009.3501 [hep-ph]}
  \BibitemShut {NoStop}%
\bibitem [{\citenamefont {Hlozek}\ \emph
  {et~al.}(2015{\natexlab{b}})\citenamefont {Hlozek}, \citenamefont {Grin},
  \citenamefont {Marsh},\ and\ \citenamefont {Ferreira}}]{axionCAMB}%
  \BibitemOpen
  \bibfield  {author} {\bibinfo {author} {\bibfnamefont {R.}~\bibnamefont
  {Hlozek}}, \bibinfo {author} {\bibfnamefont {D.}~\bibnamefont {Grin}},
  \bibinfo {author} {\bibfnamefont {D.~J.~E.}\ \bibnamefont {Marsh}}, \ and\
  \bibinfo {author} {\bibfnamefont {P.~G.}\ \bibnamefont {Ferreira}},\ }\href
  {\doibase 10.1103/PhysRevD.91.103512} {\bibfield  {journal} {\bibinfo
  {journal} {Physical Review D}\ }\textbf {\bibinfo {volume} {91}},\ \bibinfo
  {pages} {103512} (\bibinfo {year} {2015}{\natexlab{b}})}\BibitemShut
  {NoStop}%
\bibitem [{\citenamefont {Widrow}\ and\ \citenamefont
  {Kaiser}(1993)}]{Widrow:1993qq}%
  \BibitemOpen
  \bibfield  {author} {\bibinfo {author} {\bibfnamefont {L.~M.}\ \bibnamefont
  {Widrow}}\ and\ \bibinfo {author} {\bibfnamefont {N.}~\bibnamefont
  {Kaiser}},\ }\href@noop {} {\bibfield  {journal} {\bibinfo  {journal}
  {Astrophys. J. Lett.}\ }\textbf {\bibinfo {volume} {416}},\ \bibinfo {pages}
  {L71} (\bibinfo {year} {1993})}\BibitemShut {NoStop}%
\bibitem [{\citenamefont {Schive}\ \emph
  {et~al.}(2014{\natexlab{a}})\citenamefont {Schive}, \citenamefont {Chiueh},\
  and\ \citenamefont {Broadhurst}}]{schive_soliton_1}%
  \BibitemOpen
  \bibfield  {author} {\bibinfo {author} {\bibfnamefont {H.-Y.}\ \bibnamefont
  {Schive}}, \bibinfo {author} {\bibfnamefont {T.}~\bibnamefont {Chiueh}}, \
  and\ \bibinfo {author} {\bibfnamefont {T.}~\bibnamefont {Broadhurst}},\
  }\href {\doibase 10.1038/nphys2996} {\bibfield  {journal} {\bibinfo
  {journal} {Nature Physics}\ }\textbf {\bibinfo {volume} {10}},\ \bibinfo
  {pages} {496} (\bibinfo {year} {2014}{\natexlab{a}})},\ \bibinfo {note}
  {arXiv:1406.6586}\BibitemShut {NoStop}%
\bibitem [{\citenamefont {Mocz}\ \emph
  {et~al.}(2017{\natexlab{a}})\citenamefont {Mocz}, \citenamefont
  {Vogelsberger}, \citenamefont {Robles}, \citenamefont {Zavala}, \citenamefont
  {Boylan-Kolchin}, \citenamefont {Fialkov},\ and\ \citenamefont
  {Hernquist}}]{soliton_velocity}%
  \BibitemOpen
  \bibfield  {author} {\bibinfo {author} {\bibfnamefont {P.}~\bibnamefont
  {Mocz}}, \bibinfo {author} {\bibfnamefont {M.}~\bibnamefont {Vogelsberger}},
  \bibinfo {author} {\bibfnamefont {V.~H.}\ \bibnamefont {Robles}}, \bibinfo
  {author} {\bibfnamefont {J.}~\bibnamefont {Zavala}}, \bibinfo {author}
  {\bibfnamefont {M.}~\bibnamefont {Boylan-Kolchin}}, \bibinfo {author}
  {\bibfnamefont {A.}~\bibnamefont {Fialkov}}, \ and\ \bibinfo {author}
  {\bibfnamefont {L.}~\bibnamefont {Hernquist}},\ }\href {\doibase
  10.1093/mnras/stx1887} {\bibfield  {journal} {\bibinfo  {journal} {Monthly
  Notices of the Royal Astronomical Society}\ }\textbf {\bibinfo {volume}
  {471}},\ \bibinfo {pages} {4559} (\bibinfo {year}
  {2017}{\natexlab{a}})}\BibitemShut {NoStop}%
\bibitem [{\citenamefont {Levkov}\ \emph {et~al.}(2018)\citenamefont {Levkov},
  \citenamefont {Panin},\ and\ \citenamefont {Tkachev}}]{Levkov:2018kau}%
  \BibitemOpen
  \bibfield  {author} {\bibinfo {author} {\bibfnamefont {D.}~\bibnamefont
  {Levkov}}, \bibinfo {author} {\bibfnamefont {A.}~\bibnamefont {Panin}}, \
  and\ \bibinfo {author} {\bibfnamefont {I.}~\bibnamefont {Tkachev}},\ }\href
  {\doibase 10.1103/PhysRevLett.121.151301} {\bibfield  {journal} {\bibinfo
  {journal} {Phys. Rev. Lett.}\ }\textbf {\bibinfo {volume} {121}},\ \bibinfo
  {pages} {151301} (\bibinfo {year} {2018})},\ \Eprint
  {http://arxiv.org/abs/1804.05857} {arXiv:1804.05857 [astro-ph.CO]}
  \BibitemShut {NoStop}%
\bibitem [{\citenamefont {Mocz}\ \emph
  {et~al.}(2017{\natexlab{b}})\citenamefont {Mocz}, \citenamefont
  {Vogelsberger}, \citenamefont {Robles}, \citenamefont {Zavala}, \citenamefont
  {Boylan-Kolchin}, \citenamefont {Fialkov},\ and\ \citenamefont
  {Hernquist}}]{Mocz:2017wlg}%
  \BibitemOpen
  \bibfield  {author} {\bibinfo {author} {\bibfnamefont {P.}~\bibnamefont
  {Mocz}}, \bibinfo {author} {\bibfnamefont {M.}~\bibnamefont {Vogelsberger}},
  \bibinfo {author} {\bibfnamefont {V.~H.}\ \bibnamefont {Robles}}, \bibinfo
  {author} {\bibfnamefont {J.}~\bibnamefont {Zavala}}, \bibinfo {author}
  {\bibfnamefont {M.}~\bibnamefont {Boylan-Kolchin}}, \bibinfo {author}
  {\bibfnamefont {A.}~\bibnamefont {Fialkov}}, \ and\ \bibinfo {author}
  {\bibfnamefont {L.}~\bibnamefont {Hernquist}},\ }\href {\doibase
  10.1093/mnras/stx1887} {\bibfield  {journal} {\bibinfo  {journal} {Monthly
  Notices of the Royal Astronomical Society}\ }\textbf {\bibinfo {volume}
  {471}},\ \bibinfo {pages} {4559} (\bibinfo {year} {2017}{\natexlab{b}})},\
  \Eprint
  {http://arxiv.org/abs/https://academic.oup.com/mnras/article-pdf/471/4/4559/19609125/stx1887.pdf}
  {https://academic.oup.com/mnras/article-pdf/471/4/4559/19609125/stx1887.pdf}
  \BibitemShut {NoStop}%
\bibitem [{\citenamefont {Dalal}\ and\ \citenamefont
  {Kravtsov}(2022)}]{Dalal:2022rmp}%
  \BibitemOpen
  \bibfield  {author} {\bibinfo {author} {\bibfnamefont {N.}~\bibnamefont
  {Dalal}}\ and\ \bibinfo {author} {\bibfnamefont {A.}~\bibnamefont
  {Kravtsov}},\ }\href {\doibase 10.1103/PhysRevD.106.063517} {\bibfield
  {journal} {\bibinfo  {journal} {Phys. Rev. D}\ }\textbf {\bibinfo {volume}
  {106}},\ \bibinfo {pages} {063517} (\bibinfo {year} {2022})},\ \Eprint
  {http://arxiv.org/abs/2203.05750} {arXiv:2203.05750 [astro-ph.CO]}
  \BibitemShut {NoStop}%
\bibitem [{\citenamefont {Mocz}\ \emph {et~al.}(2019)\citenamefont {Mocz} \emph
  {et~al.}}]{Mocz:2019pyf}%
  \BibitemOpen
  \bibfield  {author} {\bibinfo {author} {\bibfnamefont {P.}~\bibnamefont
  {Mocz}} \emph {et~al.},\ }\href {\doibase 10.1103/PhysRevLett.123.141301}
  {\bibfield  {journal} {\bibinfo  {journal} {Phys. Rev. Lett.}\ }\textbf
  {\bibinfo {volume} {123}},\ \bibinfo {pages} {141301} (\bibinfo {year}
  {2019})},\ \Eprint {http://arxiv.org/abs/1910.01653} {arXiv:1910.01653
  [astro-ph.GA]} \BibitemShut {NoStop}%
\bibitem [{\citenamefont {Gough}\ and\ \citenamefont
  {Uhlemann}(2022)}]{Gough:2022pof}%
  \BibitemOpen
  \bibfield  {author} {\bibinfo {author} {\bibfnamefont {A.}~\bibnamefont
  {Gough}}\ and\ \bibinfo {author} {\bibfnamefont {C.}~\bibnamefont
  {Uhlemann}},\ }\href {\doibase 10.21105/astro.2206.11918} {\  (\bibinfo
  {year} {2022}),\ 10.21105/astro.2206.11918},\ \Eprint
  {http://arxiv.org/abs/2206.11918} {arXiv:2206.11918 [astro-ph.CO]}
  \BibitemShut {NoStop}%
\bibitem [{\citenamefont {Nori}\ and\ \citenamefont
  {Baldi}(2018)}]{Nori:2018hud}%
  \BibitemOpen
  \bibfield  {author} {\bibinfo {author} {\bibfnamefont {M.}~\bibnamefont
  {Nori}}\ and\ \bibinfo {author} {\bibfnamefont {M.}~\bibnamefont {Baldi}},\
  }\href {\doibase 10.1093/mnras/sty1224} {\bibfield  {journal} {\bibinfo
  {journal} {Mon. Not. Roy. Astron. Soc.}\ }\textbf {\bibinfo {volume} {478}},\
  \bibinfo {pages} {3935} (\bibinfo {year} {2018})},\ \Eprint
  {http://arxiv.org/abs/1801.08144} {arXiv:1801.08144 [astro-ph.CO]}
  \BibitemShut {NoStop}%
\bibitem [{\citenamefont {Veltmaat}\ and\ \citenamefont
  {Niemeyer}(2016)}]{Veltmaat:2016rxo}%
  \BibitemOpen
  \bibfield  {author} {\bibinfo {author} {\bibfnamefont {J.}~\bibnamefont
  {Veltmaat}}\ and\ \bibinfo {author} {\bibfnamefont {J.~C.}\ \bibnamefont
  {Niemeyer}},\ }\href {\doibase 10.1103/PhysRevD.94.123523} {\bibfield
  {journal} {\bibinfo  {journal} {Phys. Rev. D}\ }\textbf {\bibinfo {volume}
  {94}},\ \bibinfo {pages} {123523} (\bibinfo {year} {2016})},\ \Eprint
  {http://arxiv.org/abs/1608.00802} {arXiv:1608.00802 [astro-ph.CO]}
  \BibitemShut {NoStop}%
\bibitem [{\citenamefont {Chen}\ \emph {et~al.}(2021)\citenamefont {Chen},
  \citenamefont {Du}, \citenamefont {Lentz}, \citenamefont {Marsh},\ and\
  \citenamefont {Niemeyer}}]{Chen:2020cef}%
  \BibitemOpen
  \bibfield  {author} {\bibinfo {author} {\bibfnamefont {J.}~\bibnamefont
  {Chen}}, \bibinfo {author} {\bibfnamefont {X.}~\bibnamefont {Du}}, \bibinfo
  {author} {\bibfnamefont {E.~W.}\ \bibnamefont {Lentz}}, \bibinfo {author}
  {\bibfnamefont {D.~J.~E.}\ \bibnamefont {Marsh}}, \ and\ \bibinfo {author}
  {\bibfnamefont {J.~C.}\ \bibnamefont {Niemeyer}},\ }\href {\doibase
  10.1103/PhysRevD.104.083022} {\bibfield  {journal} {\bibinfo  {journal}
  {Phys. Rev. D}\ }\textbf {\bibinfo {volume} {104}},\ \bibinfo {pages}
  {083022} (\bibinfo {year} {2021})},\ \Eprint
  {http://arxiv.org/abs/2011.01333} {arXiv:2011.01333 [astro-ph.CO]}
  \BibitemShut {NoStop}%
\bibitem [{\citenamefont {Almgren}\ \emph {et~al.}(2013)\citenamefont
  {Almgren}, \citenamefont {Bell}, \citenamefont {Lijewski}, \citenamefont
  {Lukic},\ and\ \citenamefont {Van~Andel}}]{Almgren:2013sz}%
  \BibitemOpen
  \bibfield  {author} {\bibinfo {author} {\bibfnamefont {A.}~\bibnamefont
  {Almgren}}, \bibinfo {author} {\bibfnamefont {J.}~\bibnamefont {Bell}},
  \bibinfo {author} {\bibfnamefont {M.}~\bibnamefont {Lijewski}}, \bibinfo
  {author} {\bibfnamefont {Z.}~\bibnamefont {Lukic}}, \ and\ \bibinfo {author}
  {\bibfnamefont {E.}~\bibnamefont {Van~Andel}},\ }\href {\doibase
  10.1088/0004-637X/765/1/39} {\bibfield  {journal} {\bibinfo  {journal}
  {Astrophys. J.}\ }\textbf {\bibinfo {volume} {765}},\ \bibinfo {pages} {39}
  (\bibinfo {year} {2013})},\ \Eprint {http://arxiv.org/abs/1301.4498}
  {arXiv:1301.4498 [astro-ph.IM]} \BibitemShut {NoStop}%
\bibitem [{\citenamefont {{May}}\ and\ \citenamefont
  {{Springel}}(2021)}]{May_FDM_cosmo_sims}%
  \BibitemOpen
  \bibfield  {author} {\bibinfo {author} {\bibfnamefont {S.}~\bibnamefont
  {{May}}}\ and\ \bibinfo {author} {\bibfnamefont {V.}~\bibnamefont
  {{Springel}}},\ }\href {\doibase 10.1093/mnras/stab1764} {\bibfield
  {journal} {\bibinfo  {journal} {Monthly Notices of the Royal Astronomical
  Society}\ }\textbf {\bibinfo {volume} {506}},\ \bibinfo {pages} {2603}
  (\bibinfo {year} {2021})},\ \Eprint {http://arxiv.org/abs/2101.01828}
  {arXiv:2101.01828 [astro-ph.CO]} \BibitemShut {NoStop}%
\bibitem [{\citenamefont {Mead}\ \emph {et~al.}(2020)\citenamefont {Mead},
  \citenamefont {Tröster}, \citenamefont {Heymans}, \citenamefont
  {Van~Waerbeke},\ and\ \citenamefont
  {McCarthy}}]{correction_for_correct_two_halo_term}%
  \BibitemOpen
  \bibfield  {author} {\bibinfo {author} {\bibfnamefont {A.~J.}\ \bibnamefont
  {Mead}}, \bibinfo {author} {\bibfnamefont {T.}~\bibnamefont {Tröster}},
  \bibinfo {author} {\bibfnamefont {C.}~\bibnamefont {Heymans}}, \bibinfo
  {author} {\bibfnamefont {L.}~\bibnamefont {Van~Waerbeke}}, \ and\ \bibinfo
  {author} {\bibfnamefont {I.~G.}\ \bibnamefont {McCarthy}},\ }\href {\doibase
  10.1051/0004-6361/202038308} {\bibfield  {journal} {\bibinfo  {journal}
  {Astronomy \& Astrophysics}\ }\textbf {\bibinfo {volume} {641}},\ \bibinfo
  {pages} {A130} (\bibinfo {year} {2020})}\BibitemShut {NoStop}%
\bibitem [{\citenamefont {Navarro}\ \emph {et~al.}(1997)\citenamefont
  {Navarro}, \citenamefont {Frenk},\ and\ \citenamefont {White}}]{NFW_profile}%
  \BibitemOpen
  \bibfield  {author} {\bibinfo {author} {\bibfnamefont {J.~F.}\ \bibnamefont
  {Navarro}}, \bibinfo {author} {\bibfnamefont {C.~S.}\ \bibnamefont {Frenk}},
  \ and\ \bibinfo {author} {\bibfnamefont {S.~D.~M.}\ \bibnamefont {White}},\
  }\href {\doibase 10.1086/304888} {\bibfield  {journal} {\bibinfo  {journal}
  {The Astrophysical Journal}\ }\textbf {\bibinfo {volume} {490}},\ \bibinfo
  {pages} {493} (\bibinfo {year} {1997})},\ \bibinfo {note} {arXiv:
  astro-ph/9611107}\BibitemShut {NoStop}%
\bibitem [{\citenamefont {Scoccimarro}\ \emph {et~al.}(2001)\citenamefont
  {Scoccimarro}, \citenamefont {Sheth}, \citenamefont {Hui},\ and\
  \citenamefont {Jain}}]{fourier_NFW_profile}%
  \BibitemOpen
  \bibfield  {author} {\bibinfo {author} {\bibfnamefont {R.}~\bibnamefont
  {Scoccimarro}}, \bibinfo {author} {\bibfnamefont {R.~K.}\ \bibnamefont
  {Sheth}}, \bibinfo {author} {\bibfnamefont {L.}~\bibnamefont {Hui}}, \ and\
  \bibinfo {author} {\bibfnamefont {B.}~\bibnamefont {Jain}},\ }\href {\doibase
  10.1086/318261} {\bibfield  {journal} {\bibinfo  {journal} {The Astrophysical
  Journal}\ }\textbf {\bibinfo {volume} {546}},\ \bibinfo {pages} {20}
  (\bibinfo {year} {2001})},\ \bibinfo {note} {arXiv:
  astro-ph/0006319v2}\BibitemShut {NoStop}%
\bibitem [{\citenamefont {Press}\ and\ \citenamefont
  {Schechter}(1974)}]{Press:1973iz}%
  \BibitemOpen
  \bibfield  {author} {\bibinfo {author} {\bibfnamefont {W.~H.}\ \bibnamefont
  {Press}}\ and\ \bibinfo {author} {\bibfnamefont {P.}~\bibnamefont
  {Schechter}},\ }\href {\doibase 10.1086/152650} {\bibfield  {journal}
  {\bibinfo  {journal} {\apj}\ }\textbf {\bibinfo {volume} {187}},\ \bibinfo
  {pages} {425} (\bibinfo {year} {1974})}\BibitemShut {NoStop}%
%%CITATION = ASJOA,187,425;%%
\bibitem [{\citenamefont {Sheth}\ and\ \citenamefont
  {Tormen}(2002)}]{sheth_tormen_multiplicity_function}%
  \BibitemOpen
  \bibfield  {author} {\bibinfo {author} {\bibfnamefont {R.~K.}\ \bibnamefont
  {Sheth}}\ and\ \bibinfo {author} {\bibfnamefont {G.}~\bibnamefont {Tormen}},\
  }\href {\doibase 10.1046/j.1365-8711.2002.04950.x} {\bibfield  {journal}
  {\bibinfo  {journal} {Monthly Notices of the Royal Astronomical Society}\
  }\textbf {\bibinfo {volume} {329}},\ \bibinfo {pages} {61} (\bibinfo {year}
  {2002})},\ \bibinfo {note} {arXiv:astro-ph/0105113}\BibitemShut {NoStop}%
\bibitem [{\citenamefont {Bryan}\ and\ \citenamefont
  {Norman}(1998)}]{virial_overdensity_formula}%
  \BibitemOpen
  \bibfield  {author} {\bibinfo {author} {\bibfnamefont {G.~L.}\ \bibnamefont
  {Bryan}}\ and\ \bibinfo {author} {\bibfnamefont {M.~L.}\ \bibnamefont
  {Norman}},\ }\href {\doibase 10.1086/305262} {\bibfield  {journal} {\bibinfo
  {journal} {The Astrophysical Journal}\ }\textbf {\bibinfo {volume} {495}},\
  \bibinfo {pages} {80} (\bibinfo {year} {1998})}\BibitemShut {NoStop}%
\bibitem [{\citenamefont {Bullock}\ \emph {et~al.}(2001)\citenamefont
  {Bullock}, \citenamefont {Kolatt}, \citenamefont {Sigad}, \citenamefont
  {Somerville}, \citenamefont {Kravtsov}, \citenamefont {Klypin}, \citenamefont
  {Primack},\ and\ \citenamefont {Dekel}}]{concentration_parameter}%
  \BibitemOpen
  \bibfield  {author} {\bibinfo {author} {\bibfnamefont {J.~S.}\ \bibnamefont
  {Bullock}}, \bibinfo {author} {\bibfnamefont {T.~S.}\ \bibnamefont {Kolatt}},
  \bibinfo {author} {\bibfnamefont {Y.}~\bibnamefont {Sigad}}, \bibinfo
  {author} {\bibfnamefont {R.~S.}\ \bibnamefont {Somerville}}, \bibinfo
  {author} {\bibfnamefont {A.~V.}\ \bibnamefont {Kravtsov}}, \bibinfo {author}
  {\bibfnamefont {A.~A.}\ \bibnamefont {Klypin}}, \bibinfo {author}
  {\bibfnamefont {J.~R.}\ \bibnamefont {Primack}}, \ and\ \bibinfo {author}
  {\bibfnamefont {A.}~\bibnamefont {Dekel}},\ }\href {\doibase
  10.1046/j.1365-8711.2001.04068.x} {\bibfield  {journal} {\bibinfo  {journal}
  {Monthly Notices of the Royal Astronomical Society}\ }\textbf {\bibinfo
  {volume} {321}},\ \bibinfo {pages} {559} (\bibinfo {year} {2001})},\ \bibinfo
  {note} {arXiv:astro-ph/9908159}\BibitemShut {NoStop}%
\bibitem [{\citenamefont {Singh}\ and\ \citenamefont
  {Ma}(2003)}]{neutrino_clustering_around_chalos_1}%
  \BibitemOpen
  \bibfield  {author} {\bibinfo {author} {\bibfnamefont {S.}~\bibnamefont
  {Singh}}\ and\ \bibinfo {author} {\bibfnamefont {C.-P.}\ \bibnamefont {Ma}},\
  }\href {\doibase 10.1103/PhysRevD.67.023506} {\bibfield  {journal} {\bibinfo
  {journal} {Phys. Rev. D}\ }\textbf {\bibinfo {volume} {67}},\ \bibinfo
  {pages} {023506} (\bibinfo {year} {2003})}\BibitemShut {NoStop}%
\bibitem [{\citenamefont {Ringwald}\ and\ \citenamefont
  {Wong}(2004)}]{neutrino_clustering_around_chalos_2}%
  \BibitemOpen
  \bibfield  {author} {\bibinfo {author} {\bibfnamefont {A.}~\bibnamefont
  {Ringwald}}\ and\ \bibinfo {author} {\bibfnamefont {Y.~Y.~Y.}\ \bibnamefont
  {Wong}},\ }\href {\doibase 10.1088/1475-7516/2004/12/005} {\bibfield
  {journal} {\bibinfo  {journal} {Journal of Cosmology and Astroparticle
  Physics}\ }\textbf {\bibinfo {volume} {2004}},\ \bibinfo {pages} {005}
  (\bibinfo {year} {2004})}\BibitemShut {NoStop}%
\bibitem [{\citenamefont {Villaescusa-Navarro}\ \emph
  {et~al.}(2018)\citenamefont {Villaescusa-Navarro}, \citenamefont {Genel},
  \citenamefont {Castorina}, \citenamefont {Obuljen}, \citenamefont {Spergel},
  \citenamefont {Hernquist}, \citenamefont {Nelson}, \citenamefont {Carucci},
  \citenamefont {Pillepich}, \citenamefont {Marinacci}, \citenamefont {Diemer},
  \citenamefont {Vogelsberger}, \citenamefont {Weinberger},\ and\ \citenamefont
  {Pakmor}}]{villaescusa_biased_tracer_HI}%
  \BibitemOpen
  \bibfield  {author} {\bibinfo {author} {\bibfnamefont {F.}~\bibnamefont
  {Villaescusa-Navarro}}, \bibinfo {author} {\bibfnamefont {S.}~\bibnamefont
  {Genel}}, \bibinfo {author} {\bibfnamefont {E.}~\bibnamefont {Castorina}},
  \bibinfo {author} {\bibfnamefont {A.}~\bibnamefont {Obuljen}}, \bibinfo
  {author} {\bibfnamefont {D.~N.}\ \bibnamefont {Spergel}}, \bibinfo {author}
  {\bibfnamefont {L.}~\bibnamefont {Hernquist}}, \bibinfo {author}
  {\bibfnamefont {D.}~\bibnamefont {Nelson}}, \bibinfo {author} {\bibfnamefont
  {I.~P.}\ \bibnamefont {Carucci}}, \bibinfo {author} {\bibfnamefont
  {A.}~\bibnamefont {Pillepich}}, \bibinfo {author} {\bibfnamefont
  {F.}~\bibnamefont {Marinacci}}, \bibinfo {author} {\bibfnamefont
  {B.}~\bibnamefont {Diemer}}, \bibinfo {author} {\bibfnamefont
  {M.}~\bibnamefont {Vogelsberger}}, \bibinfo {author} {\bibfnamefont
  {R.}~\bibnamefont {Weinberger}}, \ and\ \bibinfo {author} {\bibfnamefont
  {R.}~\bibnamefont {Pakmor}},\ }\href {\doibase 10.3847/1538-4357/aadba0}
  {\bibfield  {journal} {\bibinfo  {journal} {The Astrophysical Journal}\
  }\textbf {\bibinfo {volume} {866}},\ \bibinfo {pages} {135} (\bibinfo {year}
  {2018})},\ \bibinfo {note} {arXiv:1804.09180}\BibitemShut {NoStop}%
\bibitem [{\citenamefont {Villaescusa-Navarro}\ \emph
  {et~al.}(2014)\citenamefont {Villaescusa-Navarro}, \citenamefont {Marulli},
  \citenamefont {Viel}, \citenamefont {Branchini}, \citenamefont {Castorina},
  \citenamefont {Sefusatti},\ and\ \citenamefont
  {Saito}}]{HMF_halobias_fct_of_Mc_sim_1}%
  \BibitemOpen
  \bibfield  {author} {\bibinfo {author} {\bibfnamefont {F.}~\bibnamefont
  {Villaescusa-Navarro}}, \bibinfo {author} {\bibfnamefont {F.}~\bibnamefont
  {Marulli}}, \bibinfo {author} {\bibfnamefont {M.}~\bibnamefont {Viel}},
  \bibinfo {author} {\bibfnamefont {E.}~\bibnamefont {Branchini}}, \bibinfo
  {author} {\bibfnamefont {E.}~\bibnamefont {Castorina}}, \bibinfo {author}
  {\bibfnamefont {E.}~\bibnamefont {Sefusatti}}, \ and\ \bibinfo {author}
  {\bibfnamefont {S.}~\bibnamefont {Saito}},\ }\href {\doibase
  10.1088/1475-7516/2014/03/011} {\bibfield  {journal} {\bibinfo  {journal}
  {Journal of Cosmology and Astroparticle Physics}\ }\textbf {\bibinfo {volume}
  {2014}},\ \bibinfo {pages} {011} (\bibinfo {year} {2014})},\ \bibinfo {note}
  {arXiv:1311.0866}\BibitemShut {NoStop}%
\bibitem [{\citenamefont {Castorina}\ \emph {et~al.}(2014)\citenamefont
  {Castorina}, \citenamefont {Sefusatti}, \citenamefont {Sheth}, \citenamefont
  {Villaescusa-Navarro},\ and\ \citenamefont
  {Viel}}]{HMF_halobias_fct_of_Mc_sim_2}%
  \BibitemOpen
  \bibfield  {author} {\bibinfo {author} {\bibfnamefont {E.}~\bibnamefont
  {Castorina}}, \bibinfo {author} {\bibfnamefont {E.}~\bibnamefont
  {Sefusatti}}, \bibinfo {author} {\bibfnamefont {R.~K.}\ \bibnamefont
  {Sheth}}, \bibinfo {author} {\bibfnamefont {F.}~\bibnamefont
  {Villaescusa-Navarro}}, \ and\ \bibinfo {author} {\bibfnamefont
  {M.}~\bibnamefont {Viel}},\ }\href {\doibase 10.1088/1475-7516/2014/02/049}
  {\bibfield  {journal} {\bibinfo  {journal} {Journal of Cosmology and
  Astroparticle Physics}\ }\textbf {\bibinfo {volume} {2014}},\ \bibinfo
  {pages} {049} (\bibinfo {year} {2014})},\ \bibinfo {note}
  {arXiv:1311.1212}\BibitemShut {NoStop}%
\bibitem [{\citenamefont {Costanzi}\ \emph {et~al.}(2013)\citenamefont
  {Costanzi}, \citenamefont {Villaescusa-Navarro}, \citenamefont {Viel},
  \citenamefont {Xia}, \citenamefont {Borgani}, \citenamefont {Castorina},\
  and\ \citenamefont {Sefusatti}}]{HMF_halobias_fct_of_Mc_sim_3}%
  \BibitemOpen
  \bibfield  {author} {\bibinfo {author} {\bibfnamefont {M.}~\bibnamefont
  {Costanzi}}, \bibinfo {author} {\bibfnamefont {F.}~\bibnamefont
  {Villaescusa-Navarro}}, \bibinfo {author} {\bibfnamefont {M.}~\bibnamefont
  {Viel}}, \bibinfo {author} {\bibfnamefont {J.-Q.}\ \bibnamefont {Xia}},
  \bibinfo {author} {\bibfnamefont {S.}~\bibnamefont {Borgani}}, \bibinfo
  {author} {\bibfnamefont {E.}~\bibnamefont {Castorina}}, \ and\ \bibinfo
  {author} {\bibfnamefont {E.}~\bibnamefont {Sefusatti}},\ }\href {\doibase
  10.1088/1475-7516/2013/12/012} {\bibfield  {journal} {\bibinfo  {journal}
  {Journal of Cosmology and Astroparticle Physics}\ }\textbf {\bibinfo {volume}
  {2013}},\ \bibinfo {pages} {012} (\bibinfo {year} {2013})},\ \bibinfo {note}
  {arXiv:1311.1514}\BibitemShut {NoStop}%
\bibitem [{\citenamefont {Schive}\ \emph
  {et~al.}(2014{\natexlab{b}})\citenamefont {Schive}, \citenamefont {Liao},
  \citenamefont {Woo}, \citenamefont {Wong}, \citenamefont {Chiueh},
  \citenamefont {Broadhurst},\ and\ \citenamefont {Hwang}}]{schive_soliton_2}%
  \BibitemOpen
  \bibfield  {author} {\bibinfo {author} {\bibfnamefont {H.-Y.}\ \bibnamefont
  {Schive}}, \bibinfo {author} {\bibfnamefont {M.-H.}\ \bibnamefont {Liao}},
  \bibinfo {author} {\bibfnamefont {T.-P.}\ \bibnamefont {Woo}}, \bibinfo
  {author} {\bibfnamefont {S.-K.}\ \bibnamefont {Wong}}, \bibinfo {author}
  {\bibfnamefont {T.}~\bibnamefont {Chiueh}}, \bibinfo {author} {\bibfnamefont
  {T.}~\bibnamefont {Broadhurst}}, \ and\ \bibinfo {author} {\bibfnamefont
  {W.-Y.~P.}\ \bibnamefont {Hwang}},\ }\href {\doibase
  10.1103/PhysRevLett.113.261302} {\bibfield  {journal} {\bibinfo  {journal}
  {Physical Review Letters}\ }\textbf {\bibinfo {volume} {113}},\ \bibinfo
  {pages} {261302} (\bibinfo {year} {2014}{\natexlab{b}})},\ \bibinfo {note}
  {arXiv:1407.7762}\BibitemShut {NoStop}%
\bibitem [{\citenamefont {Niemeyer}(2020)}]{soliton_velocity_virial_vel_1}%
  \BibitemOpen
  \bibfield  {author} {\bibinfo {author} {\bibfnamefont {J.~C.}\ \bibnamefont
  {Niemeyer}},\ }\href {\doibase 10.1016/j.ppnp.2020.103787} {\bibfield
  {journal} {\bibinfo  {journal} {Progress in Particle and Nuclear Physics}\
  }\textbf {\bibinfo {volume} {113}},\ \bibinfo {pages} {103787} (\bibinfo
  {year} {2020})}\BibitemShut {NoStop}%
\bibitem [{\citenamefont {Eggemeier}\ \emph {et~al.}(2022)\citenamefont
  {Eggemeier}, \citenamefont {Schwabe}, \citenamefont {Niemeyer},\ and\
  \citenamefont {Easther}}]{soliton_velocity_virial_vel_2}%
  \BibitemOpen
  \bibfield  {author} {\bibinfo {author} {\bibfnamefont {B.}~\bibnamefont
  {Eggemeier}}, \bibinfo {author} {\bibfnamefont {B.}~\bibnamefont {Schwabe}},
  \bibinfo {author} {\bibfnamefont {J.~C.}\ \bibnamefont {Niemeyer}}, \ and\
  \bibinfo {author} {\bibfnamefont {R.}~\bibnamefont {Easther}},\ }\href
  {\doibase 10.1103/PhysRevD.105.023516} {\bibfield  {journal} {\bibinfo
  {journal} {Physical Review D}\ }\textbf {\bibinfo {volume} {105}},\ \bibinfo
  {pages} {023516} (\bibinfo {year} {2022})}\BibitemShut {NoStop}%
\bibitem [{\citenamefont {Marsh}(2016{\natexlab{b}})}]{Marsh:2016vgj}%
  \BibitemOpen
  \bibfield  {author} {\bibinfo {author} {\bibfnamefont {D.~J.~E.}\
  \bibnamefont {Marsh}},\ }\href@noop {} {\  (\bibinfo {year}
  {2016}{\natexlab{b}})},\ \Eprint {http://arxiv.org/abs/1605.05973}
  {arXiv:1605.05973 [astro-ph.CO]} \BibitemShut {NoStop}%
\bibitem [{\citenamefont {Hannestad}\ \emph {et~al.}(2020)\citenamefont
  {Hannestad}, \citenamefont {Upadhye},\ and\ \citenamefont
  {Wong}}]{spoon_neutrino_halo_model}%
  \BibitemOpen
  \bibfield  {author} {\bibinfo {author} {\bibfnamefont {S.}~\bibnamefont
  {Hannestad}}, \bibinfo {author} {\bibfnamefont {A.}~\bibnamefont {Upadhye}},
  \ and\ \bibinfo {author} {\bibfnamefont {Y.~Y.}\ \bibnamefont {Wong}},\
  }\href {\doibase 10.1088/1475-7516/2020/11/062} {\bibfield  {journal}
  {\bibinfo  {journal} {Journal of Cosmology and Astroparticle Physics}\
  }\textbf {\bibinfo {volume} {2020}},\ \bibinfo {pages} {062} (\bibinfo {year}
  {2020})}\BibitemShut {NoStop}%
\bibitem [{\citenamefont {Brandbyge}\ \emph {et~al.}(2008)\citenamefont
  {Brandbyge}, \citenamefont {Hannestad}, \citenamefont {Haugbølle},\ and\
  \citenamefont {Thomsen}}]{spoon_neutrino_sim_1}%
  \BibitemOpen
  \bibfield  {author} {\bibinfo {author} {\bibfnamefont {J.}~\bibnamefont
  {Brandbyge}}, \bibinfo {author} {\bibfnamefont {S.}~\bibnamefont
  {Hannestad}}, \bibinfo {author} {\bibfnamefont {T.}~\bibnamefont
  {Haugbølle}}, \ and\ \bibinfo {author} {\bibfnamefont {B.}~\bibnamefont
  {Thomsen}},\ }\href {\doibase 10.1088/1475-7516/2008/08/020} {\bibfield
  {journal} {\bibinfo  {journal} {Journal of Cosmology and Astroparticle
  Physics}\ }\textbf {\bibinfo {volume} {2008}},\ \bibinfo {pages} {020}
  (\bibinfo {year} {2008})}\BibitemShut {NoStop}%
\bibitem [{\citenamefont {Bird}\ \emph {et~al.}(2012)\citenamefont {Bird},
  \citenamefont {Viel},\ and\ \citenamefont {Haehnelt}}]{spoon_neutrino_sim_2}%
  \BibitemOpen
  \bibfield  {author} {\bibinfo {author} {\bibfnamefont {S.}~\bibnamefont
  {Bird}}, \bibinfo {author} {\bibfnamefont {M.}~\bibnamefont {Viel}}, \ and\
  \bibinfo {author} {\bibfnamefont {M.~G.}\ \bibnamefont {Haehnelt}},\ }\href
  {\doibase 10.1111/j.1365-2966.2011.20222.x} {\bibfield  {journal} {\bibinfo
  {journal} {Monthly Notices of the Royal Astronomical Society}\ }\textbf
  {\bibinfo {volume} {420}},\ \bibinfo {pages} {2551} (\bibinfo {year}
  {2012})}\BibitemShut {NoStop}%
\bibitem [{\citenamefont {Nori}\ \emph {et~al.}(2019)\citenamefont {Nori},
  \citenamefont {Murgia}, \citenamefont {Ir\v{s}i\v{c}}, \citenamefont
  {Baldi},\ and\ \citenamefont {Viel}}]{Nori:2018pka}%
  \BibitemOpen
  \bibfield  {author} {\bibinfo {author} {\bibfnamefont {M.}~\bibnamefont
  {Nori}}, \bibinfo {author} {\bibfnamefont {R.}~\bibnamefont {Murgia}},
  \bibinfo {author} {\bibfnamefont {V.}~\bibnamefont {Ir\v{s}i\v{c}}}, \bibinfo
  {author} {\bibfnamefont {M.}~\bibnamefont {Baldi}}, \ and\ \bibinfo {author}
  {\bibfnamefont {M.}~\bibnamefont {Viel}},\ }\href {\doibase
  10.1093/mnras/sty2888} {\bibfield  {journal} {\bibinfo  {journal} {Mon. Not.
  Roy. Astron. Soc.}\ }\textbf {\bibinfo {volume} {482}},\ \bibinfo {pages}
  {3227} (\bibinfo {year} {2019})},\ \Eprint {http://arxiv.org/abs/1809.09619}
  {arXiv:1809.09619 [astro-ph.CO]} \BibitemShut {NoStop}%
\bibitem [{\citenamefont {Freundlich}\ \emph {et~al.}(2016)\citenamefont
  {Freundlich}, \citenamefont {El-Zant},\ and\ \citenamefont
  {Combes}}]{baryonic_feedback}%
  \BibitemOpen
  \bibfield  {author} {\bibinfo {author} {\bibfnamefont {J.}~\bibnamefont
  {Freundlich}}, \bibinfo {author} {\bibfnamefont {A.}~\bibnamefont {El-Zant}},
  \ and\ \bibinfo {author} {\bibfnamefont {F.}~\bibnamefont {Combes}},\
  }\href@noop {} {\bibfield  {journal} {\bibinfo  {journal} {SF2A-2016:
  Proceedings of the Annual meeting of the French Society of Astronomy and
  Astrophysics}\ ,\ \bibinfo {pages} {153}} (\bibinfo {year}
  {2016})}\BibitemShut {NoStop}%
\bibitem [{\citenamefont {Mead}\ \emph
  {et~al.}(2015{\natexlab{b}})\citenamefont {Mead}, \citenamefont {Peacock},
  \citenamefont {Heymans}, \citenamefont {Joudaki},\ and\ \citenamefont
  {Heavens}}]{halo_model_max_k_1}%
  \BibitemOpen
  \bibfield  {author} {\bibinfo {author} {\bibfnamefont {A.~J.}\ \bibnamefont
  {Mead}}, \bibinfo {author} {\bibfnamefont {J.~A.}\ \bibnamefont {Peacock}},
  \bibinfo {author} {\bibfnamefont {C.}~\bibnamefont {Heymans}}, \bibinfo
  {author} {\bibfnamefont {S.}~\bibnamefont {Joudaki}}, \ and\ \bibinfo
  {author} {\bibfnamefont {A.~F.}\ \bibnamefont {Heavens}},\ }\href {\doibase
  10.1093/mnras/stv2036} {\bibfield  {journal} {\bibinfo  {journal} {Monthly
  Notices of the Royal Astronomical Society}\ }\textbf {\bibinfo {volume}
  {454}},\ \bibinfo {pages} {1958} (\bibinfo {year}
  {2015}{\natexlab{b}})}\BibitemShut {NoStop}%
\bibitem [{\citenamefont {Schive}\ \emph {et~al.}(2016)\citenamefont {Schive},
  \citenamefont {Chiueh}, \citenamefont {Broadhurst},\ and\ \citenamefont
  {Huang}}]{Schive:2015kza}%
  \BibitemOpen
  \bibfield  {author} {\bibinfo {author} {\bibfnamefont {H.-Y.}\ \bibnamefont
  {Schive}}, \bibinfo {author} {\bibfnamefont {T.}~\bibnamefont {Chiueh}},
  \bibinfo {author} {\bibfnamefont {T.}~\bibnamefont {Broadhurst}}, \ and\
  \bibinfo {author} {\bibfnamefont {K.-W.}\ \bibnamefont {Huang}},\ }\href
  {\doibase 10.3847/0004-637X/818/1/89} {\bibfield  {journal} {\bibinfo
  {journal} {Astrophys. J.}\ }\textbf {\bibinfo {volume} {818}},\ \bibinfo
  {pages} {89} (\bibinfo {year} {2016})},\ \Eprint
  {http://arxiv.org/abs/1508.04621} {arXiv:1508.04621 [astro-ph.GA]}
  \BibitemShut {NoStop}%
\bibitem [{\citenamefont {Corasaniti}\ \emph {et~al.}(2017)\citenamefont
  {Corasaniti}, \citenamefont {Agarwal}, \citenamefont {Marsh},\ and\
  \citenamefont {Das}}]{Corasaniti:2016epp}%
  \BibitemOpen
  \bibfield  {author} {\bibinfo {author} {\bibfnamefont {P.~S.}\ \bibnamefont
  {Corasaniti}}, \bibinfo {author} {\bibfnamefont {S.}~\bibnamefont {Agarwal}},
  \bibinfo {author} {\bibfnamefont {D.~J.~E.}\ \bibnamefont {Marsh}}, \ and\
  \bibinfo {author} {\bibfnamefont {S.}~\bibnamefont {Das}},\ }\href {\doibase
  10.1103/PhysRevD.95.083512} {\bibfield  {journal} {\bibinfo  {journal} {Phys.
  Rev. D}\ }\textbf {\bibinfo {volume} {95}},\ \bibinfo {pages} {083512}
  (\bibinfo {year} {2017})},\ \Eprint {http://arxiv.org/abs/1611.05892}
  {arXiv:1611.05892 [astro-ph.CO]} \BibitemShut {NoStop}%
\bibitem [{\citenamefont {Luu}\ \emph {et~al.}(2020)\citenamefont {Luu},
  \citenamefont {Tye},\ and\ \citenamefont {Broadhurst}}]{Luu:2018afg}%
  \BibitemOpen
  \bibfield  {author} {\bibinfo {author} {\bibfnamefont {H.~N.}\ \bibnamefont
  {Luu}}, \bibinfo {author} {\bibfnamefont {S.~H.~H.}\ \bibnamefont {Tye}}, \
  and\ \bibinfo {author} {\bibfnamefont {T.}~\bibnamefont {Broadhurst}},\
  }\href {\doibase 10.1016/j.dark.2020.100636} {\bibfield  {journal} {\bibinfo
  {journal} {Phys. Dark Univ.}\ }\textbf {\bibinfo {volume} {30}},\ \bibinfo
  {pages} {100636} (\bibinfo {year} {2020})},\ \Eprint
  {http://arxiv.org/abs/1811.03771} {arXiv:1811.03771 [astro-ph.GA]}
  \BibitemShut {NoStop}%
\bibitem [{\citenamefont {Gosenca}\ \emph {et~al.}(2023)\citenamefont
  {Gosenca}, \citenamefont {Eberhardt}, \citenamefont {Wang}, \citenamefont
  {Eggemeier}, \citenamefont {Kendall}, \citenamefont {Zagorac},\ and\
  \citenamefont {Easther}}]{Gosenca:2023yjc}%
  \BibitemOpen
  \bibfield  {author} {\bibinfo {author} {\bibfnamefont {M.}~\bibnamefont
  {Gosenca}}, \bibinfo {author} {\bibfnamefont {A.}~\bibnamefont {Eberhardt}},
  \bibinfo {author} {\bibfnamefont {Y.}~\bibnamefont {Wang}}, \bibinfo {author}
  {\bibfnamefont {B.}~\bibnamefont {Eggemeier}}, \bibinfo {author}
  {\bibfnamefont {E.}~\bibnamefont {Kendall}}, \bibinfo {author} {\bibfnamefont
  {J.~L.}\ \bibnamefont {Zagorac}}, \ and\ \bibinfo {author} {\bibfnamefont
  {R.}~\bibnamefont {Easther}},\ }\href@noop {} {\  (\bibinfo {year} {2023})},\
  \Eprint {http://arxiv.org/abs/2301.07114} {arXiv:2301.07114 [astro-ph.CO]}
  \BibitemShut {NoStop}%
\bibitem [{\citenamefont {Dentler}\ \emph {et~al.}(2022)\citenamefont
  {Dentler}, \citenamefont {Marsh}, \citenamefont {Hlo\v{z}ek}, \citenamefont
  {Lagu\"e}, \citenamefont {Rogers},\ and\ \citenamefont
  {Grin}}]{Dentler:2021zij}%
  \BibitemOpen
  \bibfield  {author} {\bibinfo {author} {\bibfnamefont {M.}~\bibnamefont
  {Dentler}}, \bibinfo {author} {\bibfnamefont {D.~J.~E.}\ \bibnamefont
  {Marsh}}, \bibinfo {author} {\bibfnamefont {R.}~\bibnamefont {Hlo\v{z}ek}},
  \bibinfo {author} {\bibfnamefont {A.}~\bibnamefont {Lagu\"e}}, \bibinfo
  {author} {\bibfnamefont {K.~K.}\ \bibnamefont {Rogers}}, \ and\ \bibinfo
  {author} {\bibfnamefont {D.}~\bibnamefont {Grin}},\ }\href {\doibase
  10.1093/mnras/stac1946} {\bibfield  {journal} {\bibinfo  {journal} {Mon. Not.
  Roy. Astron. Soc.}\ }\textbf {\bibinfo {volume} {515}},\ \bibinfo {pages}
  {5646} (\bibinfo {year} {2022})},\ \Eprint {http://arxiv.org/abs/2111.01199}
  {arXiv:2111.01199 [astro-ph.CO]} \BibitemShut {NoStop}%
\bibitem [{\citenamefont {Grin}\ \emph {et~al.}(2019)\citenamefont {Grin},
  \citenamefont {Amin}, \citenamefont {Gluscevic}, \citenamefont {Hlǒzek},
  \citenamefont {Marsh}, \citenamefont {Poulin}, \citenamefont
  {Prescod-Weinstein},\ and\ \citenamefont {Smith}}]{Grin:2019mub}%
  \BibitemOpen
  \bibfield  {author} {\bibinfo {author} {\bibfnamefont {D.}~\bibnamefont
  {Grin}}, \bibinfo {author} {\bibfnamefont {M.~A.}\ \bibnamefont {Amin}},
  \bibinfo {author} {\bibfnamefont {V.}~\bibnamefont {Gluscevic}}, \bibinfo
  {author} {\bibfnamefont {R.}~\bibnamefont {Hlǒzek}}, \bibinfo {author}
  {\bibfnamefont {D.~J.~E.}\ \bibnamefont {Marsh}}, \bibinfo {author}
  {\bibfnamefont {V.}~\bibnamefont {Poulin}}, \bibinfo {author} {\bibfnamefont
  {C.}~\bibnamefont {Prescod-Weinstein}}, \ and\ \bibinfo {author}
  {\bibfnamefont {T.~L.}\ \bibnamefont {Smith}},\ }\href@noop {} {\  (\bibinfo
  {year} {2019})},\ \Eprint {http://arxiv.org/abs/1904.09003} {arXiv:1904.09003
  [astro-ph.CO]} \BibitemShut {NoStop}%
\bibitem [{\citenamefont {Ade}\ \emph {et~al.}(2019)\citenamefont {Ade} \emph
  {et~al.}}]{SimonsObservatory:2018koc}%
  \BibitemOpen
  \bibfield  {author} {\bibinfo {author} {\bibfnamefont {P.}~\bibnamefont
  {Ade}} \emph {et~al.} (\bibinfo {collaboration} {Simons Observatory}),\
  }\href {\doibase 10.1088/1475-7516/2019/02/056} {\bibfield  {journal}
  {\bibinfo  {journal} {JCAP}\ }\textbf {\bibinfo {volume} {02}},\ \bibinfo
  {pages} {056} (\bibinfo {year} {2019})},\ \Eprint
  {http://arxiv.org/abs/1808.07445} {arXiv:1808.07445 [astro-ph.CO]}
  \BibitemShut {NoStop}%
\bibitem [{\citenamefont {Amendola}\ \emph {et~al.}(2018)\citenamefont
  {Amendola} \emph {et~al.}}]{Amendola:2016saw}%
  \BibitemOpen
  \bibfield  {author} {\bibinfo {author} {\bibfnamefont {L.}~\bibnamefont
  {Amendola}} \emph {et~al.},\ }\href {\doibase 10.1007/s41114-017-0010-3}
  {\bibfield  {journal} {\bibinfo  {journal} {Living Rev. Rel.}\ }\textbf
  {\bibinfo {volume} {21}},\ \bibinfo {pages} {2} (\bibinfo {year} {2018})},\
  \Eprint {http://arxiv.org/abs/1606.00180} {arXiv:1606.00180 [astro-ph.CO]}
  \BibitemShut {NoStop}%
\bibitem [{\citenamefont {{Hunter}}(2007)}]{matplotlib}%
  \BibitemOpen
  \bibfield  {author} {\bibinfo {author} {\bibfnamefont {J.~D.}\ \bibnamefont
  {{Hunter}}},\ }\href@noop {} {\bibfield  {journal} {\bibinfo  {journal}
  {Computing in Science Engineering}\ }\textbf {\bibinfo {volume} {9}},\
  \bibinfo {pages} {90} (\bibinfo {year} {2007})}\BibitemShut {NoStop}%
\bibitem [{\citenamefont {Harris}\ \emph {et~al.}(2020)\citenamefont {Harris},
  \citenamefont {Millman}, \citenamefont {van~der Walt}, \citenamefont
  {Gommers}, \citenamefont {Virtanen}, \citenamefont {Cournapeau},
  \citenamefont {Wieser}, \citenamefont {Taylor}, \citenamefont {Berg},
  \citenamefont {Smith}, \citenamefont {Kern}, \citenamefont {Picus},
  \citenamefont {Hoyer}, \citenamefont {van Kerkwijk}, \citenamefont {Brett},
  \citenamefont {Haldane}, \citenamefont {del R{'{\i}}o}, \citenamefont
  {Wiebe}, \citenamefont {Peterson}, \citenamefont {G{'{e}}rard-Marchant},
  \citenamefont {Sheppard}, \citenamefont {Reddy}, \citenamefont {Weckesser},
  \citenamefont {Abbasi}, \citenamefont {Gohlke},\ and\ \citenamefont
  {Oliphant}}]{numpy}%
  \BibitemOpen
  \bibfield  {author} {\bibinfo {author} {\bibfnamefont {C.~R.}\ \bibnamefont
  {Harris}}, \bibinfo {author} {\bibfnamefont {K.~J.}\ \bibnamefont {Millman}},
  \bibinfo {author} {\bibfnamefont {S.~J.}\ \bibnamefont {van~der Walt}},
  \bibinfo {author} {\bibfnamefont {R.}~\bibnamefont {Gommers}}, \bibinfo
  {author} {\bibfnamefont {P.}~\bibnamefont {Virtanen}}, \bibinfo {author}
  {\bibfnamefont {D.}~\bibnamefont {Cournapeau}}, \bibinfo {author}
  {\bibfnamefont {E.}~\bibnamefont {Wieser}}, \bibinfo {author} {\bibfnamefont
  {J.}~\bibnamefont {Taylor}}, \bibinfo {author} {\bibfnamefont
  {S.}~\bibnamefont {Berg}}, \bibinfo {author} {\bibfnamefont {N.~J.}\
  \bibnamefont {Smith}}, \bibinfo {author} {\bibfnamefont {R.}~\bibnamefont
  {Kern}}, \bibinfo {author} {\bibfnamefont {M.}~\bibnamefont {Picus}},
  \bibinfo {author} {\bibfnamefont {S.}~\bibnamefont {Hoyer}}, \bibinfo
  {author} {\bibfnamefont {M.~H.}\ \bibnamefont {van Kerkwijk}}, \bibinfo
  {author} {\bibfnamefont {M.}~\bibnamefont {Brett}}, \bibinfo {author}
  {\bibfnamefont {A.}~\bibnamefont {Haldane}}, \bibinfo {author} {\bibfnamefont
  {J.~F.}\ \bibnamefont {del R{'{\i}}o}}, \bibinfo {author} {\bibfnamefont
  {M.}~\bibnamefont {Wiebe}}, \bibinfo {author} {\bibfnamefont
  {P.}~\bibnamefont {Peterson}}, \bibinfo {author} {\bibfnamefont
  {P.}~\bibnamefont {G{'{e}}rard-Marchant}}, \bibinfo {author} {\bibfnamefont
  {K.}~\bibnamefont {Sheppard}}, \bibinfo {author} {\bibfnamefont
  {T.}~\bibnamefont {Reddy}}, \bibinfo {author} {\bibfnamefont
  {W.}~\bibnamefont {Weckesser}}, \bibinfo {author} {\bibfnamefont
  {H.}~\bibnamefont {Abbasi}}, \bibinfo {author} {\bibfnamefont
  {C.}~\bibnamefont {Gohlke}}, \ and\ \bibinfo {author} {\bibfnamefont {T.~E.}\
  \bibnamefont {Oliphant}},\ }\href {\doibase 10.1038/s41586-020-2649-2}
  {\bibfield  {journal} {\bibinfo  {journal} {Nature}\ }\textbf {\bibinfo
  {volume} {585}},\ \bibinfo {pages} {357} (\bibinfo {year}
  {2020})}\BibitemShut {NoStop}%
\bibitem [{\citenamefont {{Virtanen}}\ \emph {et~al.}(2020)\citenamefont
  {{Virtanen}}, \citenamefont {{Gommers}}, \citenamefont {{Oliphant}},
  \citenamefont {{Haberland}}, \citenamefont {{Reddy}}, \citenamefont
  {{Cournapeau}}, \citenamefont {{Burovski}}, \citenamefont {{Peterson}},
  \citenamefont {{Weckesser}}, \citenamefont {{Bright}}, \citenamefont {{van
  der Walt}}, \citenamefont {{Brett}}, \citenamefont {{Wilson}}, \citenamefont
  {{Jarrod Millman}}, \citenamefont {{Mayorov}}, \citenamefont {{Nelson}},
  \citenamefont {{Jones}}, \citenamefont {{Kern}}, \citenamefont {{Larson}},
  \citenamefont {{Carey}}, \citenamefont {{Polat}}, \citenamefont {{Feng}},
  \citenamefont {{Moore}}, \citenamefont {{Vand erPlas}}, \citenamefont
  {{Laxalde}}, \citenamefont {{Perktold}}, \citenamefont {{Cimrman}},
  \citenamefont {{Henriksen}}, \citenamefont {{Quintero}}, \citenamefont
  {{Harris}}, \citenamefont {{Archibald}}, \citenamefont {{Ribeiro}},
  \citenamefont {{Pedregosa}}, \citenamefont {{van Mulbregt}},\ and\
  \citenamefont {{Contributors}}}]{scipy}%
  \BibitemOpen
  \bibfield  {author} {\bibinfo {author} {\bibfnamefont {P.}~\bibnamefont
  {{Virtanen}}}, \bibinfo {author} {\bibfnamefont {R.}~\bibnamefont
  {{Gommers}}}, \bibinfo {author} {\bibfnamefont {T.~E.}\ \bibnamefont
  {{Oliphant}}}, \bibinfo {author} {\bibfnamefont {M.}~\bibnamefont
  {{Haberland}}}, \bibinfo {author} {\bibfnamefont {T.}~\bibnamefont
  {{Reddy}}}, \bibinfo {author} {\bibfnamefont {D.}~\bibnamefont
  {{Cournapeau}}}, \bibinfo {author} {\bibfnamefont {E.}~\bibnamefont
  {{Burovski}}}, \bibinfo {author} {\bibfnamefont {P.}~\bibnamefont
  {{Peterson}}}, \bibinfo {author} {\bibfnamefont {W.}~\bibnamefont
  {{Weckesser}}}, \bibinfo {author} {\bibfnamefont {J.}~\bibnamefont
  {{Bright}}}, \bibinfo {author} {\bibfnamefont {S.~J.}\ \bibnamefont {{van der
  Walt}}}, \bibinfo {author} {\bibfnamefont {M.}~\bibnamefont {{Brett}}},
  \bibinfo {author} {\bibfnamefont {J.}~\bibnamefont {{Wilson}}}, \bibinfo
  {author} {\bibfnamefont {K.}~\bibnamefont {{Jarrod Millman}}}, \bibinfo
  {author} {\bibfnamefont {N.}~\bibnamefont {{Mayorov}}}, \bibinfo {author}
  {\bibfnamefont {A.~R.~J.}\ \bibnamefont {{Nelson}}}, \bibinfo {author}
  {\bibfnamefont {E.}~\bibnamefont {{Jones}}}, \bibinfo {author} {\bibfnamefont
  {R.}~\bibnamefont {{Kern}}}, \bibinfo {author} {\bibfnamefont
  {E.}~\bibnamefont {{Larson}}}, \bibinfo {author} {\bibfnamefont
  {C.}~\bibnamefont {{Carey}}}, \bibinfo {author} {\bibfnamefont
  {{\.I}.}~\bibnamefont {{Polat}}}, \bibinfo {author} {\bibfnamefont
  {Y.}~\bibnamefont {{Feng}}}, \bibinfo {author} {\bibfnamefont {E.~W.}\
  \bibnamefont {{Moore}}}, \bibinfo {author} {\bibfnamefont {J.}~\bibnamefont
  {{Vand erPlas}}}, \bibinfo {author} {\bibfnamefont {D.}~\bibnamefont
  {{Laxalde}}}, \bibinfo {author} {\bibfnamefont {J.}~\bibnamefont
  {{Perktold}}}, \bibinfo {author} {\bibfnamefont {R.}~\bibnamefont
  {{Cimrman}}}, \bibinfo {author} {\bibfnamefont {I.}~\bibnamefont
  {{Henriksen}}}, \bibinfo {author} {\bibfnamefont {E.~A.}\ \bibnamefont
  {{Quintero}}}, \bibinfo {author} {\bibfnamefont {C.~R.}\ \bibnamefont
  {{Harris}}}, \bibinfo {author} {\bibfnamefont {A.~M.}\ \bibnamefont
  {{Archibald}}}, \bibinfo {author} {\bibfnamefont {A.~H.}\ \bibnamefont
  {{Ribeiro}}}, \bibinfo {author} {\bibfnamefont {F.}~\bibnamefont
  {{Pedregosa}}}, \bibinfo {author} {\bibfnamefont {P.}~\bibnamefont {{van
  Mulbregt}}}, \ and\ \bibinfo {author} {\bibfnamefont {S.~.~.}\ \bibnamefont
  {{Contributors}}},\ }\href {\doibase
  https://doi.org/10.1038/s41592-019-0686-2} {\bibfield  {journal} {\bibinfo
  {journal} {Nature Methods}\ }\textbf {\bibinfo {volume} {17}},\ \bibinfo
  {pages} {261} (\bibinfo {year} {2020})}\BibitemShut {NoStop}%
\bibitem [{\citenamefont {{Astropy Collaboration}}\ \emph
  {et~al.}(2022)\citenamefont {{Astropy Collaboration}}, \citenamefont
  {{Price-Whelan}}, \citenamefont {{Lim}}, \citenamefont {{Earl}},
  \citenamefont {{Starkman}}, \citenamefont {{Bradley}}, \citenamefont
  {{Shupe}}, \citenamefont {{Patil}}, \citenamefont {{Corrales}}, \citenamefont
  {{Brasseur}}, \citenamefont {{N{"o}the}}, \citenamefont {{Donath}},
  \citenamefont {{Tollerud}}, \citenamefont {{Morris}}, \citenamefont
  {{Ginsburg}}, \citenamefont {{Vaher}}, \citenamefont {{Weaver}},
  \citenamefont {{Tocknell}}, \citenamefont {{Jamieson}}, \citenamefont {{van
  Kerkwijk}}, \citenamefont {{Robitaille}}, \citenamefont {{Merry}},
  \citenamefont {{Bachetti}}, \citenamefont {{G{"u}nther}}, \citenamefont
  {{Aldcroft}}, \citenamefont {{Alvarado-Montes}}, \citenamefont {{Archibald}},
  \citenamefont {{B{'o}di}}, \citenamefont {{Bapat}}, \citenamefont
  {{Barentsen}}, \citenamefont {{Baz{'a}n}}, \citenamefont {{Biswas}},
  \citenamefont {{Boquien}}, \citenamefont {{Burke}}, \citenamefont {{Cara}},
  \citenamefont {{Cara}}, \citenamefont {{Conroy}}, \citenamefont {{Conseil}},
  \citenamefont {{Craig}}, \citenamefont {{Cross}}, \citenamefont {{Cruz}},
  \citenamefont {{D'Eugenio}}, \citenamefont {{Dencheva}}, \citenamefont
  {{Devillepoix}}, \citenamefont {{Dietrich}}, \citenamefont {{Eigenbrot}},
  \citenamefont {{Erben}}, \citenamefont {{Ferreira}}, \citenamefont
  {{Foreman-Mackey}}, \citenamefont {{Fox}}, \citenamefont {{Freij}},
  \citenamefont {{Garg}}, \citenamefont {{Geda}}, \citenamefont {{Glattly}},
  \citenamefont {{Gondhalekar}}, \citenamefont {{Gordon}}, \citenamefont
  {{Grant}}, \citenamefont {{Greenfield}}, \citenamefont {{Groener}},
  \citenamefont {{Guest}}, \citenamefont {{Gurovich}}, \citenamefont
  {{Handberg}}, \citenamefont {{Hart}}, \citenamefont {{Hatfield-Dodds}},
  \citenamefont {{Homeier}}, \citenamefont {{Hosseinzadeh}}, \citenamefont
  {{Jenness}}, \citenamefont {{Jones}}, \citenamefont {{Joseph}}, \citenamefont
  {{Kalmbach}}, \citenamefont {{Karamehmetoglu}}, \citenamefont
  {{Ka{l}uszy{'n}ski}}, \citenamefont {{Kelley}}, \citenamefont {{Kern}},
  \citenamefont {{Kerzendorf}}, \citenamefont {{Koch}}, \citenamefont
  {{Kulumani}}, \citenamefont {{Lee}}, \citenamefont {{Ly}}, \citenamefont
  {{Ma}}, \citenamefont {{MacBride}}, \citenamefont {{Maljaars}}, \citenamefont
  {{Muna}}, \citenamefont {{Murphy}}, \citenamefont {{Norman}}, \citenamefont
  {{O'Steen}}, \citenamefont {{Oman}}, \citenamefont {{Pacifici}},
  \citenamefont {{Pascual}}, \citenamefont {{Pascual-Granado}}, \citenamefont
  {{Patil}}, \citenamefont {{Perren}}, \citenamefont {{Pickering}},
  \citenamefont {{Rastogi}}, \citenamefont {{Roulston}}, \citenamefont
  {{Ryan}}, \citenamefont {{Rykoff}}, \citenamefont {{Sabater}}, \citenamefont
  {{Sakurikar}}, \citenamefont {{Salgado}}, \citenamefont {{Sanghi}},
  \citenamefont {{Saunders}}, \citenamefont {{Savchenko}}, \citenamefont
  {{Schwardt}}, \citenamefont {{Seifert-Eckert}}, \citenamefont {{Shih}},
  \citenamefont {{Jain}}, \citenamefont {{Shukla}}, \citenamefont {{Sick}},
  \citenamefont {{Simpson}}, \citenamefont {{Singanamalla}}, \citenamefont
  {{Singer}}, \citenamefont {{Singhal}}, \citenamefont {{Sinha}}, \citenamefont
  {{Sip{H{o}}cz}}, \citenamefont {{Spitler}}, \citenamefont {{Stansby}},
  \citenamefont {{Streicher}}, \citenamefont {{{{S}}umak}}, \citenamefont
  {{Swinbank}}, \citenamefont {{Taranu}}, \citenamefont {{Tewary}},
  \citenamefont {{Tremblay}}, \citenamefont {{Val-Borro}}, \citenamefont {{Van
  Kooten}}, \citenamefont {{Vasovi{'c}}}, \citenamefont {{Verma}},
  \citenamefont {{de Miranda Cardoso}}, \citenamefont {{Williams}},
  \citenamefont {{Wilson}}, \citenamefont {{Winkel}}, \citenamefont
  {{Wood-Vasey}}, \citenamefont {{Xue}}, \citenamefont {{Yoachim}},
  \citenamefont {{Zhang}}, \citenamefont {{Zonca}},\ and\ \citenamefont
  {{Astropy Project Contributors}}}]{astropy}%
  \BibitemOpen
  \bibfield  {author} {\bibinfo {author} {\bibnamefont {{Astropy
  Collaboration}}}, \bibinfo {author} {\bibfnamefont {A.~M.}\ \bibnamefont
  {{Price-Whelan}}}, \bibinfo {author} {\bibfnamefont {P.~L.}\ \bibnamefont
  {{Lim}}}, \bibinfo {author} {\bibfnamefont {N.}~\bibnamefont {{Earl}}},
  \bibinfo {author} {\bibfnamefont {N.}~\bibnamefont {{Starkman}}}, \bibinfo
  {author} {\bibfnamefont {L.}~\bibnamefont {{Bradley}}}, \bibinfo {author}
  {\bibfnamefont {D.~L.}\ \bibnamefont {{Shupe}}}, \bibinfo {author}
  {\bibfnamefont {A.~A.}\ \bibnamefont {{Patil}}}, \bibinfo {author}
  {\bibfnamefont {L.}~\bibnamefont {{Corrales}}}, \bibinfo {author}
  {\bibfnamefont {C.~E.}\ \bibnamefont {{Brasseur}}}, \bibinfo {author}
  {\bibfnamefont {M.}~\bibnamefont {{N{"o}the}}}, \bibinfo {author}
  {\bibfnamefont {A.}~\bibnamefont {{Donath}}}, \bibinfo {author}
  {\bibfnamefont {E.}~\bibnamefont {{Tollerud}}}, \bibinfo {author}
  {\bibfnamefont {B.~M.}\ \bibnamefont {{Morris}}}, \bibinfo {author}
  {\bibfnamefont {A.}~\bibnamefont {{Ginsburg}}}, \bibinfo {author}
  {\bibfnamefont {E.}~\bibnamefont {{Vaher}}}, \bibinfo {author} {\bibfnamefont
  {B.~A.}\ \bibnamefont {{Weaver}}}, \bibinfo {author} {\bibfnamefont
  {J.}~\bibnamefont {{Tocknell}}}, \bibinfo {author} {\bibfnamefont
  {W.}~\bibnamefont {{Jamieson}}}, \bibinfo {author} {\bibfnamefont {M.~H.}\
  \bibnamefont {{van Kerkwijk}}}, \bibinfo {author} {\bibfnamefont {T.~P.}\
  \bibnamefont {{Robitaille}}}, \bibinfo {author} {\bibfnamefont
  {B.}~\bibnamefont {{Merry}}}, \bibinfo {author} {\bibfnamefont
  {M.}~\bibnamefont {{Bachetti}}}, \bibinfo {author} {\bibfnamefont {H.~M.}\
  \bibnamefont {{G{"u}nther}}}, \bibinfo {author} {\bibfnamefont {T.~L.}\
  \bibnamefont {{Aldcroft}}}, \bibinfo {author} {\bibfnamefont {J.~A.}\
  \bibnamefont {{Alvarado-Montes}}}, \bibinfo {author} {\bibfnamefont {A.~M.}\
  \bibnamefont {{Archibald}}}, \bibinfo {author} {\bibfnamefont
  {A.}~\bibnamefont {{B{'o}di}}}, \bibinfo {author} {\bibfnamefont
  {S.}~\bibnamefont {{Bapat}}}, \bibinfo {author} {\bibfnamefont
  {G.}~\bibnamefont {{Barentsen}}}, \bibinfo {author} {\bibfnamefont
  {J.}~\bibnamefont {{Baz{'a}n}}}, \bibinfo {author} {\bibfnamefont
  {M.}~\bibnamefont {{Biswas}}}, \bibinfo {author} {\bibfnamefont
  {M.}~\bibnamefont {{Boquien}}}, \bibinfo {author} {\bibfnamefont {D.~J.}\
  \bibnamefont {{Burke}}}, \bibinfo {author} {\bibfnamefont {D.}~\bibnamefont
  {{Cara}}}, \bibinfo {author} {\bibfnamefont {M.}~\bibnamefont {{Cara}}},
  \bibinfo {author} {\bibfnamefont {K.~E.}\ \bibnamefont {{Conroy}}}, \bibinfo
  {author} {\bibfnamefont {S.}~\bibnamefont {{Conseil}}}, \bibinfo {author}
  {\bibfnamefont {M.~W.}\ \bibnamefont {{Craig}}}, \bibinfo {author}
  {\bibfnamefont {R.~M.}\ \bibnamefont {{Cross}}}, \bibinfo {author}
  {\bibfnamefont {K.~L.}\ \bibnamefont {{Cruz}}}, \bibinfo {author}
  {\bibfnamefont {F.}~\bibnamefont {{D'Eugenio}}}, \bibinfo {author}
  {\bibfnamefont {N.}~\bibnamefont {{Dencheva}}}, \bibinfo {author}
  {\bibfnamefont {H.~A.~R.}\ \bibnamefont {{Devillepoix}}}, \bibinfo {author}
  {\bibfnamefont {J.~P.}\ \bibnamefont {{Dietrich}}}, \bibinfo {author}
  {\bibfnamefont {A.~D.}\ \bibnamefont {{Eigenbrot}}}, \bibinfo {author}
  {\bibfnamefont {T.}~\bibnamefont {{Erben}}}, \bibinfo {author} {\bibfnamefont
  {L.}~\bibnamefont {{Ferreira}}}, \bibinfo {author} {\bibfnamefont
  {D.}~\bibnamefont {{Foreman-Mackey}}}, \bibinfo {author} {\bibfnamefont
  {R.}~\bibnamefont {{Fox}}}, \bibinfo {author} {\bibfnamefont
  {N.}~\bibnamefont {{Freij}}}, \bibinfo {author} {\bibfnamefont
  {S.}~\bibnamefont {{Garg}}}, \bibinfo {author} {\bibfnamefont
  {R.}~\bibnamefont {{Geda}}}, \bibinfo {author} {\bibfnamefont
  {L.}~\bibnamefont {{Glattly}}}, \bibinfo {author} {\bibfnamefont
  {Y.}~\bibnamefont {{Gondhalekar}}}, \bibinfo {author} {\bibfnamefont {K.~D.}\
  \bibnamefont {{Gordon}}}, \bibinfo {author} {\bibfnamefont {D.}~\bibnamefont
  {{Grant}}}, \bibinfo {author} {\bibfnamefont {P.}~\bibnamefont
  {{Greenfield}}}, \bibinfo {author} {\bibfnamefont {A.~M.}\ \bibnamefont
  {{Groener}}}, \bibinfo {author} {\bibfnamefont {S.}~\bibnamefont {{Guest}}},
  \bibinfo {author} {\bibfnamefont {S.}~\bibnamefont {{Gurovich}}}, \bibinfo
  {author} {\bibfnamefont {R.}~\bibnamefont {{Handberg}}}, \bibinfo {author}
  {\bibfnamefont {A.}~\bibnamefont {{Hart}}}, \bibinfo {author} {\bibfnamefont
  {Z.}~\bibnamefont {{Hatfield-Dodds}}}, \bibinfo {author} {\bibfnamefont
  {D.}~\bibnamefont {{Homeier}}}, \bibinfo {author} {\bibfnamefont
  {G.}~\bibnamefont {{Hosseinzadeh}}}, \bibinfo {author} {\bibfnamefont
  {T.}~\bibnamefont {{Jenness}}}, \bibinfo {author} {\bibfnamefont {C.~K.}\
  \bibnamefont {{Jones}}}, \bibinfo {author} {\bibfnamefont {P.}~\bibnamefont
  {{Joseph}}}, \bibinfo {author} {\bibfnamefont {J.~B.}\ \bibnamefont
  {{Kalmbach}}}, \bibinfo {author} {\bibfnamefont {E.}~\bibnamefont
  {{Karamehmetoglu}}}, \bibinfo {author} {\bibfnamefont {M.}~\bibnamefont
  {{Ka{l}uszy{'n}ski}}}, \bibinfo {author} {\bibfnamefont {M.~S.~P.}\
  \bibnamefont {{Kelley}}}, \bibinfo {author} {\bibfnamefont {N.}~\bibnamefont
  {{Kern}}}, \bibinfo {author} {\bibfnamefont {W.~E.}\ \bibnamefont
  {{Kerzendorf}}}, \bibinfo {author} {\bibfnamefont {E.~W.}\ \bibnamefont
  {{Koch}}}, \bibinfo {author} {\bibfnamefont {S.}~\bibnamefont {{Kulumani}}},
  \bibinfo {author} {\bibfnamefont {A.}~\bibnamefont {{Lee}}}, \bibinfo
  {author} {\bibfnamefont {C.}~\bibnamefont {{Ly}}}, \bibinfo {author}
  {\bibfnamefont {Z.}~\bibnamefont {{Ma}}}, \bibinfo {author} {\bibfnamefont
  {C.}~\bibnamefont {{MacBride}}}, \bibinfo {author} {\bibfnamefont {J.~M.}\
  \bibnamefont {{Maljaars}}}, \bibinfo {author} {\bibfnamefont
  {D.}~\bibnamefont {{Muna}}}, \bibinfo {author} {\bibfnamefont {N.~A.}\
  \bibnamefont {{Murphy}}}, \bibinfo {author} {\bibfnamefont {H.}~\bibnamefont
  {{Norman}}}, \bibinfo {author} {\bibfnamefont {R.}~\bibnamefont {{O'Steen}}},
  \bibinfo {author} {\bibfnamefont {K.~A.}\ \bibnamefont {{Oman}}}, \bibinfo
  {author} {\bibfnamefont {C.}~\bibnamefont {{Pacifici}}}, \bibinfo {author}
  {\bibfnamefont {S.}~\bibnamefont {{Pascual}}}, \bibinfo {author}
  {\bibfnamefont {J.}~\bibnamefont {{Pascual-Granado}}}, \bibinfo {author}
  {\bibfnamefont {R.~R.}\ \bibnamefont {{Patil}}}, \bibinfo {author}
  {\bibfnamefont {G.~I.}\ \bibnamefont {{Perren}}}, \bibinfo {author}
  {\bibfnamefont {T.~E.}\ \bibnamefont {{Pickering}}}, \bibinfo {author}
  {\bibfnamefont {T.}~\bibnamefont {{Rastogi}}}, \bibinfo {author}
  {\bibfnamefont {B.~R.}\ \bibnamefont {{Roulston}}}, \bibinfo {author}
  {\bibfnamefont {D.~F.}\ \bibnamefont {{Ryan}}}, \bibinfo {author}
  {\bibfnamefont {E.~S.}\ \bibnamefont {{Rykoff}}}, \bibinfo {author}
  {\bibfnamefont {J.}~\bibnamefont {{Sabater}}}, \bibinfo {author}
  {\bibfnamefont {P.}~\bibnamefont {{Sakurikar}}}, \bibinfo {author}
  {\bibfnamefont {J.}~\bibnamefont {{Salgado}}}, \bibinfo {author}
  {\bibfnamefont {A.}~\bibnamefont {{Sanghi}}}, \bibinfo {author}
  {\bibfnamefont {N.}~\bibnamefont {{Saunders}}}, \bibinfo {author}
  {\bibfnamefont {V.}~\bibnamefont {{Savchenko}}}, \bibinfo {author}
  {\bibfnamefont {L.}~\bibnamefont {{Schwardt}}}, \bibinfo {author}
  {\bibfnamefont {M.}~\bibnamefont {{Seifert-Eckert}}}, \bibinfo {author}
  {\bibfnamefont {A.~Y.}\ \bibnamefont {{Shih}}}, \bibinfo {author}
  {\bibfnamefont {A.~S.}\ \bibnamefont {{Jain}}}, \bibinfo {author}
  {\bibfnamefont {G.}~\bibnamefont {{Shukla}}}, \bibinfo {author}
  {\bibfnamefont {J.}~\bibnamefont {{Sick}}}, \bibinfo {author} {\bibfnamefont
  {C.}~\bibnamefont {{Simpson}}}, \bibinfo {author} {\bibfnamefont
  {S.}~\bibnamefont {{Singanamalla}}}, \bibinfo {author} {\bibfnamefont
  {L.~P.}\ \bibnamefont {{Singer}}}, \bibinfo {author} {\bibfnamefont
  {J.}~\bibnamefont {{Singhal}}}, \bibinfo {author} {\bibfnamefont
  {M.}~\bibnamefont {{Sinha}}}, \bibinfo {author} {\bibfnamefont {B.~M.}\
  \bibnamefont {{Sip{H{o}}cz}}}, \bibinfo {author} {\bibfnamefont {L.~R.}\
  \bibnamefont {{Spitler}}}, \bibinfo {author} {\bibfnamefont {D.}~\bibnamefont
  {{Stansby}}}, \bibinfo {author} {\bibfnamefont {O.}~\bibnamefont
  {{Streicher}}}, \bibinfo {author} {\bibfnamefont {J.}~\bibnamefont
  {{{{S}}umak}}}, \bibinfo {author} {\bibfnamefont {J.~D.}\ \bibnamefont
  {{Swinbank}}}, \bibinfo {author} {\bibfnamefont {D.~S.}\ \bibnamefont
  {{Taranu}}}, \bibinfo {author} {\bibfnamefont {N.}~\bibnamefont {{Tewary}}},
  \bibinfo {author} {\bibfnamefont {G.~R.}\ \bibnamefont {{Tremblay}}},
  \bibinfo {author} {\bibfnamefont {M.~d.}\ \bibnamefont {{Val-Borro}}},
  \bibinfo {author} {\bibfnamefont {S.~J.}\ \bibnamefont {{Van Kooten}}},
  \bibinfo {author} {\bibfnamefont {Z.}~\bibnamefont {{Vasovi{'c}}}}, \bibinfo
  {author} {\bibfnamefont {S.}~\bibnamefont {{Verma}}}, \bibinfo {author}
  {\bibfnamefont {J.~V.}\ \bibnamefont {{de Miranda Cardoso}}}, \bibinfo
  {author} {\bibfnamefont {P.~K.~G.}\ \bibnamefont {{Williams}}}, \bibinfo
  {author} {\bibfnamefont {T.~J.}\ \bibnamefont {{Wilson}}}, \bibinfo {author}
  {\bibfnamefont {B.}~\bibnamefont {{Winkel}}}, \bibinfo {author}
  {\bibfnamefont {W.~M.}\ \bibnamefont {{Wood-Vasey}}}, \bibinfo {author}
  {\bibfnamefont {R.}~\bibnamefont {{Xue}}}, \bibinfo {author} {\bibfnamefont
  {P.}~\bibnamefont {{Yoachim}}}, \bibinfo {author} {\bibfnamefont
  {C.}~\bibnamefont {{Zhang}}}, \bibinfo {author} {\bibfnamefont
  {A.}~\bibnamefont {{Zonca}}}, \ and\ \bibinfo {author} {\bibnamefont
  {{Astropy Project Contributors}}},\ }\href {\doibase
  10.3847/1538-4357/ac7c74} {\bibfield  {journal} {\bibinfo  {journal} {apj}\
  }\textbf {\bibinfo {volume} {935}},\ \bibinfo {eid} {167} (\bibinfo {year}
  {2022})},\ \Eprint {http://arxiv.org/abs/2206.14220} {arXiv:2206.14220
  [astro-ph.IM]} \BibitemShut {NoStop}%
\bibitem [{\citenamefont {Mead}\ \emph
  {et~al.}(2021{\natexlab{b}})\citenamefont {Mead}, \citenamefont {Brieden},
  \citenamefont {Tröster},\ and\ \citenamefont {Heymans}}]{HMCode_mead_2020}%
  \BibitemOpen
  \bibfield  {author} {\bibinfo {author} {\bibfnamefont {A.}~\bibnamefont
  {Mead}}, \bibinfo {author} {\bibfnamefont {S.}~\bibnamefont {Brieden}},
  \bibinfo {author} {\bibfnamefont {T.}~\bibnamefont {Tröster}}, \ and\
  \bibinfo {author} {\bibfnamefont {C.}~\bibnamefont {Heymans}},\ }\href
  {\doibase 10.1093/mnras/stab082} {\bibfield  {journal} {\bibinfo  {journal}
  {Monthly Notices of the Royal Astronomical Society}\ }\textbf {\bibinfo
  {volume} {502}},\ \bibinfo {pages} {1401} (\bibinfo {year}
  {2021}{\natexlab{b}})},\ \bibinfo {note} {arXiv: 2009.01858}\BibitemShut
  {NoStop}%
\bibitem [{\citenamefont {Mead}\ \emph {et~al.}(2016)\citenamefont {Mead},
  \citenamefont {Heymans}, \citenamefont {Lombriser}, \citenamefont {Peacock},
  \citenamefont {Steele},\ and\ \citenamefont {Winther}}]{HMCode_mead_2016}%
  \BibitemOpen
  \bibfield  {author} {\bibinfo {author} {\bibfnamefont {A.~J.}\ \bibnamefont
  {Mead}}, \bibinfo {author} {\bibfnamefont {C.}~\bibnamefont {Heymans}},
  \bibinfo {author} {\bibfnamefont {L.}~\bibnamefont {Lombriser}}, \bibinfo
  {author} {\bibfnamefont {J.~A.}\ \bibnamefont {Peacock}}, \bibinfo {author}
  {\bibfnamefont {O.~I.}\ \bibnamefont {Steele}}, \ and\ \bibinfo {author}
  {\bibfnamefont {H.~A.}\ \bibnamefont {Winther}},\ }\href {\doibase
  10.1093/mnras/stw681} {\bibfield  {journal} {\bibinfo  {journal} {Monthly
  Notices of the Royal Astronomical Society}\ }\textbf {\bibinfo {volume}
  {459}},\ \bibinfo {pages} {1468} (\bibinfo {year} {2016})}\BibitemShut
  {NoStop}%
\bibitem [{\citenamefont {Copeland}\ \emph {et~al.}(2020)\citenamefont
  {Copeland}, \citenamefont {Taylor},\ and\ \citenamefont
  {Hall}}]{halo_bloating_term_kspace_prof}%
  \BibitemOpen
  \bibfield  {author} {\bibinfo {author} {\bibfnamefont {D.}~\bibnamefont
  {Copeland}}, \bibinfo {author} {\bibfnamefont {A.}~\bibnamefont {Taylor}}, \
  and\ \bibinfo {author} {\bibfnamefont {A.}~\bibnamefont {Hall}},\ }\href
  {\doibase 10.1093/mnras/staa314} {\bibfield  {journal} {\bibinfo  {journal}
  {Monthly Notices of the Royal Astronomical Society}\ }\textbf {\bibinfo
  {volume} {493}},\ \bibinfo {pages} {1640} (\bibinfo {year}
  {2020})}\BibitemShut {NoStop}%
\bibitem [{\citenamefont {Smith}\ \emph {et~al.}(2003)\citenamefont {Smith},
  \citenamefont {Peacock}, \citenamefont {Jenkins}, \citenamefont {White},
  \citenamefont {Frenk}, \citenamefont {Pearce}, \citenamefont {Thomas},
  \citenamefont {Efstathiou},\ and\ \citenamefont
  {Couchman}}]{one_halo_term_damping_reason}%
  \BibitemOpen
  \bibfield  {author} {\bibinfo {author} {\bibfnamefont {R.~E.}\ \bibnamefont
  {Smith}}, \bibinfo {author} {\bibfnamefont {J.~A.}\ \bibnamefont {Peacock}},
  \bibinfo {author} {\bibfnamefont {A.}~\bibnamefont {Jenkins}}, \bibinfo
  {author} {\bibfnamefont {S.~D.~M.}\ \bibnamefont {White}}, \bibinfo {author}
  {\bibfnamefont {C.~S.}\ \bibnamefont {Frenk}}, \bibinfo {author}
  {\bibfnamefont {F.~R.}\ \bibnamefont {Pearce}}, \bibinfo {author}
  {\bibfnamefont {P.~A.}\ \bibnamefont {Thomas}}, \bibinfo {author}
  {\bibfnamefont {G.}~\bibnamefont {Efstathiou}}, \ and\ \bibinfo {author}
  {\bibfnamefont {H.~M.~P.}\ \bibnamefont {Couchman}},\ }\href {\doibase
  10.1046/j.1365-8711.2003.06503.x} {\bibfield  {journal} {\bibinfo  {journal}
  {Monthly Notices of the Royal Astronomical Society}\ }\textbf {\bibinfo
  {volume} {341}},\ \bibinfo {pages} {1311} (\bibinfo {year}
  {2003})}\BibitemShut {NoStop}%
\bibitem [{\citenamefont {Emberson}\ \emph {et~al.}(2017)\citenamefont
  {Emberson} \emph {et~al.}}]{Emberson:2016ecv}%
  \BibitemOpen
  \bibfield  {author} {\bibinfo {author} {\bibfnamefont {J.~D.}\ \bibnamefont
  {Emberson}} \emph {et~al.},\ }\href {\doibase 10.1088/1674-4527/17/8/85}
  {\bibfield  {journal} {\bibinfo  {journal} {Res. Astron. Astrophys.}\
  }\textbf {\bibinfo {volume} {17}},\ \bibinfo {pages} {085} (\bibinfo {year}
  {2017})},\ \Eprint {http://arxiv.org/abs/1611.01545} {arXiv:1611.01545
  [astro-ph.CO]} \BibitemShut {NoStop}%
\bibitem [{\citenamefont {Villaescusa-Navarro}\ \emph
  {et~al.}(2013)\citenamefont {Villaescusa-Navarro}, \citenamefont {Bird},
  \citenamefont {Peña-Garay},\ and\ \citenamefont
  {Viel}}]{neutrino_halo_profile}%
  \BibitemOpen
  \bibfield  {author} {\bibinfo {author} {\bibfnamefont {F.}~\bibnamefont
  {Villaescusa-Navarro}}, \bibinfo {author} {\bibfnamefont {S.}~\bibnamefont
  {Bird}}, \bibinfo {author} {\bibfnamefont {C.}~\bibnamefont {Peña-Garay}}, \
  and\ \bibinfo {author} {\bibfnamefont {M.}~\bibnamefont {Viel}},\ }\href
  {\doibase 10.1088/1475-7516/2013/03/019} {\bibfield  {journal} {\bibinfo
  {journal} {Journal of Cosmology and Astroparticle Physics}\ }\textbf
  {\bibinfo {volume} {2013}},\ \bibinfo {pages} {019} (\bibinfo {year}
  {2013})}\BibitemShut {NoStop}%
\bibitem [{\citenamefont {Schmidt}(2016)}]{correction_HMF}%
  \BibitemOpen
  \bibfield  {author} {\bibinfo {author} {\bibfnamefont {F.}~\bibnamefont
  {Schmidt}},\ }\href {\doibase 10.1103/PhysRevD.93.063512} {\bibfield
  {journal} {\bibinfo  {journal} {Physical Review D}\ }\textbf {\bibinfo
  {volume} {93}},\ \bibinfo {pages} {063512} (\bibinfo {year}
  {2016})}\BibitemShut {NoStop}%
\end{thebibliography}%

\end{document}